\documentclass[aps,prd,a4paper,final,showpacs,twocolumn]{revtex4}
\usepackage{amsmath}
\usepackage{amssymb}
\usepackage{float}
\usepackage[dvips]{graphics}
\usepackage[pdftex]{graphicx}
\newcommand{\ba}{\left( \begin{array}{cc}}
\newcommand{\ea} {\end{array} \right)}

\newcommand{\be}{\begin{eqnarray}}
\newcommand{\ee}{\end{eqnarray}}
\newcommand{\dd}{\, {\rm d}}

\newcommand{\M}{\mathcal{M}}
\newcommand{\dM}{\partial\mathcal{M}}

\begin{document}

\title{A Testable Solution of the Cosmological Constant and Coincidence
Problems}
\author{Douglas J. Shaw}
\email{d.shaw@damtp.ac.uk}
\author{John D. Barrow}
\email{j.d.barrow@damtp.cam.ac.uk}
\affiliation{DAMTP, Centre for Mathematical Sciences, Cambridge CB3 0WA,
United Kingdom}
\date{\today}

\begin{abstract}
We present a new solution to the cosmological constant (CC) and coincidence problems in which the observed value of the CC, $\Lambda$, is linked to other observable properties of the universe.  This is achieved by promoting the CC from a parameter which must to specified, to a field which can take many possible values. The observed value of $\Lambda \approx (9.3\,{\rm Gyrs})^{-2}$  ($\approx 10^{-120}$ in Planck units) is determined by a new constraint equation which follows from the application of a causally restricted variation principle. 
When applied to our visible universe, the model makes a testable prediction for the dimensionless spatial curvature of $\Omega_{\rm k0} = -0.0056 (\zeta_{\rm b}/0.5)$; where $\zeta_{\rm b} \sim 1/2$ is a QCD parameter. Requiring that a classical history exist, our model determines the probability of observing a given $\Lambda$. The observed CC value, which we successfully predict, is typical within our model even before the effects of anthropic selection are included. When anthropic selection effects are accounted for, we find that the observed coincidence between $t_{\Lambda} = \Lambda^{-1/2}$ and the age of the universe, $t_{\rm U}$, is a typical occurrence in our model.   In contrast to multiverse explanations of the CC problems, our solution is independent of the choice of a prior weighting of different $\Lambda$-values and does not rely on anthropic selection effects.    Our model includes no unnatural small parameters and does not require the introduction of  new dynamical scalar fields or modifications to general relativity, and it can be tested by astronomical observations in the near future. 
\end{abstract}

\maketitle



\section{Introduction}

The cosmological constant (CC), $\lambda $, was first introduced by Einstein
in 1917 \cite{Einstein:1917aa} to ensure that his new general theory of
relativity admitted a static cosmological solution. The introduction of $%
\lambda $ required only the addition of the divergence-free term $-\lambda
g_{\mu \nu }$ to the original field equations: 
\begin{equation*}
G^{\mu \nu }=R^{\mu \nu }-\frac{1}{2}Rg^{\mu \nu }=8\pi GT^{\mu \nu
}\rightarrow G^{\mu \nu }=8\pi GT^{\mu \nu }-\lambda g^{\mu \nu },
\end{equation*}%
where $R_{\mu \nu }$ is the Ricci curvature of $g_{\mu \nu }$, and $T_{\mu
\nu }$ is the energy-momentum tensor of matter. It was not an easy matter to
unambiguously interpret the astronomical data concerning galaxy motions and
attribute them to systematic recession rather than a steady lateral drift.
The first observations of galaxy redshifts were made by Slipher in 1912 \cite%
{Slipher:1913ab}. By 1917, Slipher had measured the redshifts of 25 spiral
galaxies; all but four of them were found to be receding from us \cite%
{Slipher:1917ab}. In 1917 de Sitter \cite{deSit} found an empty expanding
solution with a $\lambda $ term present and in the early 1920s, Friedmann
discovered a class of homogeneous and isotropic cosmological solutions of
general relativity without a $\lambda $ term. These cosmological models were
not static but could either expand or contract. Lema\^{\i}tre found a wide
range of expanding and contracting universes, both with and without $\lambda 
$ in 1927 and also predicted the theoretical relationship between
distance and redshift in an expanding universe \cite{Lemaitre:1927ab, Kragh:1996ab}.
Notably, Lema\^{\i}tre proposed that an expanding universe could explain the
velocities of galaxies first measured by Slipher and first deduced what
became known as `Hubble's Law' (unfortunately the translation \cite%
{Lemaitre:1927ab} omitted the crucial footnote where it appears in the
original). Two years later, Hubble and Humason empirically derived the
redshift-distance relation \cite{Hubble:1929ab}. This led to the static
universe model, which first motivated Einstein to introduce $\lambda $,
being abandoned in favour of the now familiar expanding universe cosmology.
Lema\^{\i}tre had also demonstrated the instability of the static universe
model with respect to conformal perturbations. Unaware of Lema\^{\i}tre's
work, Eddington had also proved the instability of the static universe
against density perturbations \cite{edd} (the full stability analysis was
only completed in 2003 and can be found in \cite{barellis}). However, some
scientists, notably Eddington, believed that $\lambda $ was a n essential
part of general relativity because it offered a possible link between
gravitation and microphysics \cite{edd2, edd3}.

Whilst the original motivation for a CC evaporated, it was later appreciated
that there were other, more fundamental reasons for its presence (see e.g.
Ref. \cite{Zeldovich:1967gd} for a discussion of this and for a modern
review see Ref. \cite{Weinberg:1988cp}). Quantum fluctuations result in a
vacuum energy, $\rho _{\mathrm{vac}}$, which contributes to the expected
value of the energy momentum tensor of matter: 
\begin{equation*}
\left\langle T^{\mu \nu }\right\rangle =T_{\mathrm{m}}^{\mu \nu }-\rho _{%
\mathrm{vac}}g^{\mu \nu },
\end{equation*}%
where $T_{\mathrm{m}}^{\mu \nu }$ vanishes \emph{in vacuo}. The quantum
expectation of the energy-momentum tensor, $\left\langle T^{\mu \nu
}\right\rangle $, acts as a source for the Einstein tensor. Hence, we have: 
\begin{equation*}
G^{\mu \nu }=8\pi GT_{\mathrm{m}}^{\mu \nu }-\Lambda g^{\mu \nu },\qquad
\Lambda =\lambda +8\pi G\rho _{\mathrm{vac}}.
\end{equation*}%
It is clear from this that the vacuum energy, $\rho _{\mathrm{vac}}$,
provides a contribution, $8\pi G\rho _{\mathrm{vac}}$, to the effective
cosmological constant, $\Lambda $. Even if the `bare' cosmological constant
is assumed to vanish, $\lambda =0$, the effective cosmological constant will
generally be non-zero. Requiring that $\Lambda =0$ means there must be an
exact cancellation of the `bare' cosmological constant, $\lambda $, and the
vacuum energy stress, $8\pi G\rho _{\mathrm{vac}}$.

Formally, the value of $\rho _{\mathrm{vac}}$ predicted by a general quantum
field theory in a flat Minkowski space background is infinite. If we assume
that the field theory is only valid up to some energy scale $M_{\ast }$,
then there is a contribution to $\rho _{\mathrm{vac}}$ of $O(M_{\ast }^{4})$%
. Collider experiments have established that the standard model is accurate
up to energy scales $M_{\ast }\gtrsim O(M_{\mathrm{EW}})$ where $M_{\mathrm{%
EW}}\approx 246\,\mathrm{GeV}$ is the electroweak scale. We would therefore
expect $\rho _{\mathrm{vac}}$ to be at least $O(M_{\mathrm{EW}}^{4})$.

In the absence of any new physics between the electroweak and the Planck
scale, $M_{\mathrm{pl}}=2.4\times 10^{18}\,\mathrm{GeV}$, where quantum
fluctuations in the gravitational field can no longer be safely neglected,
we would expect $\rho _{\mathrm{vac}}\sim O(M_{\mathrm{pl}}^{4})$.
Astrophysical observations do, however, strongly suggest there exists some
new form of dark, weakly interacting matter, beyond that described by the
standard model. The most developed theoretical extensions of the standard
model, which include candidates for this dark matter, introduce an
additional supersymmetry (SUSY) between fermions and bosons. If
supersymmetry were an unbroken symmetry of Nature, the quantum contributions
to the vacuum energy would all exactly cancel leaving $\rho _{\mathrm{vac}%
}=0\Rightarrow \Lambda =\lambda $. However, our universe is not
supersymmetric today, and so SUSY must have been broken at some energy
scale $M_{\mathrm{SUSY}},$ where $1\,\mathrm{TeV}\lesssim M_{\mathrm{SUSY}%
}\lesssim M_{\mathrm{pl}}$ and so we expect $\rho _{\mathrm{vac}}\sim
O(M_{\mathrm{SUSY}}^{4})$.

Given the standard model of particle physics and reasonable extensions of
it, a $\rho _{\mathrm{vac}}$ somewhere between $M_{\mathrm{EW}}^{4}$ and $M_{%
\mathrm{pl}}^{4}$ appears unavoidable. Furthermore, in the absence of exact
cancellations, we would expect the effective vacuum energy,
\begin{equation*}
\rho _{\mathrm{vac}}^{\mathrm{eff}}=\frac{\lambda }{8\pi G}+\rho _{\mathrm{%
vac}}\equiv \frac{\Lambda }{8\pi G},
\end{equation*}%
to be no smaller than $\rho _{\mathrm{vac}}$, giving an estimate of $\rho _{%
\mathrm{vac}}^{\mathrm{eff}}\gtrsim O(M_{\mathrm{EW}}^{4})$. This cannot,
however, be the case.

The expansion rate of our universe is sensitive to  $\rho _{\mathrm{vac}%
}^{\mathrm{eff}}$, or equivalently $\Lambda $, through Einstein's equations.
Measurements of this expansion rate have established that $\left( \rho _{%
\mathrm{vac}}^{\mathrm{eff}}\right) ^{1/4}\approx 2.4\pm 0.3\times 10^{-12}\,%
\mathrm{GeV}$ \cite{Komatsu:2010fb} and this implies that $\rho _{\mathrm{vac%
}}^{\mathrm{eff}}$ is some $10^{60}-10^{120}$ times smaller than the
expected contribution from quantum fluctuations.

This gives rise to \emph{the cosmological constant problem}: ``Why is the
measured effective vacuum energy or cosmological constant so much smaller
than the expected contributions to it from quantum fluctuations?''
Equivalently, assuming the estimate of $\rho _{\mathrm{vac}}$ from quantum
fluctuations is accurate: ``Why does the approximate equality $\lambda \approx
-8\pi G\rho _{\mathrm{vac}}$ hold good to an accuracy of somewhere between $%
60$ to $120$ decimal places?'' A fuller exposition and
review of the cosmological constant problem and earlier attempts at its
solution can be found in Weinberg in Ref. \cite{Weinberg:1988cp}.

Observations of the cosmic microwave background (CMB) \cite{Komatsu:2010fb},
Type Ia Supernovae (SNe Ia) \cite%
{Riess:1998cb,Perlmutter:1998np,Hicken:2009dk,Hicken:2009df}, and large
scale structure (LSS) \cite{AdelmanMcCarthy:2005se,Tegmark:2006az} all
strongly prefer a small (but non-zero) value for $\rho _{\mathrm{vac}}^{%
\mathrm{eff}}$: specifically, $\rho _{\mathrm{vac}}^{\mathrm{eff}}=(3.8\pm
0.2)\times 10^{-6}\,\mathrm{GeV}\,\mathrm{cm}^{-3}$ \cite{Komatsu:2010fb}.
This presents an additional conundrum; it is easier to conceive a
situation where $|\rho _{\mathrm{vac}}+\Lambda /8\pi G|$ is exactly zero,
than one in which the cancellation between the two terms is very nearly
exact. This is related to the \emph{coincidence problem} which we describe
in more detail below.

The presence of an (effective) cosmological constant, $\Lambda $, introduces
a fixed time-scale: $t_{\Lambda }=\Lambda ^{-1/2}$. Curiously, the observed
value of $t_{\Lambda }\approx 9.3\,\mathrm{Gyrs}$ is of the same order as
the age of universe today $t_{\mathrm{U}}\approx 13.7\,\mathrm{Gyrs}$. This
gives rise to the \emph{coincidence problem}: \textquotedblleft Why is $%
t_{\Lambda }\sim t_{U}$ today?\textquotedblright\ The epoch at which we observe the universe is conditioned by the
requirement that the universe be old enough for typical stars to have
experienced a period of stable hydrogen burning and then produce the
heavier elements required for biological complexity \cite{dicke}. The
characteristic time-scale, $t_{\ast }$, over which this occurs is determined
by a combination of the constants of nature: $t_{\ast }\sim \alpha _{\mathrm{%
em}}^{2}/Gm_{\mathrm{p}}m_{\mathrm{e}}=5.7\,\mathrm{Gyrs}$ \cite{btip}.
Naturally, one expects that $t_{\mathrm{U}}\sim O(1)t_{\ast }$, which is
indeed the case. Thus, the coincidence problem can be alternatively viewed
as the coincidence of two fundamental time scales, $t_{\Lambda }$ and $%
t_{\ast }$, determined entirely by fundamental constants of Nature. The
coincidence problem is then simply \textquotedblleft Why is $t_{\Lambda }\sim
t_{\ast }$?\textquotedblright

The coincidence problem is puzzling because it implies that we live at a
special epoch $t_{U}$ when, by chance $t_{\Lambda }\sim O(t_{U}\sim t_{\ast
})$, or that there is some deep reason, related to the solution of the
cosmological constant problem, why $\Lambda $ is such that $t_{\Lambda }\sim
t_{\ast }(\sim t_{U})$.

Recently, in the field of cosmology, there has been more literature
addressing the coincidence problem than the cosmological constant problem.
It is generally assumed (or perhaps hoped) that there is a dynamical mechanism that
ensures that $\rho _{\mathrm{vac}}^{\mathrm{eff}}=\rho _{\mathrm{vac}%
}+\lambda /8\pi G$ vanishes exactly. The observed effective cosmological
constant then comes about due to some other mechanism e.g. the energy
density of a slowly rolling scalar field. In dark energy models, for
instance, the effective cosmological constant is not actually constant.
Instead there is a additional field (the eponymous dark energy) whose energy
density has caused the expansion of the universe to accelerate in such a way
that is, up to current measurement accuracy, indistinguishable from the
effect of a cosmological constant. Whilst some dark energy models can
alleviate the coincidence problem, they invariably feature a high degree of
fine tuning to ensure that the transition to a dark energy dominated
expansion occurs at a time scale $\sim O(t_{\ast })$.

Although in principle it seems natural for $\Lambda $ to be significantly
larger than $t_{\mathrm{U}}^{-2}$ (i.e. $t_{\Lambda }\ll t_{\mathrm{U}}$),
first Barrow and Tipler \cite{btip}, and then Weinberg \cite{sw} and
Efstathiou \cite{ef}, showed that 'observers' similar to ourselves could not
exist if this were the case. Our existence requires that small
inhomogeneities in the early universe are able to grow by gravitational
instability so as to form galaxies and stars. If $\Lambda $ is too large
this cannot occur: gravitational instabilities turn off once the universe
starts accelerating. The requirement that galaxies and stars exist places an
anthropic upper-bound on observable values of $\Lambda $ equivalent to $%
t_{\Lambda }\gtrsim 0.7\,\mathrm{Gyrs}$ \cite{btip}. If there is only one
universe, with one value of $\Lambda $, the anthropic constraint on $\Lambda 
$ brings us no closer to understanding why $\Lambda $ is so small (although
if some 'constants vary cosmologically there is the possibility that a small
non-zero $\Lambda $ might be anthropically necessary in order to switch off
variations in constants before they stop atoms from existing, see ref \cite%
{bms}). However, if there are many possible universes (or a 'multiverse')
each with different values of $\Lambda $, then our universe could only ever
be in the (possibly small) subset of universes where $t_{\Lambda }\gtrsim
0.7\,\mathrm{Gyrs}$.

If we knew the prior probability distribution, $f_{\mathrm{prior}}(\Lambda )$%
, of values of $\Lambda $ in such a multiverse, one could then calculate the
conditional probability of finding $t_{\Lambda }\sim t_{\mathrm{U}}$ given
the requirement that observers such as ourselves exist. Weinberg \cite{sw}
noted that if $f_{\mathrm{prior}}(\Lambda )\approx \mathrm{const}$ for $%
t_{\Lambda }\gtrsim 0.7\,\mathrm{Gyrs}$, we would typically expect $t_{%
\mathrm{\Lambda }}\sim \mathrm{few}\times 0.7\,\mathrm{Gyrs}$. The observed
value of $t_{\Lambda }\sim 9.3\,\mathrm{Gyrs}$ would then look fairly
reasonable, and one could argue that the cosmological constant and
coincidence problems had been solved. This $f_{\mathrm{prior}}(\Lambda )$
corresponds to an approximately uniform distribution of $\Lambda $ values
smaller than the anthropic upper-bound. Such an $f(\Lambda )$ is not,
however, the only reasonable possibility for the prior distribution. If, for
instance $\Lambda =M_{\mathrm{pl}}e^{\phi },$ and values of $\phi $ were
uniformly distributed in the multiverse, one would naturally expect $\Lambda 
$ to be much smaller than the anthropic upper-bound i.e. $t_{\Lambda }\gg t_{%
\mathrm{U}}$ (and $f_{\mathrm{prior}}(\Lambda )\propto \Lambda ^{-1}$).
Before we had observations consistent with a non-zero value of $\Lambda $,
Coleman \cite{col1, col2} and Hawking \cite{Hawking:1984hk}, and later Ng
and van Dam \cite{Ng:1990ab}, used euclidean approaches to quantum gravity
to argue that the distribution of $\Lambda $ values should be strongly
peaked about $\Lambda =0$ (i.e. $f_{\mathrm{prior}}(\Lambda )=\exp (3\pi
/G\Lambda )$) with a form that is interestingly characteristic of a
Fisher-Tippett extreme-value distribution \cite{bb}. Again, this would make $%
t_{\Lambda }\sim t_{\mathrm{U}}$ seem highly unnatural.

Ultimately, we would like to calculate $f_{\mathrm{prior}}(\Lambda )$ from
some fundamental theory. Currently, the notion of a multiverse with
different values of $\Lambda $ seems to have a natural realization in the
some $10^{500}$ different vacua of string theory (see e.g. Ref. \cite%
{Polchinski:2006gy}). A derivation of $f_{\mathrm{prior}}(\Lambda )$ in this
landscape of string vacua for those vacua compatible with life still
represents a major theoretical challenge. Common criticisms of anthropic
selection in a multiverse as a explanation of the CC problems are that it is
not clear that observers similar to ourselves are the only potential
observers we should consider when restricting possible values of $\Lambda $;
or that this explanation, as it is currently understood, makes no sharp
predictions that can be tested by observations.

Ideally, we would like to find explanations of the cosmological constant and
coincidence problems that are natural, in the sense of requiring little or
no fine tuning, and are, at least in principle, falsifiable by future
observations. In this paper we propose such a solution. Formally, we propose
a paradigm which can be applied to a variety of models, including extensions
of general relativity and extra dimensions, and sometimes in a number of
different ways. This paradigm establishes a new field equation for the bare
cosmological constant $\lambda $ which determines its value in terms of
other properties of the observed universe. Crucially, one finds the
effective cosmological constant, $\Lambda $, which is a sum of the bare
cosmological constant and quantum fluctuations, to be of the observed order
of magnitude $\Lambda \sim O(t_{U}^{-2})$. When our proposal is applied to
general relativity, $\Lambda $ is not seen to evolve (i.e. it is constant
throughout the universe). Hence, the resulting cosmology is indistinguishable
from general relativity with the value of $\Lambda $ put in by hand.
However, any given application of our theory produces a firm prediction
for $\Lambda $ in terms of other measurable quantities. If the actual value
of $\Lambda $ deviates from this predicted value then that particular
application of the paradigm is ruled out. It should be stressed that our
paradigm is equally applicable to models where general relativity is
modified in some way, or where there are more than four dimensions.
In such theories, the order of magnitude of the predicted effective
cosmological constant is generally the same as it is in 3+1 general
relativity.


The rest of this paper is laid out as follows: We specify and describe our
new scheme to solve the cosmological constant problems in \S \ref%
{sec:paradigm}. In \S \ref{sec:cosmology}, we apply it to a
realistic model of our universe. We find that the predicted value of $%
\Lambda $ depends in detail on the spatial curvature and energy density of
baryonic matter. Given the measured value of $\Lambda $, this results in a
prediction for the spatial curvature of the observable universe if our
scenario is the correct explanation for the observed value of $\Lambda $. 

In inflationary scenarios, different regions of the universe undergo different
amounts of inflation (measured by the number of e-folds, $N$). The
observed spatial curvature scales as $\exp (-2N)$ following inflation, and so the
spatial curvature would be different in each bubble universe according to
the amount of inflation it experiences. Our model therefore provides a link
between the probability of living in a bubble universe where  a
given value of the cosmological constant is observed and the duration of inflation in
that bubble. In \S \ref{sec:cosmology:natural} we calculate the probability
of living in a bubble universe where $t_{\Lambda }$ coincides with $t_{%
\mathrm{U}}$ and find that, in our model it is indeed a typical occurrence.
Our conclusions, together with a list of answers to some possible questions
about our scheme and its application to cosmology, are found in \S \ref%
{sec:conclusion}. Some detailed background calculations are presented in the
appendices. We have provided a condensed presentation of our proposal in Ref. \cite{shortversion}.  We work throughout with a metric signature $(-+++)$ and units
where $c=\hbar =1$; we denote $\kappa =8\pi G$.

\section{A Proposal for Solving the CC Problems}

\label{sec:paradigm} In this section we propose a new approach to
solve the cosmological constant (CC) problems without fine tuning.

\paragraph*{Preliminaries:}

We begin with some preliminary definitions. We will take the total action of
the universe defined on a manifold $\mathcal{M}$, and with effective
cosmological constant $\Lambda ,$ to be $I_{\mathrm{tot}}[g_{\mu \nu },\Psi
^{a},\Lambda ;\mathcal{M}]$, where $\Psi ^{a}$ are the matter fields and $%
g_{\mu \nu }$ is the metric field. We define $\partial \mathcal{M}=\partial 
\mathcal{M}_{I}\cup \partial \mathcal{M}_{u}$ where $\partial \mathcal{M}_{I}
$ denotes some initial hypersurface, and $\partial \mathcal{M}_{u}$ denotes
the rest of $\partial \mathcal{M}$.

As usual, provided certain quantities are held fixed on $\partial \mathcal{M}
$, the classical field equations result from the requirement that $I_{%
\mathrm{tot}}[g_{\mu \nu },\Psi ^{a},\Lambda ;\mathcal{M}]$ be stationary
with respect to small variations in $g_{\mu \nu }$ and $\Psi ^{a}$. We
represent the classical field equations for $g_{\mu \nu }$ and $\Psi ^{a}$
by $E^{\mu \nu }=0$ and $\Phi _{a}=0,$ respectively. The quantities that
must be held fixed on $\partial \mathcal{M}$ depend on the surface terms in $%
I_{\mathrm{tot}}$. For instance, it is well known that we can introduce the
Gibbons-Hawking-York (GHY) surface term on $\partial \mathcal{M}$ \cite%
{York:1972sj, Gibbons:1976ue} into $I_{\mathrm{tot}}$ so that the only
quantities that need to be fixed on the boundary are the fields $\Psi ^{a}$
and the induced 3-metric, $\gamma _{\mu \nu }$, on $\partial \mathcal{M}$.

In general, the quantities that must be held fixed cannot be freely
specified on $\partial \mathcal{M}$. The classical fields generally imply
consistency conditions that must be satisfied by these quantities. This is
particularly the case if some parts of $\partial \mathcal{M}$ are causally
connected to other parts. For instance, if $\partial \mathcal{M}_{I}$
represents a Cauchy surface for $\partial \mathcal{M}_{u}$, then the $\gamma
_{\mu \nu }$ and $\Psi ^{a}$ on $\partial \mathcal{M}_{u}$ will be at least
partially determined by the specification of the initial data on $\partial 
\mathcal{M}_{I}$ and by the field equations $E^{\mu \nu }=0$ and $\Phi
_{a}=0 $.

We define $\left\{ Q^{A}\right\} $ to be a minimal set of quantities that
need be freely specified on $\partial \mathcal{M}$ and held fixed, such that 
$I_{\mathrm{tot}}$ is a stationary point with respect to variations in $%
g_{\mu \nu }$ and $\Psi ^{a}$. For definiteness, we consider the total
action with GHY surface term and focus on the variation of the metric. For
an unconstrained metric variation we have: 
\begin{eqnarray}
\delta I_{\mathrm{tot}}&=&\frac{1}{2\kappa }\int_{\partial \mathcal{M}}\sqrt{%
|\gamma |}\,\mathrm{d}^{3}x\,N^{\mu \nu }\delta \gamma _{\mu \nu } \nonumber \\ &&+\frac{1}{%
2\kappa }\int_{\mathcal{M}}\sqrt{-g}\,\mathrm{d}^{4}x\,E^{\mu \nu }\delta
g_{\mu \nu }, \nonumber
\end{eqnarray}
for some tensor $N^{\mu \nu }$. We hold some $\left\{ Q^{A}\right\} $ fixed
and decompose the variations in $g_{\mu \nu }$ into $g_{\mu \nu }=g_{\mu \nu
}^{(0)}+\delta g_{\mu \nu }^{(\mathcal{M})}+\delta g_{\mu \nu }^{(\partial 
\mathcal{M})}$. We define $\delta \gamma _{\mu \nu }^{(\mathcal{M})}$ and $%
\delta \gamma _{\mu \nu }^{(\partial \mathcal{M})}$ respectively to be the
projections of $\delta g_{\mu \nu }^{(\mathcal{M})}$ and $\delta g_{\mu \nu
}^{(\partial \mathcal{M})}$ onto $\partial \mathcal{M}$. The decomposition
of the metric variation is performed so that $\delta \gamma _{\mu \nu }^{(%
\mathcal{M})}=0$ and a priori $\delta \gamma _{\mu \nu }^{(\partial \mathcal{%
M})}\neq 0$. We write $\bar{g}_{\mu \nu }=g_{\mu \nu }^{(0)}+\delta g_{\mu
\nu }^{(\partial \mathcal{M})}$. Minimizing the action with respect to
fluctuations $\delta g_{\mu \nu }^{(\mathcal{M})}$ that vanish when
projected onto $\partial \mathcal{M}$ requires: $E^{\mu \nu }[\bar{g}_{\mu
\nu }]=0$. This equation, combined with the fixed $\left\{ Q^{A}\right\} $,
constrains the form of $\delta \gamma _{\mu \nu }^{(\partial \mathcal{M})}$.
We require that fixing the set $\left\{ Q^{A}\right\} $ and imposing $%
E^{\mu \nu }[\bar{g}_{\mu \nu }]=0$ are sufficient to determine that, with
fixed $\Lambda $, $N^{\mu \nu }\delta \gamma _{\mu \nu }=N^{\mu \nu }\delta
\gamma _{\mu \nu }^{(\partial \mathcal{M})}\cong 0$, where $\cong 0$ here
indicates that $N^{\mu \nu }\delta \gamma _{\mu \nu }$ is a total derivative
and hence $\delta I_{\mathrm{tot}}=0$ when $E^{\mu \nu }=0$. Usually this
implies that $E^{\mu \nu }=0$ and the fixed $\left\{ Q^{A}\right\} $
completely fix the induced metric ($\gamma _{\mu \nu }$) up to
diffeomorphisms of $\partial \mathcal{M}$. This is just a restatement of the
usual variational principle.

The gravitational field equations, $E^{\mu \nu }=0$, depend on the
(effective) cosmological constant $\Lambda $. It follows that the metric $%
\gamma _{\mu \nu }$ on $\partial \mathcal{M}$ determined by $\left\{
Q^{A}\right\} $ and $E^{\mu \nu }=0$ depends on $\Lambda $. Usually, $%
\Lambda $ is treated as a fixed parameter, either put in by hand or picked
from a distribution of different values in a multiverse. With $\Lambda $
fixed, the $\left\{ Q^{A}\right\} $ and the equations $E^{\mu \nu }=0$ then
fix $\gamma _{\mu \nu },$ and hence imply $\delta \gamma _{\mu \nu }=0$.
However, if $\Lambda $ is varied by some small amount $\delta \Lambda $, one
would have: 
\begin{equation*}
\delta \gamma _{\mu \nu }=\mathcal{H}_{\mu \nu }\delta \Lambda ,
\end{equation*}%
where 
\begin{equation*}
\mathcal{H}_{\mu \nu }=\left. \frac{\delta \gamma _{\mu \nu }}{\delta
\Lambda }\right\vert _{E^{\mu \nu }=\Phi _{a}=0;\left\{ Q^{A}\right\} }.
\end{equation*}%
Similarly, for the matter fields, we can define $\mathcal{P}^{a}$ by: 
\begin{equation*}
\mathcal{P}^{a}=\left. \frac{\delta \Psi ^{a}}{\delta \Lambda }\right\vert
_{E^{\mu \nu }=\Phi _{a}=0;\left\{ Q^{A}\right\} }.
\end{equation*}

\paragraph*{A New Proposal:}

Given the definitions above, our proposal for solving the CC problems is as
follows:

\begin{itemize}
\item We promote the bare cosmological constant, $\lambda $, from a fixed
parameter to a field (albeit one that is constant in space and time).
Quantum mechanically, the partition function of the universe (see \S \ref%
{sec:prop:part} below) includes a sum over all possible values of $\lambda $
in addition to the usual sum over configurations of $g_{\mu \nu }$ and $\Psi
^{a}$. The effective cosmological constant, $\Lambda $, is equal to $\lambda
+\mathrm{const}$ and so a sum over all possible values of $\lambda $ is
equivalent to a sum over all $\Lambda $. This sum over $\Lambda $ is defined
up to an unknown weighting function, $\mu \lbrack \Lambda ]$, which is
similar to the prior weighting of different $\Lambda $ in multiverse models.

\item We  sum over configurations of $g_{\mu\nu}$ and $\Psi^{a}$ keeping some data $\left\lbrace Q^{A}\right\rbrace$ fixed on the boundary, $%
\partial\mathcal{M}$, of the manifold $\mathcal{M}$ on which the action, $I_{%
\mathrm{tot}}$, is defined.

\item The classical field equations are found by requiring that $\delta I_{%
\mathrm{tot}}=0$ with respect to variations in the fields that preserve $%
\left\{ Q^{A}\right\} $. For variations of $g_{\mu \nu }$ and $\Psi ^{a},$
this gives respectively $E^{\mu \nu }=0$ and $\Phi _{a}=0$. The classical
value of the effective cosmological constant is now determined by the
requirement that $I_{\mathrm{tot}}$ be stationary with respect to variations
in $\Lambda $ i.e.
\begin{equation*}
\left. \frac{\delta I_{\mathrm{tot}}}{\delta \Lambda }\right\vert _{\left\{
Q^{A}\right\} }=0.
\end{equation*}%
Crucially, this new field equation for $\Lambda $ includes contributions
from the variation of the boundary values of $g_{\mu \nu }$ and $\Psi ^{a}$
with respect to $\Lambda $ (with fixed $\left\{ Q^{A}\right\} $). This
provides a non-trivial equation for the classical value of the effective
cosmological constant in $\mathcal{M}$. Note that this classical field
equation for $\Lambda $ is independent of the prior weighting $\mu \lbrack
\Lambda ]$.

\item The classical value of the effective CC, $\Lambda ,$ that is
determined in this way does \emph{not} depend on the quantum vacuum energy.
It is instead determined by $\mathcal{M}$, and the fixed quantities $\left\{
Q^{A}\right\} $ (which could be taken as the initial conditions). Because $%
\Lambda $ is no longer determined by $\rho _{\mathrm{vac}}$, the quantum
cosmological constant problem is evaded in our proposal.

\item Finally, we construct a concrete application by demanding that, for a
given observer, \emph{the sum of different configurations of the partition
function depends only on the potential configurations in the observer's
causal past}. This implies that $\mathcal{M}$ is the causal past of the
observer. Given this choice, or similar choices, for $\mathcal{M}$, an order
of magnitude estimate for the classical value of $\Lambda $ seen by an
observer at a time when the age of the universe is $t_{U}$ is always $%
\Lambda \sim O(t_{U}^{-2})$ and a solution of the coincidence problem is
ensured.
\end{itemize}

We define $I_{\mathrm{class}}(\Lambda ;\mathcal{M})$ to be the value of $I_{%
\mathrm{tot}}[g_{\mu \nu },\Psi ^{a},\Lambda ;\mathcal{M}]$ evaluated with $%
g_{\mu \nu }$ and $\Psi ^{a}$ obeying their classical field equations for
fixed boundary / initial conditions, $\left\{ Q^{A}\right\} $. We show below
that the field equation for the effective CC, $\Lambda $, is then given
succinctly by 
\begin{equation}
\frac{\,\mathrm{d}I_{\mathrm{class}}(\Lambda ;\mathcal{M})}{\,\mathrm{d}%
\Lambda }=0.  \label{eq:phisimp}
\end{equation}%
In this rest of this section we present a more detailed statement of our
proposal for solving the CC problem, give the general form of the new
field equation for $\Lambda ,$ and show that the classical value of the CC
determined by this equation is typically expected to be of the observed
order of magnitude.

\subsection{Partition Function of the Universe}

\label{sec:prop:part} A relatively simple and revealing statement of our
proposed paradigm for determining the effective CC can be given in terms of
the partition function (or quantum state), $Z$, of the universe. $Z$ is
given by a sum over all possible configurations of fields, consistent with
certain fixed quantities on the boundary (i.e. the $\left\{ Q^{A}\right\} $)
and weighted by $\exp (iI_{\mathrm{tot}}),$ where $I_{\mathrm{tot}}$ is the
total action. The action is defined on a manifold $\mathcal{M}$ with
boundary $\partial \mathcal{M}$. The fields are the metric, $g_{\mu \nu }$,
and supporting matter fields, $\Psi ^{a}$.

In the usual approach the bare CC, $\lambda $, is not a field, in the sense
that its different configurations are summed; rather, it is a fixed
parameter which determines $I_{\mathrm{tot}}$. With fixed $\lambda $ the
partition function is $Z[\lambda ;\mathcal{M}]\equiv Z_{\Lambda }[\mathcal{M}%
]$ where $\Lambda =\lambda +\mathrm{const}$ and: 
\begin{eqnarray}
Z_{\Lambda }[\mathcal{M}] &=&\sum_{\substack{ g_{\mu \nu },\Psi ^{a}  \\ 
\mathrm{fixed}\,\,\left\{ Q^{A}\right\} }}e^{iI_{\mathrm{tot}}[g_{\mu \nu
},\Psi ^{a},\Lambda ;\mathcal{M}]} \nonumber \\ &&\times \,\,\left[\mathrm{%
gauge\,\,fixing\,\,terms}\right] . \nonumber
\end{eqnarray}

In the classical limit $Z_{\Lambda }[\mathcal{M}]$ is dominated by
configurations $g_{\mu \nu }$ and $\Psi ^{a}$ that are compatible with the
fixed $\left\{ Q^{A}\right\} $ for which $I_{\mathrm{tot}}$ is stationary.
We assume there are $N$ such classical solutions (not related by gauge
transformations) and for the $\alpha ^{\mathrm{th}}$ solutions : $I_{\mathrm{%
tot}}[g_{\mu \nu },\Psi ^{a},\Lambda ;\mathcal{M}]=I_{\mathrm{class}%
}^{(\alpha )}[\Lambda ;\mathcal{M}]$. In this limit, we have 
\begin{equation*}
Z_{\Lambda }[\mathcal{M}]\approx \sum_{\alpha =1}^{N}e^{iI_{\mathrm{class}%
}^{(\alpha )}[\Lambda ;\mathcal{M}]}.
\end{equation*}%
We demand that the quantities that must be held fixed on $\partial \mathcal{M%
}$, i.e. the $\left\{ Q^{A}\right\} $, are independent and can be freely
specified. Hence, when they are held fixed, the stationary points of $I_{%
\mathrm{tot}}$ correspond to configurations obeying the usual classical
field equations, $E^{\mu \nu }=\Phi _{a}=0$.

We propose to promote $\lambda $ from a fixed parameter to a `field' whose
different configurations are summed over in the partition function. The
introduction of a sum over $\lambda $ is only defined up to some arbitrary
weighting function $\mu _{\lambda }\equiv \mu \lbrack \Lambda ]$. The total
partition function for the scenario we have proposed is then simply given
by: 
\begin{eqnarray}
Z[\mathcal{M}]&=&\sum_{\Lambda }\mu \lbrack \Lambda ]Z_{\Lambda }[\mathcal{M}%
] \nonumber \\ &=&\sum_{\substack{ \Lambda ,g_{\mu \nu },\Psi ^{a}  \\ \mathrm{fixed}%
\,\,\left\{ Q^{A}\right\} }}\mu \lbrack \Lambda ]e^{iI_{\mathrm{tot}}[g_{\mu
\nu },\Psi ^{a},\Lambda ;\mathcal{M}]} \nonumber \\ &&\times\,\,\left[\mathrm{%
gauge\,\,fixing\,\,terms}\right] .\nonumber
\end{eqnarray}%
In principle, the weighting $\mu \lbrack \Lambda ]$ should be determined by
a fundamental theory or some symmetry principle. Crucially, we shall see
that our results are independent of this $\mu \lbrack \Lambda ]$ and so we do not need to concern ourselves with its precise form. This is in
contrast to multiverse scenarios, where the extent to which the observed
value of the CC is natural depends significantly on the prior weighting of
different values of $\Lambda $ in the multiverse. Here, whilst $g_{\mu \nu }$
and $\Psi ^{a}$ are space-time fields on $\mathcal{M}$, $\Lambda $ is a
space-time constant. In the different histories that are summed over, $%
\Lambda $ takes different values, but in each history it takes only one value
throughout $\mathcal{M}$. We could promote $\Lambda $ to a space-time scalar
field, $\Lambda (x^{\mu }),$ provided we introduced a delta function that
requires $\nabla _{\mu }\Lambda =0$ - see appendix \ref{app:unimodular} for
further discussion of this; taking this approach does not alter our results.
Alternatively, the variable $\lambda $ required in our model could be
associated with the squared four-form field strength $\mathbf{F}_{4}^{2}$,
where $\mathbf{F}_{4}=\mathrm{d}\wedge \mathbf{A}_{3}$ is a 4-form field
strength and $\mathbf{A}_{3}$ is a 3-form gauge field. Such a term arises
naturally in $N=8$ supergravity in 4-dimensions (see Ref. \cite{fourform}
for further details). With the inclusion of an appropriate boundary term,
the sum over different configurations of $\mathbf{A}_{3}$ reduces to the sum
over a contribution, $\lambda $, to the cosmological constant with is
constant over $\mathcal{M}$ in each history.

Taking the classical limit for $g_{\mu \nu }$ and $\Psi ^{a}$, $Z[\mathcal{M}%
]$ then reduces the partition function to: 
\begin{equation*}
Z[\mathcal{M}]\approx \sum_{\alpha =1}^{N}\sum_{\Lambda }\mu \lbrack \Lambda
]e^{iI_{\mathrm{class}}^{(\alpha )}[\Lambda ;\mathcal{M}]}.
\end{equation*}%
The sum over $\Lambda $ in the above expression is then dominated by the
value(s) of $\Lambda $ for which $\,\mathrm{d}I_{\mathrm{class}}^{(\alpha
)}/\,\mathrm{d}\Lambda =0$. This provides the classical field equation for $%
\Lambda $. In order of the universe to appear classical to an observer,
there should be a unique classical solution (once the gauge freedoms are
fixed) for $g_{\mu \nu }$, $\Psi ^{a}$ (i.e. $N=1$) and for $\Lambda $. If
there were more than one solution for $\Lambda $, some number $N_{\Lambda }$
say, the observer would see a superposition of $N_{\Lambda }$ classical
histories each with different values of $\Lambda $. If there were no
classical solutions for $\Lambda $, the universe would be observed to behave
in a fundamentally quantum manner. Provided a unique classical solution
exists for $\Lambda $, the partition function is dominated by a single
history in which the CC is $\Lambda $, as given by Eq. (\ref{eq:phisimp}),
and $I_{\mathrm{tot}}=I_{\mathrm{class}}[\Lambda ;\mathcal{M}]$. In the
classical limit we have: 
\begin{equation*}
Z[\mathcal{M}]\approx \mu \lbrack \Lambda ]e^{iI_{\mathrm{class}}[\Lambda ;%
\mathcal{M}]}.
\end{equation*}%
The expectation of an observable $\mathcal{O}(g_{\mu \nu },\Phi ^{a},\Lambda
)$ is given by: 
\begin{eqnarray}
\left\langle \mathcal{O}\right\rangle &=&\frac{1}{Z[\mathcal{M}]}\sum 
_{\substack{ \Lambda ,g_{\mu \nu },\Psi ^{a}  \\ \mathrm{fixed}\,\,\left\{
Q^{A}\right\} }}\mathcal{O}(g_{\mu \nu },\Phi ^{a},\Lambda )\mu \lbrack
\Lambda ]e^{iI_{\mathrm{tot}}[g_{\mu \nu },\Psi ^{a},\Lambda ;\mathcal{M}]}%
\nonumber \\ &&\times \,\,\left[ \mathrm{gauge\,\,fixing\,\,terms}\right] .\nonumber
\end{eqnarray}%
When Eq. (\ref{eq:phisimp}) has a solution for the CC, and a classical limit
exists, we then have: 
\begin{equation*}
\left\langle \mathcal{O}\right\rangle \approx \mathcal{O}(g_{\mu \nu },\Phi
^{a},\Lambda )\left[ \frac{\mu \lbrack \Lambda ]e^{iI_{\mathrm{class}%
}[\Lambda ;\mathcal{M}]}}{Z[\mathcal{M}]}\right] =\mathcal{O}(g_{\mu \nu
},\Phi ^{a},\Lambda ),
\end{equation*}%
which is independent of the prior weighting $\mu \lbrack \Lambda ]$.

\subsection{Field Equation for $\Lambda$}

\label{sec:prop:Leqn} We have proposed a paradigm for dynamically
determining the value of the effective cosmological $\Lambda $. In our
proposal, we\ have found that the value of $\Lambda $ is given by an
additional field equation Eq. (\ref{eq:phisimp}). In order to estimate the
order of magnitude of $\Lambda $ determined by Eq. (\ref{eq:phisimp}) it is
helpful to rewrite it in an expanded form.

The total action, $I_{\mathrm{tot}}$, is composed of the gravitational
action, $I_{\mathrm{grav}}$, the bare matter action, $I_{\mathrm{m}}$, and
the bare cosmological constant action $I_{\mathrm{CC}}[\lambda ,g_{\mu \nu };%
\mathcal{M}]$, where 
\begin{equation*}
I_{\mathrm{CC}}[\lambda ,g_{\mu \nu };\mathcal{M}]=-\frac{1}{\kappa }\int_{%
\mathcal{M}}\sqrt{-g}\,\mathrm{d}^{4}x\,\lambda .
\end{equation*}%
In this context $I_{\mathrm{m}}$ is `bare' in the sense that it includes the
contribution from the matter sector to the vacuum energy i.e. $I_{\mathrm{m}%
}=I_{\mathrm{matter}}+I_{\mathrm{vac}}$, where $I_{\mathrm{matter}}$ vanishes
in a vacuum and $I_{\mathrm{vac}}=I_{\mathrm{CC}}[\kappa \rho _{\mathrm{vac}%
},g_{\mu \nu };\mathcal{M}]$ gives the contribution from the vacuum energy.
Henceforth we refer to $I_{\mathrm{matter}}$ as the matter action and note
that it makes no contribution to the vacuum energy. With the effective CC, $%
\Lambda $, given by $\Lambda =\lambda +\kappa \rho _{\mathrm{vac}}$ we have
\begin{eqnarray}
I_{\mathrm{tot}}[g_{\mu \nu },\Psi ^{a},\Lambda ;\mathcal{M}] &=& I_{\mathrm{grav%
}}[g_{\mu \nu };\mathcal{M}]+I_{\mathrm{CC}}[\Lambda ,g_{\mu \nu };\mathcal{M%
}] \nonumber \\ &&+I_{\mathrm{matter}}[\Psi ^{a},g_{\mu \nu };\mathcal{M}]. \nonumber
\end{eqnarray}%
At this stage, for illustrative purposes, we assume that the boundary terms
in $I_{\mathrm{grav}}$ and $I_{\mathrm{matter}}$ have been chosen so that
the action is first order in the derivatives of $g_{\mu \nu }$ and $\Psi
^{a} $. With this choice, small perturbations in the fields $g_{\mu \nu }$, $%
\Psi ^{a},$ and in the bare cosmological constant $\lambda ,$ give $I_{%
\mathrm{tot}}\rightarrow I_{\mathrm{tot}}+\delta I_{\mathrm{tot}}$, where
schematically, 
\begin{eqnarray}
\delta I_{\mathrm{tot}} &=&\int_{\partial \mathcal{M}}\sqrt{|\gamma |}\,%
\mathrm{d}^{3}x\,\left[ \frac{1}{2\kappa }N^{\mu \nu }\delta \gamma _{\mu
\nu }+\Sigma _{a}\delta \Psi ^{a}\right] \\
&&+\int_{\mathcal{M}}\sqrt{-g}\,\mathrm{d}^{4}x\left[ \frac{1}{2\kappa }%
E^{\mu \nu }\delta g_{\mu \nu }+\Phi _{a}\delta \Psi ^{a}\right]  \notag \\
&&-\frac{\delta \lambda }{\kappa }\int_{\mathcal{M}}\sqrt{-g}\,\mathrm{d}%
^{4}x,  \notag
\end{eqnarray}%
for some $N^{\mu \nu }$, $\Sigma _{a}$, $E^{\mu \nu }$ and $\Phi _{a}$.
Minimizing this action with respect to variations of $g_{\mu \nu }$ and $%
\Psi ^{a}$ in the `bulk', $\mathcal{M}$, with fixed boundary values requires $%
E^{\mu \nu }=\Phi _{a}=0$. We showed above that these equations combined
with the requirement that some $\left\{ Q^{A}\right\} $ (which can be freely
specified) are fixed on the boundary, $\partial \mathcal{M}$, restrict the
variations of $\delta \gamma _{\mu \nu }$ and $\delta \Psi ^{a}$ (up to
gauge transformations) on $\partial \mathcal{M}$ to be of the form: 
\begin{equation*}
\left. \delta \gamma _{\mu \nu }\right\vert _{\partial \mathcal{M}}=\mathcal{%
H}_{\mu \nu }\delta \lambda ,\qquad \left. \Psi ^{a}\right\vert _{\partial 
\mathcal{M}}=\mathcal{P}^{a}\delta \lambda .
\end{equation*}%
Thus, with $E^{\mu \nu }=\Phi _{a}=0$, for $\delta I_{\mathrm{tot}}=0$ one
needs $\delta I_{\mathrm{tot}}/\delta \lambda =0$, which is equivalent to $\,%
\mathrm{d}I_{\mathrm{class}}/\,\mathrm{d}\lambda =0$, and which, from the
above, can be written as 
\begin{equation}
\int_{\partial \mathcal{M}}\sqrt{|\gamma |}\left[ \frac{1}{2\kappa }N^{\mu
\nu }\mathcal{H}_{\mu \nu }+\Sigma _{a}\mathcal{P}^{a}\right] =\frac{1}{%
\kappa }\int_{\mathcal{M}}\sqrt{-g}\,\mathrm{d}^{4}x. \label{eq:phi}
\end{equation}%
The forms of $\mathcal{H}_{\mu \nu }$ and $\mathcal{P}^{a}$ are determined
by $E^{\mu \nu }=0$, $\Phi _{a}=0,$ and the requirement that $\left\{
Q^{A}\right\} $ are fixed. Eq. (\ref{eq:phi}) is equivalent to Eq. (\ref%
{eq:phisimp}) in the case where the boundary terms in $I_{\mathrm{tot}}$ are
chosen so that the action is first order in derivatives $g_{\mu \nu }$ and $%
\Psi ^{a}$. Although Eq.(\ref{eq:phisimp}) represents a more succinct
statement of the field equation for $\Lambda $, the expanded form given by
Eq. (\ref{eq:phi}) is more useful in estimating the order of magnitude of
the value of $\Lambda $ determined by its field equation.

\subsection{The Natural Order of Magnitude of the Effective Cosmological
Constant}

In this subsection, we estimate the order of magnitude of the classical
effective CC that arises from solutions of the $\Lambda$-field equation i.e.
Eq. (\ref{eq:phisimp}) or equivalently Eq. (\ref{eq:phi}).

We focus on a cosmological setting where $\mathcal{M}$ is taken to be the
causal past and $\partial \mathcal{M}_{u}$ is the past light cone. The
boundary is $\partial \mathcal{M}=\partial \mathcal{M}_{u}\cup \partial 
\mathcal{M}_{I}$ where $\partial \mathcal{M}_{I}$ is the initial
hypersurface. We assume that the fixed quantities, $\left\{ Q^{A}\right\} $,
are such as to fix the initial state on $\partial \mathcal{M}_{I}$; $%
\mathcal{H}_{\mu \nu }$ and $\mathcal{P}_{a}$ vanish (up to diffeomorphisms)
on $\partial \mathcal{M}_{I}$. Eq. (\ref{eq:phi}) then reads: 
\begin{equation}
\int_{\partial \mathcal{M}_{u}}\sqrt{|\gamma |}\left[ \frac{1}{2}N^{\mu \nu }%
\mathcal{H}_{\mu \nu }+\kappa \Sigma _{a}\mathcal{P}^{a}\right] =\int_{%
\mathcal{M}}\sqrt{-g}\,\mathrm{d}^{4}x. \label{eq:phi2}
\end{equation}%
Now we estimate 
\begin{equation*}
N^{\mu \nu }\mathcal{H}_{\mu \nu }=N^{\mu \nu }\frac{\delta \gamma _{\mu \nu
}}{\delta \Lambda }\sim O(\mathrm{tr}\,N/\Lambda ),
\end{equation*}%
where $\mathrm{tr}\,N=N^{\mu \nu }\gamma _{\mu \nu }$. In many theories of
gravity, including general relativity, $\mathrm{tr}\,N\sim O(\mathrm{tr}\,K)$%
, where $K_{\mu \nu }$ is the extrinsic curvature of the boundary, $\partial 
\mathcal{M}_{u}$. Cosmologically, $\mathrm{tr}\,K\sim O(H)$ where $H$ is the
Hubble parameter; $H_{0}$ is its value today. Thus, we have 
\begin{eqnarray}
\frac{1}{2}\int_{\partial \mathcal{M}}\,\mathrm{d}^{3}x\,\sqrt{|\gamma |}%
N^{\mu \nu }\mathcal{H}_{\mu \nu } &\sim& \int_{\partial \mathcal{M}}\sqrt{%
|\gamma |}\,\mathrm{d}^{3}x\,\Lambda ^{-1}H \nonumber \\ &&\sim \frac{H_{0}}{\Lambda }%
A_{\partial \mathcal{M}}, \nonumber
\end{eqnarray}%
where $A_{\partial \mathcal{M}}$ is the surface area of $\partial \mathcal{M}
$. Similarly, the contribution from the $\kappa \Sigma _{a}\mathcal{P}^{a}$
matter terms in Eq. (\ref{eq:phi2}) is generally of the same order as that
from $N^{\mu \nu }\mathcal{H}_{\mu \nu }$. The left-hand side of Eq. (\ref%
{eq:phi2}) is therefore generally $\sim O(H_{0}A_{\partial \mathcal{M}%
}/\Lambda )$. The right-hand side of Eq. (\ref{eq:phi}) is simply the
4-volume, $V_{\mathcal{M}}$, of $\mathcal{M}$. We note that typically $V_{%
\mathcal{M}}\sim t_{U}A_{\partial \mathcal{M}}$ where $t_{U}$ is the age of
the universe. Putting these estimates together in Eq. (\ref{eq:phi}) we
have 
\begin{equation*}
\frac{H_{0}}{\Lambda }A_{\partial \mathcal{M}}\sim V_{\mathcal{M}%
}\Rightarrow \Lambda \sim \frac{H_{0}A_{\partial \mathcal{M}}}{V_{\mathcal{M}%
}}\sim \frac{H_{0}}{t_{U}}.
\end{equation*}%
Using $t_{U}\sim H_{0}^{-1}$, we find the general order of magnitude
estimate is $\Lambda \sim t_{U}^{-2}$. We note that the presence of $D$ small
(or even large) extra dimensions with volume $V_{\mathrm{extra}}$ would not
change this order of magnitude estimate. The extra dimensions would result
in $A_{\partial \mathcal{M}}\rightarrow A_{\partial \mathcal{M}}V_{\mathrm{%
extra}}$, $V_{\mathcal{M}}\rightarrow V_{\mathcal{M}}V_{\mathrm{extra}}$,
but the prediction $\Lambda \sim H_{0}A_{\partial \mathcal{M}}/V_{\mathcal{M}%
}\rightarrow H_{0}A_{\partial \mathcal{M}}/V_{\mathcal{M}}\sim 1/t_{U}^{2}$
remains unchanged.

Thus, provided the field equation for $\Lambda $, Eq. (\ref{eq:phisimp}),
admits a unique classical solution, we naturally expect the magnitude of the
classical value of the effective CC, $\Lambda $, to be $O(1/t_{U}^{2})$.
Thus our proposal results in a $\Lambda $ whose expected magnitude is
naturally of the order of the observed value. Provided a specific
application of our proposal realizes a unique prediction for $\Lambda $ of
this magnitude ($\sim 1/t_{U}^{2})$, it will have simultaneously solved both
the cosmological constant and the coincidence problems (see \S \ref%
{sec:cosmology}) for an example of such an application).

Our proposal results in a situation where those classical histories that
dominate the partition function naturally have a value of the bare
cosmological constant $\lambda $ that all but exactly cancels the vacuum
energy of $\kappa \rho _{\mathrm{vac}}$ in $\mathcal{M}$. The effective CC, $%
\Lambda =\lambda +\kappa \rho _{\mathrm{vac}}$, is then determined by the
properties of $\mathcal{M}$. This is achieved without introducing ad hoc
small parameters or special fine-tunings. In this sense, the solution to the
CC problem provided by scheme could be considered natural.

\section{Application to Cosmology}\label{sec:cosmology} 
In this section we consider the application of our
proposal for solving the CC problem to cosmological models. The scheme we
laid out in the previous section is flexible in that it does not make any
specific assumptions about either the theory of gravity or the
dimensionality of the universe. There is also a freedom in how one chooses
to define the manifold $\mathcal{M}$ on which the total action, $I_{\mathrm{%
tot}}$, is defined.

In this section, for simplicity, we assume that gravitational sector is
described by unmodified general relativity, space-time has 3+1 dimensions,
and that $\mathcal{M}$ is the causal past of the observer. We take the
observer to be at a fixed point, $p_{0}$. The manifold $\mathcal{M}$ is
bounded by the past-light cone $\partial \mathcal{M}_{u}$ of $p_{0}$ and an
initial time-like hypersurface $\partial \mathcal{M}_{I}$ with given normal $%
t_{0}^{\mu }$. Our proposal requires that $\mathcal{M}$ remains fixed for
different values of the bare cosmological constant, $\lambda $; that is,
there exists a coordinate chart $\mathcal{C}=\left\{ x^{\mu }\right\} $ such
that, for all $\lambda ,$ the values of the $\left\{ x^{\mu }\right\} $ at $%
p_{0},$ and on the boundaries $\partial \mathcal{M}_{u}$ and $\partial 
\mathcal{M}_{I},$ are the same for all $\lambda $.

\begin{figure*}[tbh]
\includegraphics[width=10cm]{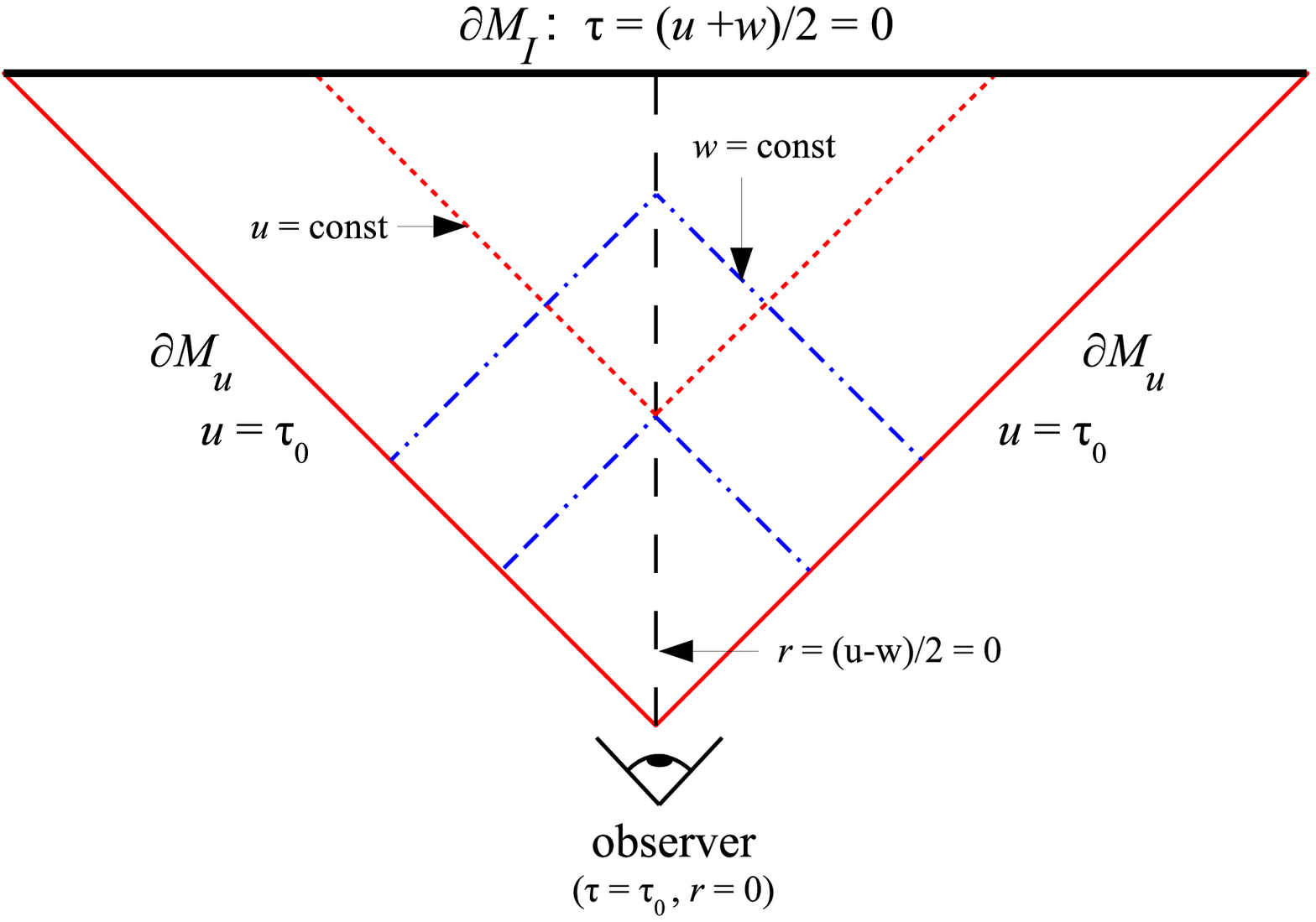}
\caption{(colour online) An illustration of the manifold $\M$ and its boundary $\dM  = \dM_{u} \cup \dM_{I}$ for the general cosmology set-up considered in \S \ref{sec:cosmology}.  Here, $\M$ is the causal past of the observer, $\dM_{u}$ is the null boundary given by the observer's past light cone and $\dM_{I}$ is a space-like boundary which represents the initial hypersurface. The model remains well-defined in the limit where $\dM_{I}$ is the initial singularity.    We pick orthogonal null coordinates $u$ and $w$ such that $\dM_{u}$ corresponds to $u=\tau_{0}$, and on $\dM_{u}$,  $-\tau_{0} < w <\tau_{0}$ for some fixed $\tau_{0}$, and the time coordinate given by $\tau=(u+w)/2$ vanishes on $\dM_{I}$.  We also define a radial coordinate $r=(u-w)/2$ and choose $w=\tau_0$ at the observer's position, so that there $\tau=\tau_0$ and $r=0$. In the figure, constant $u$ surfaces are shown as dotted red line (except $\dM_{u}$ which is a solid red line), and correspond to past-light cones of points on the line $r=0$ (the dashed black line)   which connects the observer with $\dM_{I}$.  The $w={\rm const}$ surfaces are future-directed light cones of points on $r=0$ and are shown above as dot-dashed blue lines.  Surfaces of constant $u$ and $w$ are closed 2-surfaces $S_{(u,w)}$ with intrinsic coordinates $\theta^{i}$; $i=1,2$.}
\label{figSpace}
\end{figure*}

While there is considerable freedom in the definition of the chart $\mathcal{%
C}$, a natural and simple choice results from the demand that changes in $%
\lambda $ preserve the light cone, and hence the causal, structure of
space-time. Given this choice, we define some null coordinates $u$ and $w$
such that $u=\tau_{0}$, for some $\lambda $-independent $\tau_{0}$, on $\partial 
\mathcal{M}_{u}$, and $u<\tau_{0}$ in $w$. We then define $w$ so that $w=-u$ on 
$\partial \mathcal{M}_{I}$ and $w=\tau_{0}$ at $p_{0}$. We define $u_{\mu
}=\nabla _{\mu }u$ and $w_{\mu }=\nabla _{\mu }w$. Now, $w_{\mu }w^{\mu
}=u_{\mu }u^{\mu }=0$, and we define $\sigma $ by $2e^{-2\sigma }=-u_{\mu
}w^{\mu }$. The metric can then be decomposed as 
\begin{equation*}
g_{\mu \nu }=-e^{2\sigma }u_{(\mu }w_{\nu )}+h_{\mu \nu }.
\end{equation*}%
Here, $h_{\mu \nu }=E_{\mu }^{i}E_{\nu }^{j}h_{ij}$ where $i=1,2$ for some
positive-definite 2-metric, $h_{ij}$, and some $E_{\mu }^{i}$ for which $%
w^{\mu }E_{\mu }^{i}=u^{\mu }E_{\mu }^{i}=0$. We define some intrinsic
coordinates $\theta ^{i}=\left\{ \theta ^{1},\theta ^{2}\right\} $ on the
closed 2-surfaces $S_{(u,w)}$ of constant $u$ and $w$. The 2-metric $h_{ij}$
is then defined by taking $h_{\mu }{}^{\nu }\partial _{\mu }\theta
^{i}=E_{\mu }^{i}$; $h^{ij}=h^{\mu \nu }E_{\mu }^{i}E_{\nu }^{j}$. We can
then write: 
\begin{equation*}
E_{\mu }^{i}=\partial _{\mu }\theta ^{i}+r^{i}u_{\mu }+s^{i}w_{\mu },
\end{equation*}%
for some $r^{i}$ and $s^{i}$. Our coordinate chart is then given by $%
\mathcal{C}=\left\{ u,w,\theta ^{1},\theta ^{2}\right\} $, and in $\mathcal{M%
}$ is determined by $u<\tau_{0}$, $-u<w<\tau_{0}$. In this chart: 
\begin{eqnarray}
\,\mathrm{d}s^{2} &=&g_{\mu \nu }\,\mathrm{d}x^{\mu }\,\mathrm{d}x^{\nu
}=-e^{2\sigma }\,\mathrm{d}u\,\mathrm{d}w+h_{ij}\mathrm{D}\theta ^{i}\mathrm{%
D}\theta ^{j}, \label{eq:metric:decomp} \\ \mathrm{D}\theta ^{i} &=&\,\mathrm{d}\theta ^{i}+r^{i}\,\mathrm{d}u+s^{i}\,\mathrm{d}w.  \nonumber
\end{eqnarray}%
We define a time-like coordinate $\tau =(u+w)/2$. The initial hypersurface
therefore corresponds to $\tau =0$ and the observer's position is $\tau
=\tau _{0}$. The normal to $\partial \mathcal{M}_{I}$ is
taken to be given, and this therefore partially restricts the freedom in the
definition of $u$ and $w$. We also define a space-like radial coordinate $%
r=(u-w)/2$; we then have $r=0$ at $p_{0}$.    FIG. \ref{figSpace} shows an illustration of $\M$ and its boundary.

The intrinsic three-metric on the initial hypersurface has line-element 
\begin{eqnarray}
\,\mathrm{d}s_{\tau =0}^{2}&=&e^{2\sigma _{(I)}}\left( \,\mathrm{d}r^{2}+\bar{h%
}_{ij}^{(I)} D_{(I)}\theta^{i}D_{(I)}\theta^{j}\right), \nonumber \\
D_{(I)}\theta^{i} &=& \mathrm{d}\theta ^{i}+K_{(I)}^{i}\,\mathrm{d}r, \nonumber
\end{eqnarray}%
where $\sigma _{(I)}=\left. \sigma \right\vert _{\tau =0}$, $\bar{h}%
_{ij}^{(I)}=\left. e^{-2\sigma }h_{ij}\right\vert _{\tau =0}$ and $%
K_{(I)}^{i}=\left. r^{i}-s^{i}\right\vert _{\tau =0}$. The requirement that
the initial state be fixed independently of $\lambda $ on $\partial \mathcal{%
M}_{I}$ implies that $\sigma _{I}$, $\bar{h}_{ij}^{(I)}$ and $K_{I}^{i}$ are
fixed up to $\lambda $ independent diffeomorphisms on $\partial \mathcal{M}%
_{I}$.

The surfaces $S_{(u,w)}$ of constant $u$ and $w$ (or equivalently constant $%
\tau $ and $r$) represent the intersection of a past and a future-directed
light cone. As such, the $S_{(u,w)}$ are closed two-surfaces. We define $%
\bar{\mathcal{R}}$ to be the scalar curvature of the conformal 2-metric $%
\bar{h}_{ij}$. Since $\bar{h}_{ij}$ describes a two-dimensional space, it is
completely characterized by $\bar{\mathcal{R}}$. Additionally, by the
Gauss-Bonnet theorem, we know that
\begin{eqnarray}
\langle \bar{\mathcal{R}}\rangle _{(u,w)}A_{2}(u,w) &\equiv& \int_{u,w=\mathrm{const%
}}\bar{\mathcal{R}}\sqrt{\bar{h}}\,\mathrm{d}^{2}\theta \nonumber \\ &=& 2\int_{u,w=\mathrm{%
const}}\sqrt{\bar{h}}\,\mathrm{d}^{2}\theta \equiv 2A_{2}(u,w). \nonumber
\end{eqnarray}%
where $A_{2}(u,w)$ is the surface area of the conformal 2-space described by $%
\bar{h}_{ij}$, and $\langle \bar{\mathcal{R}}\rangle_{(u,w)}$ is the
average curvature. The conformal 2-surfaces are homotopic to a 2-sphere. If
the intrinsic metric on the 2-spheres were that of a two-sphere with
conformal radius $\rho $, then we would have $A_{2}(u,w)=4\pi \rho^{2}$ and $%
\langle \bar{\mathcal{R}}\rangle_{(u,w)}=2/\rho^{2}$. This singles out a
preferred class of definitions for $u$ and $w$ on the initial hypersurface
on which $\tau =(u+w)/2=0$. We can always pick $r=(u-w)/2$
so that on $\partial \mathcal{M}_{I}$ (where $u=-w$), $A_{2}(-w,w)=4\pi
r^{2}$ or equivalently $\langle \bar{\mathcal{R}}\rangle
_{(-w,w)}=2/r^{2}$.

We note that for a 2-metric, with constant $r$, the diffeomorphism invariant
structure of $\bar{h}_{ij}^{(I)}$ is completely determined by its scalar
curvature $\bar{\mathcal{R}}_{(I)}$. Since $(u,w)=\mathrm{const}$ represents
the intersection of two light cones, the surfaces of constant $u$ and $w$
are closed, and so $\bar{\mathcal{R}}_{(I)}>0$. We note that we can always
choose $u$ and $w$ so that on $\partial \mathcal{M}_{I}$, $\bar{\mathcal{R}}%
_{(I)}=2/r^2$ with $r=u =-w$ on $\dM_{I}$.

\paragraph*{Choice of Surface Terms:}

Another freedom in our scheme is the choice of surface terms in $I_{\mathrm{%
tot}}$. Focussing on the variation of the metric, and keeping all other
fields including the CC fixed, these surface terms determine the quantities
that must be held on $\partial \mathcal{M,}$ so that $\delta I_{\mathrm{tot}%
}/\delta g_{\mu \nu }=0$ when the classical field equations hold. The metric
on and around a space-like and time-like boundary is described by the
induced metric $\gamma _{\mu \nu }$ and the extrinsic curvature, $K^{\mu \nu
}$. On a null boundary the situation is slightly more complicated and we
discuss it further below; nonetheless, there are quantities analogous to $%
\gamma _{\mu \nu }$ and $K^{\mu \nu }$. The $\gamma _{\mu \nu }$ and $K^{\mu
\nu }$ are respectively analogous to position variables and their associated
momenta. In most cases it is natural to choose the surface terms so that
(for fixed CC), the `position variable' $\gamma _{\mu \nu }$ must be held
fixed. The required surface term was first identified by York \cite%
{York:1972sj}, and then rediscovered and refined by Gibbons and Hawking \cite%
{Gibbons:1976ue}. We refer to ii as the Gibbons-Hawking-York (GHY) boundary
term. However, if $K^{\mu \nu }$, (or some components of it), diverge faster
than $\gamma _{\mu \nu }$ as one approaches the boundary, different choices
of boundary term may be required.

Metric quantities are suitably well-behaved on the null boundary $\partial%
\mathcal{M}_{u}$ and so for this boundary it is natural to pick the null
boundary analogue of the GHY term.

In the cosmological setting, $\partial \mathcal{M}_{I}$ is the initial
singularity and the intrinsic metric, $\gamma _{\mu \nu }^{(0)}$, on $%
\partial \mathcal{M}_{I}$ has vanishing determinant. More formally, taking $%
\gamma _{\mu \nu }^{(\tau )}$, to be the induced metric on surfaces of
constant $\tau $, $\lim_{\tau \rightarrow 0^{+}}\mathrm{det}\gamma ^{(\tau
)}=0$. We define $K_{(\tau )}^{\mu \nu }$ to be the extrinsic curvature on
constant $\tau $ surfaces; $K_{(\tau )}=K^{\mu }{}_{(\tau )\mu }$. Generally, 
$K$ diverges as $\tau \rightarrow 0^{+}$. The quantities $\mathrm{det}\gamma 
$ and $K$ are canonically conjugate. It is most natural to choose boundary
terms so that the most divergence of two canonically conjugate variables is
held fixed (c.f. the argument for fixing the charge rather than the chemical
potential in Ref. \cite{Sen:2008yk}). This implies that, on $\partial 
\mathcal{M}_{I}$, we choose the surface term so that $K$ rather than $%
\mathrm{det}\gamma $ is fixed. In Appendix \ref{app:surface}, we note that
this term is the `cosmological' boundary term found by York in Ref. \cite%
{York:1972sj}. Thus, we fix the surface terms in $I_{\mathrm{tot}}$ to be
York's cosmological boundary term on $\partial \mathcal{M}_{I}$, and the GHY
boundary term on $\partial \mathcal{M}_{u}$.   It should be stressed, though, this choice of York rather tha GHY boundary term on $\partial \mathcal{M}_{I}$ has no affect on the equation for $\lambda$ and is only made for the technical reason stated above.  This is because the initial state on $\dM_{\rm I}$ is fixed independently of $\lambda$ in our proposal and so for any $\dM_{I}$ boundary term, $I_{\dM_{I}}$, $\delta I_{\rm dM_{I}}/\delta \lambda = 0$.  It follows that boundary terms on $\dM_{I}$ do not contribute to the $\lambda$-equation:  $\delta I_{\rm tot}/\delta \lambda =0$.

\subsection{General Cosmology}

We begin by writing down the form of $I_{\mathrm{tot}}$ for the general
cosmological setting and considering its variation. In addition to $I_{%
\mathrm{tot}}$, our scheme requires that we specify a set of quantities $%
\left\{ Q^{A}\right\} $ that are kept fixed (for all values of the bare
cosmological constant) and which can be independently specified. By
considering the variation of the action, we present a natural choice of
these fixed quantities.

Since we have taken the gravity sector to be described (to a suitable
approximation) by unmodified general relativity, the gravitational action, $%
I_{\mathrm{grav}}$, is given by
\begin{equation*}
I_{\mathrm{grav}}[g_{\mu \nu };\mathcal{M}]=I_{\mathrm{EH}}[g_{\mu \nu };%
\mathcal{M}]+I_{\mathrm{surf}}[g_{\mu \nu };\partial \mathcal{M}],
\end{equation*}%
where $I_{\mathrm{surf}}$ are the surface terms defined on $\partial 
\mathcal{M}$ and $I_{\mathrm{EH}}$ is the Einstein-Hilbert action defined on 
$\mathcal{M}$: 
\begin{equation*}
I_{\mathrm{EH}}[g_{\mu \nu };\mathcal{M}]=\frac{1}{2\kappa }\int_{\mathcal{M}%
}\,\mathrm{d}^{4}x\,\sqrt{-g}R(g).
\end{equation*}%
We have taken the boundary to be $\partial \mathcal{M}=\partial \mathcal{M}%
_{(u)}\cup \partial \mathcal{M}_{I}$, where $\partial \mathcal{M}_{(u)}$ is
described by the vanishing of the null coordinate $u$, and so it represents
a null boundary. Our $\partial \mathcal{M}_{I}$ is the initial space-like
hypersurface given by $\tau =0$. As one approaches $\tau =0$, the
determinant of the induced metric on $\tau =\mathrm{const}$ hypersurfaces
vanishes, whilst the trace of the extrinsic curvature diverges. On $\partial 
\mathcal{M}_{(u)}$ it is natural to take the surface term to be the null
boundary analogue of the Gibbons-Hawking-York (GHY) term, $I_{\mathrm{GHY}%
}^{(u)}$ say, whereas on $\partial \mathcal{M}_{I}$ the divergence of $K$
makes $\tau \rightarrow 0$ limit of York's cosmological (YC) surface more
natural; we write this as $I_{\mathrm{YC}}^{(I)}$.

In appendix \ref{app:surface} we present a detailed rederivation and
discussion of boundary terms in general relativity for both non-null and
null boundaries. Here, we briefly review those results where they apply to
the form of the $I_{\mathrm{GHY}}^{(u)}$ and $I_{\mathrm{YC}}^{(I)}$ terms.

\subsubsection{GHY term on $\partial\mathcal{M}_{u}$ ($u=\tau_0$):}

We consider a null boundary $\partial \mathcal{M}_{(u)}$ described by some $%
u(x^{\mu })=\tau_{0}$ where, with $u_{\mu }=\nabla _{\mu }u$, we have $u^{\mu
}u_{\mu }=0$. We define $n^{\mu }=-e^{\sigma }u^{\mu }$ for some $\sigma $.
In order to describe points on $\partial \mathcal{M}_{(u)},$ we have a $%
w(x^{\mu })$ such that $w_{\mu }=\nabla _{\mu }w$ is null and $w_{\mu
}u^{\mu }=-2e^{-2\sigma }>0$. We also define $\tau =(u+w)/2$ and so, on $%
\partial \mathcal{M}_{u}$, $\tau =(\tau_{0}+w)/2$. Eq. (\ref{eq:metric:decomp})
gives the decomposition of the metric $g_{\mu \nu }$ in terms of $u$, $w$
and the intrinsic coordinates $\theta ^{i}$ on the closed 2-surfaces, $%
S_{(u,w)}$, of constant $u$ and $w$. The $h_{ij}$ is the induced 2-metric on
the $S_{(u,w)}$ and $h_{\mu \nu }=g_{\mu \nu }-e^{2\sigma }w_{(\mu }u_{\nu
)} $. We define $n^{\mu }=-e^{\sigma }u^{\mu }$, and $\bar{n}^{\mu
}=e^{\sigma }w^{\mu }$ so that $\bar{n}^{\mu }n_{\mu }=2$.

The extrinsic curvature of $S_{(u,w)}$ along $n^{\mu }$ is $\mathcal{K}^{\mu
\nu }$ and is defined by 
\begin{eqnarray}
\mathcal{K}^{\mu \nu } &=&-\frac{1}{2}h^{\mu \rho }h^{\nu \sigma }\mathcal{L}%
_{n}h_{\rho \sigma } \\
&=&e^{\sigma }h^{\mu \rho }h^{\nu \sigma }\nabla _{\rho }u_{\sigma }.  \notag
\end{eqnarray}%
Writing $e_{\mu }^{i}=\partial _{\mu }\theta ^{i}$, we define $\mathcal{K}%
^{ij}=e_{\mu }^{i}e_{\nu }^{j}\mathcal{K}^{\mu \nu }$ and the trace of the
extrinsic curvature is $\mathcal{K}=\mathcal{K}^{\mu \nu }h_{\mu \nu }=%
\mathcal{K}^{ij}h_{ij}$. We also define the \emph{inaffinity}, $\nu $, and 
\emph{twist}, $\omega ^{\mu }$, by 
\begin{equation*}
\nu =-\mathcal{L}_{n}\sigma =e^{\sigma }u^{\mu }\nabla _{\mu }\sigma ,\qquad
\omega ^{\mu }=\frac{1}{2}h^{\mu \nu }\bar{n}_{\rho }\nabla _{\nu }n^{\rho }.
\end{equation*}%
We also define $\omega ^{i}=\omega ^{\mu }e_{\mu }^{i}$.

The usual GHY term (defined on non-null boundaries) has the property that it
renders the action first order in derivatives of the metric. The variation
of the total action with respect to the metric is then free of surface terms
whenever the induced boundary 3-metric is held fixed. On a null boundary
there is no (non-singular) boundary 3-metric. In its place are $h_{\mu \nu }$
and $e^{\sigma }$. This is clear when one notes that the invariant area
element on $\partial \mathcal{M}_{(u)}$ is $e^{\sigma }\sqrt{h}\,\mathrm{d}%
\tau \,\mathrm{d}^{2}\theta $, as opposed to $\sqrt{|\gamma |}\,\mathrm{d}%
^{3}x$ on a non-null boundary. Thus, the analogue of the GHY term for a null
boundary is defined by the property that when $e^{\sigma }$ and $h_{\mu \nu
} $ are fixed, variation of the total action with respect to the metric is
free of surface terms on $\partial \mathcal{M}_{u}$.

Given this, we find in Appendix \ref{app:surface} that the GHY term for $%
\partial \mathcal{M}_{(u)}$ is: 
\begin{equation*}
I_{\mathrm{GHY}}^{(u)}=\frac{1}{\kappa }\int_{\partial \mathcal{M}%
_{u}}e^{\sigma }\sqrt{h}\,\mathrm{d}\tau \,\mathrm{d}^{2}\theta \,\left[ 
\mathcal{K}+\nu \right] .
\end{equation*}%
One finds the same boundary term if one starts with the EH action in terms
of a vierbein and adds a boundary term so that action is first order in
derivatives of the vierbein (see Appendix \ref{app:surface} for a proof of
this).

On a space-like or time-like boundary, $\Sigma $ say, it is well known that
the GHY boundary term invariant under diffeomorphisms restricted to $\Sigma $
(i.e. those under which the 4-vector normal to $\Sigma $ is invariant).
However, on a null boundary this is \emph{not} the case. $I_{\mathrm{GHY}%
}^{(u)}$ is invariant under diffeomorphisms on the two surfaces of constant $%
u$ and $w$ (i.e. $S_{(u,w)}$), but it is not invariant under
reparametrizations of the `radial' coordinate (in this case $w$ or
equivalently $\tau $, along the null hypersurface). Specifically, it is
always possible to find such diffeomorphisms that preserve the normal to $%
\partial \mathcal{M}_{(u)}$ but under which
\begin{equation*}
\left. \nu \right\vert _{\partial \mathcal{M}_{u}}\rightarrow \left. \nu
\right\vert _{\partial \mathcal{M}_{u}}+e^{\sigma }\mathcal{L}%
_{u}f_{-}(w,\theta ^{i}).
\end{equation*}%
for any $f_{-}(w,\theta ^{i})$. This means that the $I_{\mathrm{GHY}}^{(u)}$
term as defined above is ambiguous. The ambiguity in the GHY term for null
boundaries is because the 3-metric normal to the boundary is degenerate.
This means that there is no preferred normalization of the normal to the
boundary and consequently no preferred `radial' coordinate along the boundary. To unambiguously define the $I_{\mathrm{GHY}%
}^{(u)}$ one must fix this remaining gauge freedom by either picking a form
for $\nu $ on $\partial \mathcal{M}_{(u)}$ or, equivalently, by specifying a
preferred choice of $w$/$\tau $ on $\partial \mathcal{M}_{u}$. Simplifying
choices that are common in the literature, and can always be achieved, are
\begin{equation*}
\left. \nu \right\vert _{\partial \mathcal{M}_{u}}=0\,\qquad \mathrm{or}%
\qquad \left. \nu \right\vert _{\partial \mathcal{M}_{u}}=-\frac{c_{0}}{2}%
\mathcal{K},
\end{equation*}%
for some constant $c_{0}$. Each choice identifies a preferred $w$ for which
the total action is first order in derivatives of metric quantities. In this
application of our proposal to solve the CC problems, we demand that the
initial state is fixed independent of $\lambda $. However, both these
choices for $\nu $ on $\partial \mathcal{M}_{u}$ given above would single
out a preferred $w$, and hence a definition of $\sigma $ which would depend on $%
\lambda $ on $\partial \mathcal{M}_{I}$. In the scenario we consider here,
it is more natural to remove the ambiguity in $\nu $ on $\partial \mathcal{M}%
_{u}$ by fixing the definition of $\sigma $ initially (i.e. on $\partial 
\mathcal{M}_{I}$). This is achieved by picking a preferred radial coordinate 
$r$ on $\partial \mathcal{M}_{I}$. More exactly, we should specify the $r$
on $\partial \mathcal{M}_{I}$ up to residual coordinate transformations that
leave $I_{\mathrm{GHY}}^{(u)}$ invariant. We noted above that since the $%
S_{(u,w)}$, which are the surfaces of constant $r$ on $\partial \mathcal{M}%
_{I}$, are closed 2-surfaces, our $r$ represents a radial coordinate on $%
\partial \mathcal{M}_{I}$. A simple and natural definition of $r$ (up to
residual coordinate transformations) is then to pick it so that the average
scalar curvature of the conformal 2-surface at $\tau =0$ and $r=\mathrm{const%
}$ (and described by the metric $\bar{h}_{ij}=e^{-2\sigma }h_{ij})$ is $%
2/r^2$ with $r=u=-w$ on $\dM_{I}$. By the Gauss-Bonnet theorem, this is
equivalent choosing $r$ so that the surface area of the conformal 2-surface
is $4\pi r^{2}$. This choice does not uniquely determine $r$ but it
is sufficient to fix $\nu $ on $\partial \mathcal{M}_{u}$.

The choice of a preferred $r$ on $\partial \mathcal{M}_{I}$ is equivalent to
the specification of an unambiguous boundary term on $\partial \mathcal{M}%
_{u}$. Making such a specification requires that one replace $\nu $ in $I_{%
\mathrm{GHY}}^{(u)}$ by some quantity that is invariant under
diffeomorphisms that vanish normal to $\partial \mathcal{M}_{u}$ and $%
\partial \mathcal{M}_{I}$. Any such choice will then pick out a preferred
set of definitions of $w$ on $\partial \mathcal{M}_{u}$ for which $\nu =-%
\mathcal{L}_{n}\sigma $ and hence the total action is first order. This in
turns picks out a preferred set of $2r=(u-w)$ on $\partial 
\mathcal{M}_{I}$ where $w=-u$. For the application of our proposal to
cosmology in this section we fix the definition of $\nu $ so that on $%
\partial \mathcal{M}_{I}$, $\left\langle \bar{\mathcal{R}}\right\rangle
_{\tau =0}=2/r^{2}$. This is arguably the simplest choice we can
make that is consistent with the requirement that the initial state be $%
\lambda $ independent when described in the coordinate chart for which the
action is (up to a boundary term on $\partial \mathcal{M}_{I}$) first order
in metric derivatives.

\subsubsection{YC term on $\partial\mathcal{M}_{I}$ ($\protect\tau=0$):}

The initial time-like hypersurface, $\partial \mathcal{M}_{I}$, is singular,
however we may still define the surface term by taking a limit as $\tau
\rightarrow 0$ from above (i.e. $\tau \rightarrow 0^{+}$). We take $t^{\mu
}=e^{\sigma }\nabla ^{\mu }\tau $ where $\tau =(u+w)/2$. We then have $%
t^{\mu }t_{\mu }=-1$; $t^{\mu }$ is the backward pointing normal to
surfaces, $\Sigma _{\tau }$, of constant $\tau $. We can decompose the
metric into 
\begin{equation*}
g_{\mu \nu }=-t_{\mu }t_{\nu }+\gamma _{\mu \nu }.
\end{equation*}%
With $r=(u-w)/2,$ we have: 
\begin{equation*}
\gamma _{\mu \nu }\,\mathrm{d}x^{\mu }\,\mathrm{d}x^{\nu }=e^{2\sigma }\,%
\mathrm{d}r^{2}+h_{ij}\left[ \,\mathrm{d}\theta ^{i}+K^{i}\,\mathrm{d}r%
\right] \left[ \,\mathrm{d}\theta ^{j}+K^{j}\,\mathrm{d}r\right] ,
\end{equation*}%
where $K^{i}=(r^{i}-s^{i})$. The $\theta ^{i}$ are the intrinsic coordinates
on the surfaces of constant $u$ and $w$. The line element for the 4-metric
can then be written as 
\begin{equation*}
\,\mathrm{d}s^{2}=-e^{\sigma }\,\mathrm{d}\tau ^{2}+\gamma _{\alpha \beta }%
\left[ \,\mathrm{d}x^{\alpha }+N^{\alpha }\,\mathrm{d}\tau \right] \left[ \,%
\mathrm{d}x^{\beta }+N^{\beta }\,\mathrm{d}\tau \right] ,
\end{equation*}%
where 
\begin{equation*}
N_{\alpha }\,\mathrm{d}x^{\alpha }=h_{ij}(r^{j}+s^{j})\,\mathrm{d}\theta
^{i}.
\end{equation*}%
We call $N^{\alpha }$ the shift-vector. We note that we can always define
the $\theta ^{i}$ so that in $\mathcal{M}$, $N^{\alpha }=0$ i.e. $%
r^{i}=-s^{i}$. The extrinsic curvature, $K^{\mu \nu }$, of $\ \Sigma _{\tau }
$ is given by
\begin{equation*}
K^{\mu \nu }=-\frac{1}{2}\gamma ^{\mu \rho }\gamma ^{\nu \sigma }\mathcal{L}%
_{t}\gamma _{\rho \sigma },
\end{equation*}%
and the trace is $K=K^{\mu \nu }\gamma _{\mu \nu }$. With these definitions
we see in Appendix \ref{app:surface} that the cosmological boundary term of
York for $\partial \mathcal{M}_{I}$ is: 
\begin{equation*}
I_{\mathrm{YC}}^{(I)}=-\lim_{\tau \rightarrow 0^{+}}\frac{1}{3\kappa }%
\int_{\partial \mathcal{M}_{I}}\sqrt{\gamma }\,\mathrm{d}^{3}x\,K.
\end{equation*}

\subsubsection{Variation of Gravitational Action and Fixed Quantities:}

In this cosmological set-up the total gravitational action is: 
\begin{eqnarray}
I_{\mathrm{grav}} &=&\frac{1}{2\kappa }\int_{\mathcal{M}}\,\mathrm{d}^{4}x\,%
\sqrt{-g}R(g) \nonumber \\ &&+\frac{1}{\kappa }\int_{\partial \mathcal{M}_{u}}e^{\sigma }%
\sqrt{h}\,\mathrm{d}\tau \,\mathrm{d}^{2}\theta \,\left[ \mathcal{K}+\nu %
\right] \nonumber \\
&&-\lim_{\tau \rightarrow 0^{+}}\frac{1}{3\kappa }\int_{\partial \mathcal{M}%
_{I}}\sqrt{\gamma }\,\mathrm{d}^{3}x\,K.  \notag
\end{eqnarray}%
In appendix \ref{app:surface} we show that the variation of this action with
respect to the metric, $g_{\mu \nu }$, gives: 
\begin{eqnarray}
\delta I_{\mathrm{grav}} &=&-\frac{1}{2\kappa }\int_{\mathcal{M}}\,\mathrm{d}%
^{4}x\,\sqrt{-g}G^{\mu \nu }\delta g_{\mu \nu }  \label{eq:grav:var}  \\ &&-\frac{1}{2\kappa }%
\int_{\partial \mathcal{M}_{I}}\left[ \tilde{P}^{\alpha \beta }\delta \tilde{%
\gamma}_{\alpha \beta }+\frac{4}{3}\sqrt{\gamma }\delta K\right] \,\mathrm{d}%
r\,\mathrm{d}^{2}\theta , \nonumber \\
&&+\frac{1}{2\kappa }\int_{\partial \mathcal{M}_{u}}e^{\sigma }\sqrt{h}\left[
(\mathcal{K}+\nu )h^{ij}-\mathcal{K}^{ij})\delta h_{ij}\right. \nonumber \\ &&+\left. 2\mathcal{K}\delta
\sigma +2\omega _{i}\delta s^{i}\right] \,\mathrm{d}\tau \,\mathrm{d}%
^{2}\theta ,  \notag
\end{eqnarray}%
where 
\begin{eqnarray}
\tilde{\gamma}_{\alpha \beta } &=& (\mathrm{det}\,\gamma )^{-1/3}\gamma _{\alpha
\beta }, \nonumber \\ \tilde{P}^{\alpha \beta } &=&(\mathrm{det}\,\gamma )^{5/6}\left[
K^{\alpha \beta }-\frac{1}{3}K\gamma ^{\alpha \beta }\right] . \nonumber
\end{eqnarray}%
Since $\tilde{P}^{\alpha \beta }\tilde{\gamma}_{\alpha \beta }=0$, it
follows that: 
\begin{eqnarray}
\tilde{P}^{\alpha \beta }\delta \tilde{\gamma}_{\alpha \beta } &=&-\tilde{%
\gamma}_{\alpha \beta }\delta \tilde{P}^{\alpha \beta }=-\sqrt{\gamma }%
\gamma _{\alpha \beta }\delta \sigma ^{\alpha \beta }, \nonumber \\
\sigma ^{\alpha \beta } &=&K^{\alpha \beta }-\frac{1}{3}K\gamma ^{\alpha
\beta }.\nonumber
\end{eqnarray}%
We now specify the fixed quantities, $\left\{ Q^{A}\right\} $, for the
gravitational sector. We see that the boundary terms in Eq. (\ref%
{eq:grav:var}) vanish when $K$ and either $\tilde{\gamma}_{\alpha \beta }$
or $\sigma ^{\alpha \beta }$ are fixed on $\partial \mathcal{M}_{I}$ (up to
residual coordinate transformations on $\partial \mathcal{M}^{I}$) and $%
s^{i} $ , $h_{ij}$ and $\sigma $ are fixed on $\partial \mathcal{M}_{u}$.
The $\left\{ Q^{A}\right\} $ will be a subset of these quantities, and their
defining property is that they are a maximal subset which can be
independently specified.

In terms of the familiar 3+1 ADM decomposition on constant time (i.e. $\tau $%
) hypersurfaces, $\Sigma _{\tau }$, our $e^{\sigma }$ is the lapse function
and $N^{\alpha }=(0,r^{i}+s^{j})$ is the shift vector; $\gamma _{\alpha
\beta }$ is the induced metric on $\Sigma _{\tau }$. The six components of $%
\gamma _{\alpha \beta }$ are given by $\sigma $, $K^{i}=r^{i}-s^{i},$ and $%
h_{ij}$.

In general, the intrinsic coordinate-independent geometry in $\mathcal{M}$
is determined uniquely when the twelve variables $\gamma _{\alpha \beta }$
and $K^{\alpha \beta }$ are specified on $\partial \mathcal{M}_{I}$. The
Einstein equations provide four constraint equations on $\partial \mathcal{M}%
_{I}$ which reduce the number of independent functions in $\gamma _{\alpha
\beta }$ and $K^{\alpha \beta }$ to eight. When $N=e^{\sigma }$ and $%
N^{\alpha }$ in $\mathcal{M}$ are fixed, this reduces the number of
independent functions that must be specified on $\partial \mathcal{M}_{I}$
to four. Thus, to completely specify the space-time in $\mathcal{M}$, we
must fix four free functions on $\partial \mathcal{M}_{I}$ as well as
specifying $N$ and $N^{\alpha }$ in $\mathcal{M}$.

The specification of $\tilde{\gamma}_{\alpha \beta }$ (or $\sigma ^{\alpha
\beta }$) and $K$ on $\partial \mathcal{M}_{I}$ fixes six functions on the
initial hypersurface in a coordinate dependent manner. We found that the
definition of $I_{\mathrm{GHY}}^{(u)}$ resulted in the specification of a
preferred $r$ on $\partial \mathcal{M}_{I}$. There remains, however, the
freedom to define the $\theta ^{i}$ on $\partial \mathcal{M}_{I}$ which can
be used, for instance, to set $K^{i}=0$ initially. This freedom means that $%
\tilde{\gamma}_{\alpha \beta }$ (or $\sigma ^{\alpha \beta }$) and $K$ on $%
\partial \mathcal{M}_{I}$ fix four free functions on $\partial \mathcal{M}%
_{I}$ in a coordinate independent manner. The lapse function, $N$, in $%
\mathcal{M}$ to be given by $N=e^{\sigma }$. If $N^{\alpha }$ can be fixed
in $\mathcal{M}$, then $\tilde{\gamma}_{\alpha \beta }$ (or $\sigma ^{\alpha
\beta }$) and $K$ on $\partial \mathcal{M}_{I}$ are sufficient to completely
determine, via the field equations, all metric quantities in $\mathcal{M}$,
and on $\partial \mathcal{M}_{u}$. Fixing $N^{\alpha }$ in terms other
metric quantities is equivalent to fixing $r^{i}+s^{i}$ which in turn is
equivalent to specifying our coordinates $\theta ^{i}$ on $S_{(u,w)}$ which
are defined on surfaces of constant $\tau $ relative to their values on $%
\partial \mathcal{M}_{I}$. A simple choice with a geometrical basis is
demand that the $\theta ^{i}$ are Lie-propagated along $\tau ^{\mu }=\nabla
^{\mu }\tau $ from the values that are arbitrarily assigned to them on the
initial hypersurface $\partial \mathcal{M}_{I}$ i.e. $\mathcal{L}_{\tau
}\theta ^{i}=0$. This implies that: 
\begin{equation*}
\mathcal{L}_{\tau }\theta ^{i}=\frac{1}{2}\left( u^{\mu }\partial _{\mu
}\theta ^{i}+w^{\mu }\partial _{\mu }\theta ^{i}\right) =e^{-2\sigma
}(r_{i}+s_{i})=0,
\end{equation*}%
and so $r_{i}=-s_{i}$ in $\mathcal{M}$. We then have $N^{\alpha }=0$. We
make this choice in our subsequent analysis.

Our choice of fixed quantities, $\left\lbrace Q^{A}\right\rbrace$, is
therefore as follows:

\begin{itemize}
\item We assume that the initial state on $\partial \mathcal{M}_{I}$ is
fixed. Thus, the fixed $\left\{ Q^{A}\right\} $ include $K$ and either $%
\tilde{\gamma}_{\mu \nu }$ or $\sigma ^{\alpha \beta }$ on $\partial 
\mathcal{M}_{I}$.

\item These quantities are fixed with respect to a $\lambda $-independent
coordinate chart $\mathcal{C}=\left\langle u,w,\theta ^{1},\theta
^{2}\right\rangle $ defined such that $u=\tau_{0}=\mathrm{fixed}$ on $\partial 
\mathcal{M}_{I}$, $w=\tau_{0}=\mathrm{fixed}$ at $p_{0}$ and $u=-w$ on $%
\partial \mathcal{M}_{I}$. $\partial \mathcal{M}_{I}$ has fixed unit normal $%
t_{I}^{\mu }=\left. e^{-\sigma }\tau ^{\mu }\right\vert _{\partial \mathcal{M%
}_{I}}$; $\tau ^{\mu }=\nabla ^{\mu }(u+w)/2$.

\item We found that an invariant definition of the boundary term on $%
\partial \mathcal{M}_{u}$ requires us to pick out a preferred set of $%
r=(u-w)/2$ on $\partial \mathcal{M}_{I}$. Our choice of $r$
is to define it so that $e^{-2\sigma }r^{-2}h_{ij}$ has an average
scalar curvature of $2$ on $\partial \mathcal{M}_{I}$.

\item The values $\theta ^{i}$ in $\mathcal{M}$ are defined by
Lie-propagating their values on $\partial \mathcal{M}_{I}$ along $\tau ^{\mu
}$: $\mathcal{L}_{\tau }\theta ^{i}=0$ and so $r^{i}=-s^{i}$ everywhere.
\end{itemize}

Similarly, for the matter variables, we fix the initial state on $\partial 
\mathcal{M}_{I}$ and any residual gauge freedom on $\partial \mathcal{M}_{u}$%
, so that the gauge is fixed independently of $\lambda $.

Given this choice $\left\lbrace Q^{A}\right\rbrace$, the 2-metric, $h_{ij}$,
on $\partial\mathcal{M}_{u}$ is determined by the classical field equations.
Since the field equations depend on $\Lambda$, this, in turn, fixes the form
of $\mathcal{H}_{ij} \equiv \, \mathrm{d} h_{ij}/\, \mathrm{d} \lambda$.

\subsubsection{The $\Lambda $ field equation:}

The field equation for $\Lambda $ in our proposed paradigm is Eq. (\ref%
{eq:phi}). Here, the total action is
\begin{equation*}
I_{\mathrm{tot}}=I_{\mathrm{grav}}[g_{\mu \nu }]+I_{\mathrm{CC}}[\Lambda
,g_{\mu \nu };\mathcal{M}]+I_{\mathrm{matter}}[\Psi ^{a},g_{\mu \nu };%
\mathcal{M}].
\end{equation*}%
where $I_{\mathrm{grav}}=I_{\mathrm{EH}}+I_{\mathrm{GHY}}^{(u)}+I_{\mathrm{YC%
}}^{(I)}$. We assume that $I_{\mathrm{matter}}$ is of the form
\begin{equation*}
I_{\mathrm{matter}}=\int_{\mathcal{M}}\sqrt{-g}\mathcal{L}_{\mathrm{matter}%
}+\int_{\partial \mathcal{M}_{I}}\sqrt{\gamma }\,\mathrm{d}^{3}x\mathcal{L}_{%
\mathrm{mI}},
\end{equation*}%
for some $\mathcal{L}_{\mathrm{mI}}$, where $\mathcal{L}_{\mathrm{matter}}$
and $\mathcal{L}_{\mathrm{mI}}$ are at most first order in derivatives of $%
\Psi ^{a}$. The choice of fixed quantities $\left\{ Q^{A}\right\} $ define
the initial state and ensure all integrals over $\partial \mathcal{M}_{I}$
in $\delta I_{\mathrm{tot}}$ vanish. The classical field equations for
gravity and matter are
\begin{eqnarray}
E^{\mu \nu } &=&-G^{\mu \nu }+\kappa T^{\mu \nu }-\Lambda g^{\mu \nu }, 
\notag \\
\Psi ^{a} &=&-\nabla _{\mu }\Pi _{a}^{\mu }+\frac{\delta \mathcal{L}_{%
\mathrm{matter}}}{\delta \Psi ^{a}}=0,  \notag
\end{eqnarray}%
where
\begin{equation*}
T^{\mu \nu }=\frac{2}{\sqrt{-g}}\frac{\delta (\sqrt{-g}\mathcal{L}_{\mathrm{%
matter}})}{\delta g_{\mu \nu }},\qquad \Pi _{a}^{\mu }=\frac{\delta \mathcal{%
L}_{\mathrm{matter}}}{\delta (\nabla _{\mu }\Psi ^{a})}.
\end{equation*}%
Combined with these classical field equations, the fixed $\left\{
Q^{A}\right\} $ determine $h_{ij}$ and $\Psi ^{a}$ on $\partial \mathcal{M}%
_{u}$ as functions of $\Lambda $. Thus, they give 
\begin{equation*}
\mathcal{H}_{ij}=\left. \frac{\,\mathrm{d}h_{ij}}{\,\mathrm{d}\Lambda }%
\right\vert _{\partial \mathcal{M}_{u};\mathrm{fixed}\,\,\left\{
Q^{A}\right\} },\qquad \mathcal{P}^{a}=\left. \frac{\,\mathrm{d}\Psi ^{a}}{\,%
\mathrm{d}\Lambda }\right\vert _{\partial \mathcal{M}_{u};\mathrm{fixed}%
\,\,\left\{ Q^{A}\right\} }.
\end{equation*}%
When $g_{\mu \nu }$ and $\Psi ^{a}$ obey $E^{\mu \nu }=\Psi ^{a}=0$, we have 
$I_{\mathrm{tot}}=I_{\mathrm{class}}[\Lambda ;\mathcal{M}]$ and so Eq. (\ref%
{eq:phi}) reads 
\begin{eqnarray}
\frac{\,\mathrm{d}I_{\mathrm{class}}}{\,\mathrm{d}\Lambda } &=&\frac{1}{%
2\kappa }\int_{\partial \mathcal{M}_{u}}N\sqrt{h}\,\mathrm{d}\tau \,\mathrm{d%
}^{2}\theta \left[ \mathcal{N}^{ij}\mathcal{H}_{ij}+\Sigma _{a}\mathcal{P}%
^{a}\right] \nonumber \\ &&-\frac{1}{\kappa }\int_{\mathcal{M}}\sqrt{-g}\,\mathrm{d}^{4}x = 0.  \notag
\end{eqnarray}%
where $\mathcal{N}^{ij}=(\mathcal{K}+\nu )h^{ij}-\mathcal{K}^{ij}$ and $%
\Sigma _{a}=-n_{\mu }\Pi _{a}^{\mu }$.

\subsubsection{Rewriting the Classical Action}

When the classical field equations for the metric and matter variables hold,
we can rewrite $I_{\mathrm{class}}$ in a form that is particularly
instructive in the cosmological setting for both expressing and solving the $%
\lambda$ equation.

Independent of the field equations, we can rewrite $I_{\mathrm{GHY}}^{(u)}$
 as 
\begin{eqnarray}
\kappa I_{\mathrm{GHY}}^{(u)} &=&\int_{\partial \mathcal{M}_{u}}e^{\sigma }%
\sqrt{h}\left[ \mathcal{K}+\nu \right] \,\mathrm{d}\tau \,\mathrm{d}%
^{2}\theta \nonumber \\ &=&\int_{\partial \mathcal{M}_{u}}\left[ \nabla _{\mu }u^{\mu
}+u^{\mu }\nabla _{\mu }\sigma \right] \sqrt{-g}\,\mathrm{d}\tau \,\mathrm{d}%
^{2}\theta   \notag \\
&=&-\int_{\mathcal{M}}\sqrt{-g}\nabla _{\mu }\left[ t^{\mu }\mathcal{L}%
_{n}\sigma \right] \,\mathrm{d}^{4}x \nonumber \\ &&-\lim_{\tau \rightarrow
0^{+}}\int_{\partial \mathcal{M}_{I}}e^{\sigma }\sqrt{h}\mathcal{L}%
_{n}\sigma \,\mathrm{d}r\,\mathrm{d}^{2}\theta,  \notag
\end{eqnarray}%
where we have used $u_{\mu }u^{\mu }=0$, and $\sqrt{-g}=0$ on $\tau =0$ and
at $p_{0}$. We have also used $n^{\mu }=-e^{\sigma }u^{\mu }$, $t^{\mu
}=e^{\sigma }\nabla ^{\mu }\tau $ and $n^{\mu }t_{\mu }=1$. Now, 
\begin{equation*}
\mathcal{L}_{n}\sigma =n^{\mu }\nabla _{\mu }\sigma =t^{\nu }n^{\mu }\nabla
_{\mu }n_{\nu }=-n^{\mu }n^{\nu }\nabla _{\mu }t_{\nu }=-K^{r}{}_{r}-a_{r},
\end{equation*}%
where $\nabla _{\mu }t_{\nu }=K_{\mu \nu }-t_{\mu }a_{\nu }$, and $a_{\mu
}=t^{\nu }\nabla _{\nu }t^{\mu }$ is the acceleration and $K_{\mu \nu }$ the
extrinsic curvature of constant $\tau $ hypersurfaces. We have also defined $%
l^{\mu }=e^{\sigma }\nabla ^{\mu }r=e^{\sigma }\nabla ^{\mu }\left[ u-w%
\right] /2$, so that $n^{\mu }=-(t^{\mu }+r^{\mu })$ and $K^{r}{}_{r}=l^{\mu
}l^{\nu }K_{\mu \nu }$, $a_{r}=l^{\mu }a_{\mu }$.

Since 
\begin{eqnarray}
\lim_{\tau \rightarrow 0^{+}}\left[\int_{\partial \mathcal{M}_{I}}e^{\sigma }\sqrt{%
h}a_{r}\,\mathrm{d}r\,\mathrm{d}^{2}\theta =\int_{\partial \mathcal{M}_{I}}t_{\mu }n^{\mu }e^{\sigma }\sqrt{h}%
a_{r}\,\mathrm{d}r\,\mathrm{d}^{2}\theta \right],  \notag \\
=\int_{\mathcal{M}}\sqrt{-g}\,\mathrm{d}^{4}x\,\nabla _{\mu }\left[ n^{\mu
}a_{r}\right] ,  \notag
\end{eqnarray}%
we can write: 
\begin{eqnarray}
I_{\mathrm{GHY}}^{(u)}&=&\frac{1}{\kappa }\int_{\mathcal{M}}\sqrt{-g}\nabla
_{\mu }\left[ t^{\mu }K^{r}{}_{r}-l^{\mu }a_{r}\right] \,\mathrm{d}%
^{4}x \nonumber \\ &&+\lim_{\tau \rightarrow 0^{+}}\frac{1}{\kappa }\int_{\partial \mathcal{M%
}_{I}}e^{\sigma }\sqrt{h}K^{r}{}_{r}\,\mathrm{d}r\,\mathrm{d}^{2}\theta . \nonumber
\end{eqnarray}%
The Ricci tensor of $\mathcal{M}$ is $R_{\mu \nu }$, and we define $R_{\mu
\nu }^{(3)}$ to be the Ricci tensor of a 3-surface of constant $\tau $. We
define $R^{r}{}_{r}=R_{\mu \nu }l^{\mu }l^{\nu }$, $R^{(3)r}{}_{r}=R_{\mu
\nu }^{(3)}l^{\mu }l^{\nu }$. We then have: 
\begin{eqnarray}
\nabla _{\mu }\left[ t^{\mu }\mathcal{L}_{n}\sigma -l^{\mu }a_{r}\right]
&=&-R_{r}^{r}+R_{r}^{(3)r}+\mathcal{L}_{l}a_{r} \nonumber \\ && +A^{2}-\Sigma ^{2}-a_{r}%
\mathcal{K}_{l}, \nonumber\\ \Sigma ^{2} &=&2\Sigma _{\mu }\Sigma ^{\mu },\qquad \Sigma _{\mu }=h_{\mu
\nu }l_{\rho }K^{\nu \rho } ,  \notag \\
 A^{2} &=& A_{\mu }A^{\mu },\qquad A_{\mu }=h_{\mu \nu }a^{\mu },\nonumber \\  
\mathcal{K}_{l} &=&h^{\mu \nu }\nabla _{\mu }l_{\nu }=\frac{1}{2}h^{\mu \nu }%
\mathcal{L}_{l}h_{\mu \nu }.   \notag
\end{eqnarray}%
We now define: 
\begin{equation}
\Gamma \equiv R_{r}^{(3)r}+\mathcal{L}_{l}a_{r}+A^{2}-\Sigma ^{2}-a_{r}%
\mathcal{K}_{l}. \label{eq:Gamma}
\end{equation}%
We can then write: 
\begin{eqnarray}
I_{\mathrm{tot}} &=&I_{\mathrm{EH}}+I_{\mathrm{GHY}}^{(u)}+I_{\mathrm{YC}%
}^{(I)}+I_{CC}+I_{\mathrm{matter}},  \notag \\
&=&\frac{1}{\kappa }\int_{\mathcal{M}}\sqrt{-g}\left[ \frac{1}{2}%
R-R^{r}{}_{r}-\Lambda +\Gamma +\kappa \mathcal{L}_{\mathrm{matter}}\right] \,%
\mathrm{d}^{4}x  \notag \\
&&+\lim_{\tau \rightarrow 0^{+}}\frac{1}{\kappa }\int_{\partial \mathcal{M}%
_{I}}\sqrt{\gamma }\left[ K_{r}^{r}-\frac{1}{3}K+\mathcal{L}_{\mathrm{mI}}%
\right] \,\mathrm{d}^{3}x.  \notag
\end{eqnarray}%
$I_{\mathrm{class}}$ is defined to be $I_{\mathrm{tot}}$ evaluated with $%
g_{\mu \nu }$ and the matter fields obeying their classical field equations.
For $g_{\mu \nu }$ this means that we have the Einstein equation $G_{\mu \nu
}=R_{\mu \nu }-g_{\mu \nu }R/2=\kappa T^{\mu \nu }-\Lambda g_{\mu \nu }$
where $T^{\mu \nu }$ is the energy-momentum tensor that follows from varying 
$\mathcal{L}_{\mathrm{matter}}$. Substituting the Einstein equation into $I_{%
\mathrm{tot}}$ and defining $P^{r}{}_{r}=T_{\mu \nu }l^{\mu }l^{\nu }$ we
arrive at: 
\begin{eqnarray}
I_{\mathrm{class}} &=&\int_{\mathcal{M}}\sqrt{-g}\left[ \kappa ^{-1}\Gamma +(%
\mathcal{L}_{\mathrm{matter}}-P^{r}{}_{r})\right] \,\mathrm{d}^{4}x \\
&&+\lim_{\tau \rightarrow 0^{+}}\frac{1}{\kappa }\int_{\partial \mathcal{M}%
_{I}}\sqrt{\gamma }\left[ K_{r}^{r}-\frac{1}{3}K+\mathcal{L}_{\mathrm{mI}}%
\right] \,\mathrm{d}^{3}x.  \notag
\end{eqnarray}%
Since the initial state on $\partial \mathcal{M}_{I}$ is taken to be fixed,
the equation for the effective cosmological constant $\Lambda $ can be
simply written as: 
\begin{eqnarray}
\frac{\,\mathrm{d}I_{\mathrm{class}}}{\,\mathrm{d}\lambda }&=&\int_{\mathcal{M}%
}\frac{\delta }{\delta \Lambda }\left\{ \sqrt{-g}\left[ \kappa ^{-1}\Gamma +(%
\mathcal{L}_{\mathrm{matter}}-P^{r}{}_{r})\right] \right\} \,\mathrm{d}%
^{4}x \nonumber \\ &=&0. \nonumber
\end{eqnarray}%
In the above equation $\mathcal{L}_{\mathrm{matter}}$ is the effective
action for matter renormalized so that it vanishes in vacuo. In a
cosmological setting, this form of the $\Lambda $ equation is often the most straightforward 
to evaluate since it involves only scalar quantities.

\subsection{$\Lambda$ in a Realistic Cosmology}

In the previous subsection, we considered the application of our scheme for
determining $\Lambda$ in a general cosmological setting where $\mathcal{M}$
is taken to be the past light cone of the observer at some fixed external
time $\tau = \tau_{0}$.

We assume that, in appropriate coordinates $(T,X^{i})$, and except in
certain strong gravity regimes (eg. near neutron stars or black hole
horizons), the space-time is well-described, to linear order in some small $%
\Psi $ and $\Phi $, by the following line element: 
\begin{eqnarray}
\,\mathrm{d}s^{2}&=&a^{2}(T)\left[ -(1+2\Psi )\,\mathrm{d}T^{2}\right. \nonumber \\ && \left.+\frac{(1-2\Phi
)}{\left( 1+\frac{1}{4}k X^{\alpha }X^{\beta }\delta _{\alpha \beta }\right)
^{2}}\delta _{\alpha \beta }\,\mathrm{d}X^{\alpha }\,\mathrm{d}X^{\beta }%
\right], \nonumber
\end{eqnarray}%
where $\alpha ,\beta $ take values $1,2,3$; $\Psi $ and $\Phi $ are
gravitational potentials which are sourced by perturbations to the homogeneous
background. They are measured to be small ($\sim O(10^{-5})$) on average; $k$
is the intrinsic spatial curvature. Observations indicate that at the
horizon $|k X^{\alpha }X^{\beta }\delta _{\alpha \beta }|\lesssim 10^{-2}$ so
that to linear order in this and the other small quantities, $\Psi $ and $%
\Phi $: 
\begin{widetext}
\begin{equation}
\,\mathrm{d}s^{2}\approx a^{2}(T)\left[ -(1+2\Psi )\,\mathrm{d}T^{2}+\left(
1-2\Phi -\frac{1}{2}k X^{\alpha }X^{\beta }\delta _{\alpha \beta }\right)
\delta _{\alpha \beta }\,\mathrm{d}X^{\alpha }\,\mathrm{d}X^{\beta }\right] . \label{eq:perturbed}
\end{equation}
\end{widetext}
We now apply our method for solving the CC problems to a universe with line
element given by Eq. (\ref{eq:perturbed}). We transform this line element to
light-cone coordinates $(u,w,\theta ^{i})$, by taking for some small $%
\partial _{\mu }\partial _{\nu }\gamma $, 
\begin{equation*}
\tau =T+\dot{\gamma},\qquad x^{\alpha }=X^{\alpha }+\delta ^{\alpha \beta
}\gamma _{,\beta },
\end{equation*}%
here $\dot{\gamma}=\partial \gamma /\partial \tau $ and $\gamma _{,\alpha
}=\partial \gamma /\partial x^{\alpha }$. To linear order in the small
quantities, we have $\partial \gamma /\partial \tau =\partial \gamma /\partial T$ and 
$\partial \gamma /\partial x^{\alpha }=\partial \gamma /\partial X^{\alpha }$%
. Here, $\tau =(u+w)/2$ and $r=\sqrt{\delta _{\alpha \beta }x^{\alpha
}x^{\beta }}=(u-w)/2$. With such a change to linear order in
the small quantities, we obtain
\begin{eqnarray}
\,\mathrm{d}s^{2} &=&a^{2}(\tau )(1+2\Omega )\left[ -\,\mathrm{d}u\,\mathrm{d%
}w+\bar{h}_{\alpha \beta }\,\mathrm{d}x^{\alpha }\,\mathrm{d}x^{\beta }%
\right] , \\
\Omega &=&\Psi -\ddot{\gamma}-\mathcal{H}\dot{\gamma}=-\Phi -\gamma ^{\prime
\prime }-\mathcal{H}\dot{\gamma}, \\ \qquad \ddot{%
\gamma} &-& \gamma ^{\prime \prime } =\Psi +\Phi +\frac{1}{4}kr^{2},
\end{eqnarray}%
with $\mathcal{H}=a_{,\tau }/a$ and $\bar{h}_{\alpha \beta }=\bar{h}_{\alpha
\beta }^{(0)}-2\gamma _{,\alpha \beta }+2\gamma ^{\prime \prime }\delta
_{\alpha \beta }$, where $\bar{h}_{\alpha \beta }^{(0)}\,\mathrm{d}x^{\alpha
}\,\mathrm{d}x^{\beta }=r^{2}\left[ \,\mathrm{d}\theta ^{2}+\sin ^{2}\,%
\mathrm{d}\phi ^{2}\right] $ and the prime superscript indicates a partial
derivative with respect to $r$; also, $x^{\alpha }x^{\beta }\bar{h}_{\alpha
\beta }=0$.

To leading order in $\Phi $, $\Psi $ and $\gamma _{,\mu \nu },$ the line
element is simply that of a Friedmann-Robertson-Walker space-time with
curvature $k$ and $\tau $ is the conformal time coordinate: 
\begin{eqnarray}
\,\mathrm{d}s_{0}^{2} &=&a^{2}(\tau )\left[ -\,\mathrm{d}u\,\mathrm{d}%
w+r^{2}\left( \,\mathrm{d}\theta ^{2}+\sin ^{2}\theta \,\mathrm{d}\phi
^{2}\right) \right] ,  \notag \\
&=&a^{2}(\tau )\left[ -\,\mathrm{d}\tau ^{2}+\,\mathrm{d}r^{2}+r^{2}\left( \,%
\mathrm{d}\theta ^{2}+\sin ^{2}\theta \,\mathrm{d}\phi ^{2}\right) \right] .
\notag
\end{eqnarray}

\subsubsection{Initial conditions for $\protect\gamma$}

If $\gamma _{0}$ satisfies the $\gamma $ equation then so does $\gamma
_{1}\rightarrow \gamma _{0}+f_{-}(\tau -r,\theta ,\phi )+f_{+}(\tau
+r,\theta ,\phi )$, for arbitrary $f_{\pm }(x,\theta \phi )$. Changing $%
\gamma $ from $\gamma _{0}$ to $\gamma _{1}$ shifts $\sqrt{h}e^{\sigma }\nu
\,\mathrm{d}w$ on $\partial \mathcal{M}_{u},$ and hence $I_{\mathrm{GHY}%
}^{(u)}$ by a term proportional to $f_{-,xxx}(x,\theta ,\phi )$. Thus, to
fix the definition of $I_{\mathrm{GHY}}^{(u)}$, we must impose initial
conditions that fix $f_{-,xxx}(x,\theta ,\phi )$.

First, we must specify $\tau =0$ to correspond to a given timelike
hypersurface (e.g. $T=0$). This determines $\dot{\gamma}$ and hence $%
f_{+,x}(x,\theta ,\phi )$ in terms of  $f_{-,x}(x,\theta ,\phi )$. For
simplicity, we choose the fixed hypersurface to be $T=0$, although similar
choices that coincide with this choice to zeroth order in the small
quantities will give similar results to the ones we obtain below. The
boundary term $I_{\mathrm{GHY}}^{(u)}$ can then be fixed by specifying $%
\gamma ^{\prime \prime \prime }$ on $\partial \mathcal{M}_{I}$. It can be
checked that fixing $r$ so that the average curvature of the conformal
2-metric on $\partial \mathcal{M}_{I}$ is $2/r^{2}$ gives $\gamma ^{\prime \prime \prime }=0,$ and hence clearly fixes $I_{%
\mathrm{GHY}}^{(u)}$. We therefore make this choice for $r$ on $\partial 
\mathcal{M}_{I}$.

\subsubsection{Evaluation of $I_{\mathrm{class}}$}

We can calculate the $\Gamma $ quantity defined by Eq. (\ref{eq:Gamma}) for
this line element. We find that to linear order in the small quantities
\begin{equation*}
\Gamma =\frac{2k}{a^{2}(\tau )}+\frac{2\Phi ^{\prime \prime }}{a^{2}(\tau )}+%
\frac{2(\Phi -\Psi )^{\prime }}{a^{2}(\tau )r}+\frac{2}{a^{4}(\tau )r}\left[
a^{2}(\tau )\dot{\gamma}^{\prime }\right] _{,\tau }+\dots, 
\end{equation*}%
where the $\dots $ indicate terms of linear order which are total
derivatives with respect to the angular coordinates and so vanish when
integrated over $S_{(u,w)}$.

For simplicity, we take the energy-momentum tensor of matter to have a
perfect-fluid form: 
\begin{equation*}
T^{\mu \nu }=U^{\mu }U^{\nu }(\rho _{\mathrm{m}}+P_{\mathrm{m}})+P_{\mathrm{m%
}}g^{\mu \nu },
\end{equation*}%
where $U^{\mu }$ is a forward pointing time-like vector with $U^{\mu }U_{\mu
}=-1$. To leading order, we have $U^{\tau }U_{\alpha }=-v_{,\alpha }$ for $%
\alpha =1,2,3$. We write 
\begin{equation*}
\rho _{\mathrm{m}}=\bar{\rho}(\tau )+\frac{\bar{\rho}_{,\tau }}{\bar{\rho}}%
v+\delta \rho .
\end{equation*}%
We assume that at those sufficiently late times that provide the dominant
contributions to $I_{\mathrm{class}}$, the background cosmology is either
dominated by pressureless matter or $\Lambda $, and so $P_{\mathrm{m}}\ll
\rho _{\mathrm{m}}$. The dominant contribution to $P_{\mathrm{m}}$ is then
from photons (and light neutrinos) and may be approximated as homogeneous to
the order to which we work i.e. $\delta P_{\mathrm{m}}/\rho _{\mathrm{m}}\ll
\Phi ,\Psi $. We therefore take $P_{\mathrm{m}}=\bar{P}(\tau )$.

The quantities $a(\tau)$, $\Psi$ and $\Phi$ are then given by
\begin{eqnarray}
H^2 &=& \frac{1}{a^2} \mathcal{H}^2 = \left(\frac{a_{,\tau}}{a^2}\right)^2 = 
\frac{8\pi G \bar{\rho}}{3} + \frac{\Lambda}{3} - \frac{k}{a^2}, \nonumber \\
 \dot{%
\bar{\rho}} &=& -3\mathcal{H}(\bar{\rho}+\bar{P}),  \nonumber \\
\Phi &=& \Psi, \qquad \vec{\nabla}^2 \Phi = 4\pi G a^2 \delta \rho,  \nonumber \\
\ddot{\Phi} &+& 3\mathcal{H}\dot{\Phi} + \left(2\dot{\mathcal{H}}+\mathcal{H}%
^2 \right) \Phi = 0, \nonumber
\end{eqnarray}
where $\mathcal{H}= a_{,\tau}/a$, and $\vec{\nabla}^2 = \delta^{\alpha
\beta}\partial_{\alpha}\partial_{\beta}$ with $\alpha,\beta = 1,2,3$. We
also have that $P^{r}{}_{r} = T_{\mu \nu}l^{\mu}l^{\nu}$ is (to linear
order): 
\begin{eqnarray}
P^{r}_{r} = \bar{P}(\tau).
\end{eqnarray}

Thus, to linear order in $\Phi $ and $\gamma _{,\mu \nu },$ we have 
\begin{eqnarray}
I_{\mathrm{class}}&=&\int_{0}^{\tau _{0}}a^{4}(\tau )\int_{0}^{\tau _{0}-\tau
}r^{2}\,\mathrm{d}r\int \,\mathrm{d}^{2}\Omega \left[ \kappa ^{-1}\Gamma \right. \nonumber  \\ &&\left. +(%
\mathcal{L}_{\mathrm{matter}}-\bar{P})\right] , \nonumber
\end{eqnarray}%
where we have dropped the terms on $\partial \mathcal{M}_{I}$ which are
fixed with respect to $\Lambda $.

Before we consider the variation of $I_{\mathrm{class}}$ with respect to $%
\Lambda $, we must extract the dominant contribution to the effective
Lagrangian density $\mathcal{L}$ of the matter.

\subsubsection{Contributions to $\bar{P}$ and $\mathcal{L}_{\mathrm{matter}}$%
}

For fields that are truly homogeneous (to leading order), for example the
inflaton or other light scalar fields, we have $\mathcal{L}_{\mathrm{matter}%
}=\bar{P}$, and so these fields make no contribution to $I_{\mathrm{class}}$.

We first clarify the definition of the quantity $\mathcal{L}_{\rm matter}$ that appears in $I_{\rm class}$ when quantum contributions to the matter action and non-negligible.  Formally, the $\mathcal{L}_{\rm matter}$ that appears in $I_{\rm class}$ is the quantum effective matter Lagrangian, $\mathcal{L}_{\rm matter}^{\rm eff}$, rather than the classical matter action.   The since the quantum vacuum energy associated with the matter have been subsumed into the definition of $\Lambda$, this $\mathcal{L}_{\rm matter}^{\rm eff}$ vanishes, by definition, in the vacuum.  Let $n_{A}$ represent a set of conversed quantities associated with the matter species, such that in the vacuum $n_{A} = 0$. For instance, an $n_{A}$ could be baryon number.  If $\mathcal{L}_{\rm matter}^{\rm cl}$ is the classical matter action, $\mathcal{L}_{\rm matter}^{\rm eff}$ is then given by:
\begin{eqnarray}
I_{\rm m}^{\rm eff}[n_{A}] &=& \int_{\M}\sqrt{-g}\mathcal{L}_{\rm matter}^{\rm eff} \dd^4 x \\ &\equiv&  {\rm Re} \left[  -i \ln \left(\frac{Z_{\rm m}[n_{A}]}{ Z_{\rm m}[0]}\right)\right], \nonumber \\
Z_{m}[n_A] &=& \sum_{\substack{ \Psi ^{a}  \\ 
\mathrm{fixed}\,\,\left\{ n_{A}\right\} }}  e^{i \int_{\M}\sqrt{-g}\mathcal{L}_{\rm matter}^{\rm cl}[\Psi^{a},g_{\mu \nu}]\dd^4 x}. \nonumber
\end{eqnarray}
We recognize that $I_{\rm m}^{\rm eff}[n_{A}]$ is the quantum effective matter action, normalized so that it vanishes \emph{in vacuo}.

For free fields, the quantum effective action has the same structure as the classical action.  It is well known that $I_{\rm m}^{\rm eff}$ and hence $\mathcal{L}_{\mathrm{matter}}$ vanishes identically for free fundamental fermion fields.  For fermion fields, $\psi$, with energy density $\rho_{\psi}$, that are weakly coupled to a gauge fields with a coupling constant $g \ll 1$ one typically has $\mathcal{L}_{\rm matter} \sim O(g^4) \rho_{\psi} \ll \rho_{\psi}$.  

For photons, to leading order  $\mathcal{L}_{\rm matter}^{\rm eff} \approx  \mathcal{L}_{\mathrm{matter}}^{\rm cl}=-F_{\mu \nu }F^{\mu \nu }/4=(E^{2}-B^{2})/2$, and for
radiation $E^{2}=B^{2}$ and so $\mathcal{L}_{\mathrm{matter}}=0$. More
generally, for an (approximately) free field, $\phi^{A}$, with energy $\omega \gg H$,  $\mathcal{L}_{\mathrm{matter}}^{\rm eff} \approx \mathcal{L}_{\mathrm{matter}}^{\rm cl} $ and the average value of the effective Lagrangian is proportional to the
dispersion relation and so, on-shell, so we have $\mathcal{L}_{\mathrm{matter}%
}\lesssim O(H/\omega )\rho _{\mathrm{matter}}\ll \rho _{\mathrm{matter}}$
for the contribution from $\phi ^{A}$. Since the mass of dark matter
particles is $\gg H$ today, we assume there their contribution to $\mathcal{L%
}_{\mathrm{matter}}$ is much less than their energy density.

Therefore, amongst the fields that contribute to $I_{\mathrm{class}}$, the
dominant contribution to $\mathcal{L}_{\mathrm{matter}}$ at late times is
expected to come from baryonic matter. Baryons contribute most because they are
not fundamental fermions fields, but composite particles consisting of
quarks bound strongly together with gluons. 

We define $\rho_{\rm baryon}$ and $n_{\rm baryon}$ to be the baryon energy and number density respectively.  For baryonic matter we have:
\be 
Z_{\rm m} &\approx& Z_{\rm baryon}[n_{\rm baryon}] \\
&=& \sum_{\substack{ q, \bar{q}, A_{\mu}  \\ 
\mathrm{fixed}\,\,\left\{ n_{\rm baryon}\right\} }}  e^{i \int_{\M}\sqrt{-g}\mathcal{L}_{\rm QCD}^{\rm cl}[q, \bar{q}, A_{\mu}; g_{\mu \nu}]\dd^4 x}. \nonumber
\ee
where $q$ and $\bar{q}$ are the quark fields and $A_{\mu}$ is the gluon field.  $\mathcal{L}_{\rm QCD}^{\rm cl}$ is the classical action for Quantum Chromodynamics  (QCD).   Now, we have 
\begin{eqnarray}
{\rm Re}\left[-i\ln Z_{\rm m} \right] &=& -\int_{\M} \sqrt{-g}\dd^4 x\, \rho_{\rm QCD-vac} \\ &&+ I_{\rm m}^{\rm eff}, \nonumber
\end{eqnarray}
where $\rho_{\rm QCD-vac}$ is the QCD contribution to the vacuum energy density.   At late times when the baryonic matter is non-relativistic, and at sub-nuclear densities ($\rho_{\rm baryon} \ll 10^{17}\,{\rm kg}\,{\rm m}^{-3}$ on average), we have $I_{\rm m}^{\rm eff} \approx I_{\rm m}^{\rm eff}[n_{\rm baryon}]$ and
$$
I_{\rm m}^{\rm eff}[n_{\rm baryon}] \approx  -\Gamma_{\rm b} \int_{\M} \sqrt{-g} n_{\rm baryon} \dd^4 x.
$$
for some constant $\Gamma_{\rm b}$. At late times, for non-relativistic baryonic matter, $\rho_{\rm baryon} = M_{\rm N} n_{\rm baryon}$, where $M_{\rm N}$ is the nucleon mass. We define the constant $\zeta_{\rm b}= \Gamma_{\rm b}/M_{\rm N}$.  For baryonic matter we it follows that:
\be 
\mathcal{L}_{\rm matter}^{\rm eff} \approx -\zeta_{\rm b} \rho_{\rm baryon}.
\ee
In principle, $\zeta_{\rm b}$ is calculable and depends only on QCD physics.  A full calculation of $\zeta_{\rm b}$ would, however, require either the derivation of the complete
low-energy effective action for QCD, or a time consuming and technically challenging lattice QCD calculation.  Both of these are far beyond the scope of this work. 

The chiral bag model (CBM) for nucleons is described by the effective
Lagrangian: 
\begin{eqnarray}
\mathcal{L}_{\mathrm{CBM}} &=&\left( \bar{\psi}i\gamma ^{\mu }\nabla _{\mu
}\psi -B\right) \theta (R-r) \\ &&-\frac{1}{2}\bar{\psi}e^{\frac{i\vec{\tau}\cdot 
\vec{\pi}\gamma _{5}}{f_{\pi }}}\psi \delta (r-R)-\mathcal{L}_{\pi }\theta
(r-R), \nonumber \\
\mathcal{L}_{\pi } &=&-\frac{f_{\pi }^{2}}{4}\mathrm{tr}\left[ L_{\mu
}L^{\mu }\right] +\frac{1}{32e^{2}}{\mathrm{tr}\,}\left[ L_{\mu },L_{\nu }%
\right] ^{2},  \notag \\
L_{\mu } &=&(\nabla _{\mu }U)U^{\dagger },\qquad U=e^{\frac{i\vec{\tau}\cdot 
\vec{\pi}}{f_{\pi }}}.  \notag
\end{eqnarray}%
Here, $R$ is the `bag radius', and $B$ is the `bag constant' which has been
interpreted as the difference between the vacuum energy of the perturbative
and non-perturbative QCD vacuums; $\mathcal{L}_{\pi }$ is the Skyrme action.
In $r<R,$ i.e. inside the bag, just free quarks and the bag constant
contribute to the mass and the action. Outside the bag quark degrees of
freedom have been confined and mesons are the effective
degrees of freedom.  We use the CBM model to approximate the effective matter action $\mathcal{L}_{\rm matter}^{\rm eff} \approx \mathcal{L}_{\rm CBM}$.

The total energy-momentum tensor is: 
\begin{equation*}
T_{\mathrm{CBM}}^{\mu \nu }=T_{q}^{\mu \nu }-Bg^{\mu \nu }\theta
(R-r)+T_{\pi }^{\mu \nu }.
\end{equation*}%
where $T_{q}^{\mu \nu }g_{\mu \nu }=0$. The meson configuration outside the
bag is given by a static soliton solution to a first approximation and so $%
T_{\pi }^{00}=-\mathcal{L}_{\pi }\theta (r-R)$. The total nucleon mass is
given by integrating $T_{\mathrm{CBM}}^{00}$ over the spatial directions,
and so 
\begin{equation*}
M_{\mathrm{N}}=M_{\mathrm{q}}+\frac{4\pi }{3}BR^{3}+M_{\pi }.
\end{equation*}%
We calculate the expectation of $\mathcal{L}_{\mathrm{CBM}}$, for a single
nucleon, integrated over the spatial hypersurface to be: 
\begin{eqnarray}
4\pi \int r\,\mathrm{d}r\left\langle \mathcal{L}_{\mathrm{CBM}}\right\rangle
&=&-\frac{4\pi }{3}BR^{3}-M_{\pi } \\ &=&-(M_{\mathrm{N}}-M_{\mathrm{q}})=-\zeta _{%
\mathrm{b}}M_{\mathrm{N}}, \nonumber \\
\zeta _{\mathrm{b}} &=&1-\frac{M_{\mathrm{q}}}{M_{\mathrm{N}}}.
\end{eqnarray}%
In general, for a collection of baryons (specifically nucleons) with
energy density $\rho _{\mathrm{b}}$, we have: 
\begin{equation*}
\mathcal{L}_{\mathrm{baryons}}\approx \mathcal{L}_{\mathrm{CBM}}=-\zeta _{%
\mathrm{b}}\rho _{\mathrm{b}}.
\end{equation*}%
The value of $\zeta _{\mathrm{b}}$ in the CBM depends on the bag radius. Ref. \cite%
{Hosaka:1996ee} provides an excellent review of the CBM. The authors note
that the best agreement with experimental physics is found when $R\approx
0.6\,\mathrm{fm}$. For this value they have $M_{\mathrm{q}}\approx M_{\pi
}+M_{\mathrm{B}}$ where $M_{\mathrm{B}}=4\pi BR^{3}/3$. Thus we have $\zeta_{\rm b}
\approx 0.5$. This estimate will be slightly reduced when the contributions
of spin to the nucleon mass are taken into account

Henceforth, we take: 
\begin{equation*}
\mathcal{L}_{\mathrm{matter}} =-\zeta _{\mathrm{b}}\rho _{\mathrm{baryon}},
\end{equation*}%
where from the CBM we use the estimate that $\zeta _{\mathrm{b}}\approx 1/2$.

\subsubsection{$\Lambda$ Equation}

The dominant contribution to the pressure term, $\bar{P}$, at late times
will come from radiation. However since $a^{4}\bar{P}=\mathrm{const}$ for
radiation this contribution just shifts $I_{\mathrm{class}}$ by a $\lambda $%
-independent constant. Dropping such constants, and any terms that are an
order of magnitude smaller than those included, we find that to leading
order: 
\begin{eqnarray}
I_{\mathrm{class}} &\approx &\int_{0}^{\tau _{0}}a^{4}(\tau )\int_{0}^{\tau
_{0}-\tau }r^{2}\,\mathrm{d}r\int \,\mathrm{d}^{2}\Omega \left[ \kappa
^{-1}a^{-2}\bar{\Gamma} \right.  \\ && \left. -\zeta _{\mathrm{b}}\rho _{\mathrm{baryon}}\right] +%
\mathrm{const},  \notag \\
\bar{\Gamma} &=&2k+2\Phi ^{\prime \prime }+\frac{2}{r^{2}}\left[ r\dot{\gamma%
}^{\prime }\right] ^{\prime }, \\
\ddot{\gamma} &-&\gamma ^{\prime \prime }=2\Phi +kr^{2}/4. 
\end{eqnarray}%
Here, we have integrated by parts to express the $\bar{\Gamma}$ term in $I_{%
\mathrm{class}}$ in the above form. If $k r^{2}\gg \Phi $ then to leading
order in deviations from flat $\Lambda $CDM we can drop $\Phi $ in the
formulae for $\bar{\Gamma}$ and $\gamma $, leaving only the contribution
from $k$.

We then have 
\begin{equation*}
\ddot{\gamma}-\gamma ^{\prime \prime }=kr^{2}/4.
\end{equation*}%
Solving with the required boundary conditions gives: 
\begin{eqnarray}
\gamma &=&-\frac{k r^{4}}{48}+\frac{k (r-\tau )^{4}}{96}+\frac{k (r+\tau )^{4}}{96%
},\nonumber  \\
\dot{\gamma}^{\prime } &=& -\frac{k(r-\tau )^{2}}{8}+\frac{k(r+\tau )^{2}}{8}=%
\frac{kr\tau }{2}. \nonumber
\end{eqnarray}%
Inserting this expression for $\gamma $ into $\bar{\Gamma}$ gives: 
\begin{equation*}
\int_{0}^{\tau _{0}-\tau }r^{2}\bar{\Gamma}\,\mathrm{d}r=\frac{2k}{3}(\tau
_{0}-\tau )^{3}+k\tau (\tau _{0}-\tau )^{2}.
\end{equation*}%
Thus, we evaluate $I_{\mathrm{class}}$ to lowest order as: 
\begin{eqnarray}
I_{\mathrm{class}} &\approx &\frac{4\pi }{\kappa }\int_{0}^{\tau
_{0}}ka^{2}(\tau )\left[ \frac{2}{3}(\tau _{0}-\tau )^{3}+\tau (\tau
_{0}-\tau )^{2}\right] \,\mathrm{d}\tau \nonumber  \\
&&-\frac{4\pi \zeta _{\mathrm{b}}}{\kappa }\int_{0}^{\tau _{0}}a^{4}\frac{%
\kappa \rho _{\mathrm{baryon}}}{3}(\tau _{0}-\tau )^{3}\,\mathrm{d}\tau +%
\mathrm{const}.  \notag
\end{eqnarray}%
We note that $a^{4}\rho _{\mathrm{baryon}}\propto a$. To this order the only
quantity in $I_{\mathrm{class}}$ that depends on $\Lambda $ is $a(\tau )$,
since we have assumed the initial conditions that determine $k$ are fixed.
Additionally,  baryogenesis and the processes which generates the dark matter density must occur at such early times that they  will have only a negligible $\Lambda$ dependence.  This implies that $a^3 \rho_{\rm m}$ and $a^3 \rho_{\rm baryon}$ are fixed independently of $\Lambda$.  Additionally, the initial conditions fix $a^3 n_{\gamma}$, where $n_{\gamma}$ is the photon number density, independently of $\Lambda$.   Given this, the $\Lambda$ independent initial conditions for the matter sector are parametrized by the energy of matter energy per photon, $\xi = \rho_{\rm m}/n_{\gamma} = {\rm const}$ and the baryon energy per photon, $\xi_{\rm b} =\rho_{\rm baryon}/n_{\rm gamma}$; $\delta \xi/\delta \Lambda = \delta \xi_{\rm b}/\delta \Lambda =0$. The measured values $\xi$ and $\xi_{\rm b}$ are $\xi = 3.43\,{\rm eV}$ and $\xi_{\rm b} = 0.54\,{\rm eV}$..

We define 
\begin{equation*}
\mathcal{A}(\tau )=\frac{\delta \ln a}{\delta \Lambda }
\end{equation*}%
and use the Friedmann equation for the background to calculate $\mathcal{A}%
(\tau )$. Under the change $\Lambda \rightarrow \Lambda +\delta \Lambda $
the Friedmann equation is perturbed to 
\begin{equation*}
2\mathcal{H}\delta \mathcal{H}=2\mathcal{H}_{,\tau }\delta \ln a+a^{2}\delta
\Lambda /3.
\end{equation*}%
Now $\mathcal{H}=\dot{a}/a$ and so $\delta \mathcal{H}=(\delta \ln a)_{,\tau}$ and thus $\delta \mathcal{H}/\delta \Lambda = {\cal A}_{,\tau}$.  It follows that:
\begin{equation*}
\mathcal{H}\dot{\mathcal{A}}-\dot{\mathcal{H}}\mathcal{A}=\frac{a^{2}}{6}%
,\Rightarrow \left( \frac{\mathcal{A}}{\mathcal{H}}\right) _{,\tau }=\frac{%
a^{2}}{6\mathcal{H}^{2}}.
\end{equation*}%
The condition that the extrinsic curvature, $K$, of the initial hypersurface be fixed independently of $\Lambda$ is equivalent to $\delta H/\delta \Lambda = 0$ at $\tau=0$ where $H= {\cal H}/a$. This condition is equivalent $\mathcal{A}_{,\tau} - \mathcal{H}\mathcal{A} = 0$ at $\tau =0$.  Inserting this condition into the above equation for $({\cal A}/{\cal H})_{,\tau}$, we find that at $\tau =0$, ${\cal A}/{\cal H} = a^2/6{\cal H}({\cal H}^2-{\cal H}_{,\tau})$ which vanishes at $\tau =0$ as $a=1/{\cal H} =0$ there.  Thus, using this boundary condition and 
integrating the above equation for $({\cal A}/{\cal H})_{,\tau}$ we arrive at 
\begin{equation*}
\mathcal{A}(\tau )=\mathcal{H}\int_{0}^{\tau }\frac{a^{2}(\tau ^{\prime })\,%
\mathrm{d}\tau ^{\prime }}{6\mathcal{H}^{2}(\tau ^{\prime })}.
\end{equation*}

We make the definitions $H_{0}=H(\tau =\tau _{0})$, $a_{0}=a(\tau _{0})$ and 
$\Omega _{\mathrm{baryon0}}=\kappa \rho _{\mathrm{baryon}}(\tau
_{0})/3H_{0}^{2}$, and then $a^{4}\kappa \rho _{\mathrm{baryon}}=3a(\tau
)a_{0}^{3}\Omega _{\mathrm{b0}}H_{0}^{2}$. The equation for $\Lambda $ is
then given explicitly by: 
\begin{widetext}
\begin{eqnarray}
\frac{\,\mathrm{d}I_{\mathrm{class}}}{\,\mathrm{d}\Lambda } &=& \frac{8\pi k}{%
\kappa a_{0}^{2}H_{0}^{2}}\int_{0}^{\tau _{0}}a^{2}(\tau )a_{0}^{2}H_{0}^{2}%
\left[ \frac{2}{3}(\tau _{0}-\tau )^{3}+\tau (\tau _{0}-\tau )^{2}\right] 
\mathcal{A}(\tau )\,\mathrm{d}\tau  \\ &&-\frac{4\pi }{\kappa }\zeta _{\mathrm{b}}\Omega _{\mathrm{b0}}\int_{0}^{\tau
_{0}}a(\tau )a_{0}^{3}H_{0}^{2}(\tau _{0}-\tau )^{3}\mathcal{A}(\tau )\,%
\mathrm{d}\tau = 0. \nonumber
\end{eqnarray}%
We can rearrange this to give an expression for the dimensionless curvature
parameter: 
\begin{eqnarray}
-\Omega _{k0}\equiv \frac{k}{a_{0}^{2}H_{0}^{2}} &=&\frac{\zeta _{\mathrm{b}%
}\Omega _{\mathrm{b0}}}{2}\mathcal{N}(\tau _{0};\Lambda ),  \qquad \mathcal{N}(\tau _{0};\Lambda ) \equiv \frac{\int_{0}^{\tau _{0}}a(\tau
)a_{0}^{3}(\tau _{0}-\tau )^{3}\mathcal{A}(\tau )\,\mathrm{d}\tau }{%
\int_{0}^{\tau _{0}}a^{2}(\tau )a_{0}^{2}\left[ \frac{2}{3}(\tau _{0}-\tau
)^{3}+\tau (\tau _{0}-\tau )^{2}\right] \mathcal{A}(\tau )\,\mathrm{d}\tau }.
\label{eq:const}
\end{eqnarray}%
\end{widetext}
Thus, we see that our new integral constraint equation for $\Lambda $ is a
consistency condition connecting the values of $\Omega _{k0}$, $\zeta _{%
\mathrm{b}}\Omega _{\mathrm{baryon0}},$ and $\Lambda $. 

\begin{figure*}[tbh]
\includegraphics[width=8cm]{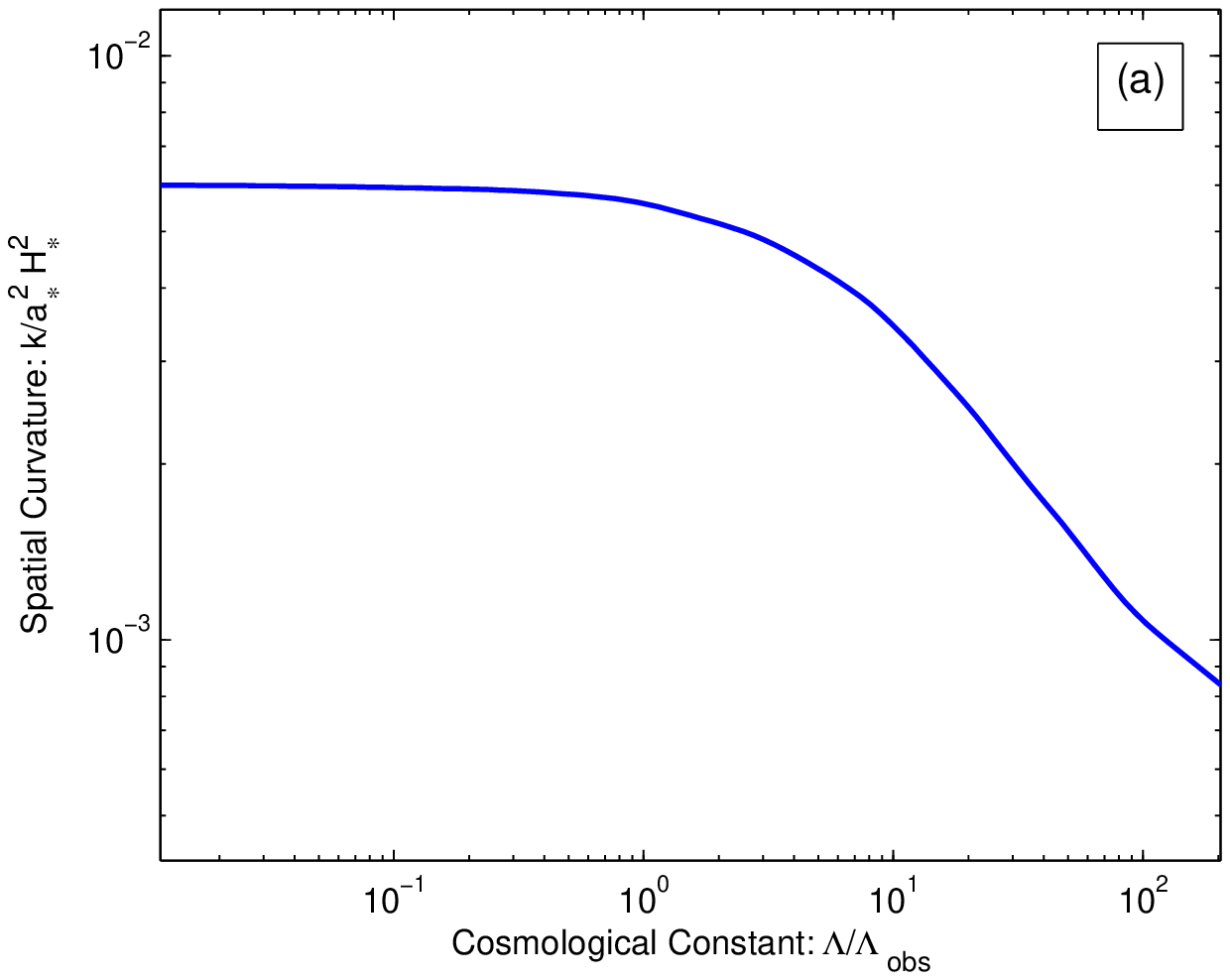}
\includegraphics[width=8cm]{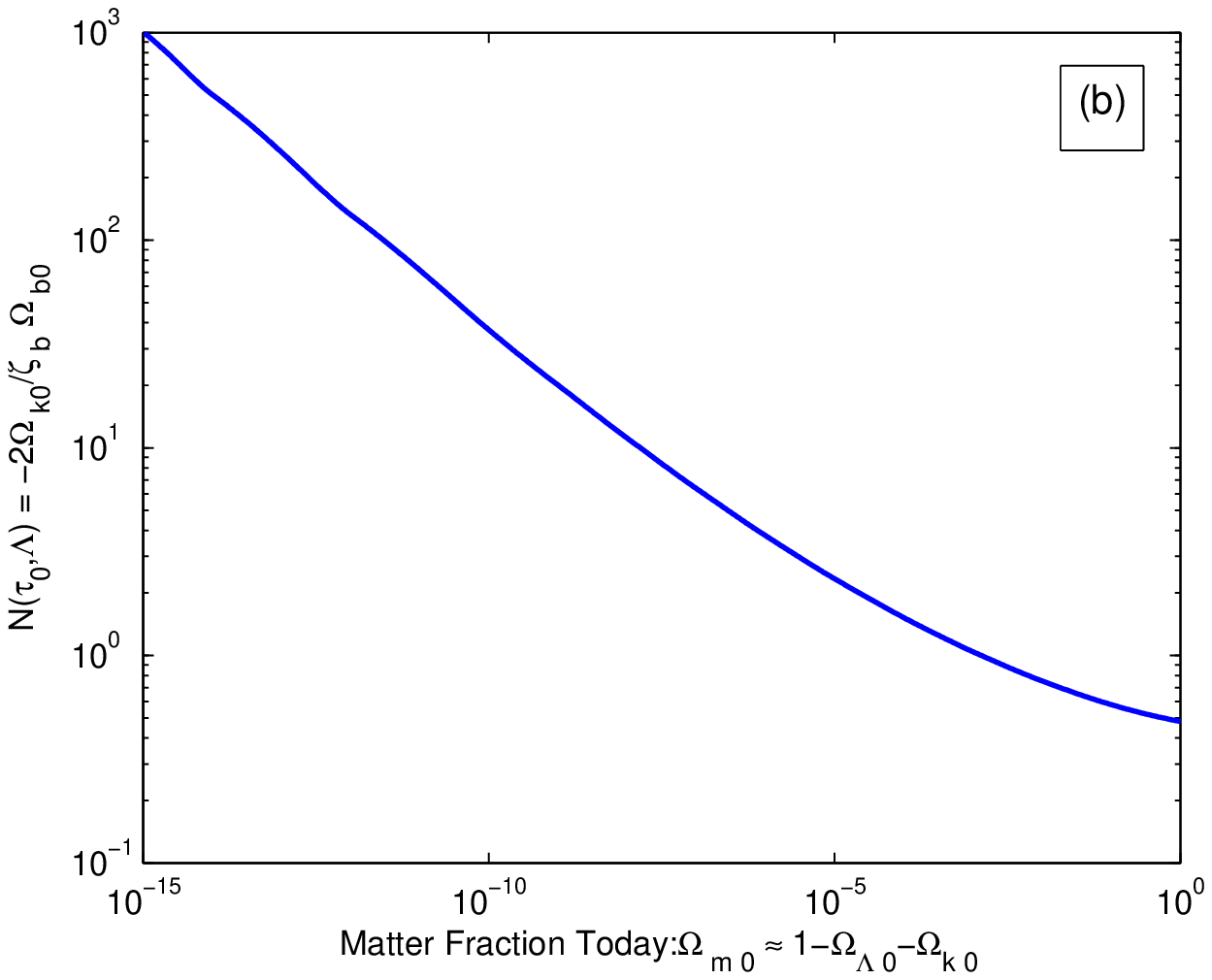}
\caption{(a) The relationship between the value of the spatial curvature $k$ and the value of $\Lambda$ that dominants the classical history.  We show $k$ in units of $a_{\star}^2 H_{\star}^2$ where $1/(a_{\star}H_{\star})$ is a fixed comoving length scale that is equal to $1/(a_{0}H_{0})$  for the observed value of $\Lambda$.  $\Lambda$ is shown relative to the value of $\Lambda$ that we observed, $\Lambda_{\rm obs}$.  Here the $k-\Lambda$ relationship predicted by our model is for observations at a time $t=t_{\rm U}\approx 13.77\,{\rm Gyrs}$.  We have also fixed the $\Lambda$ independent properties of matter by fixing the matter energy per photon, $\xi$, and the baryon energy per photon, $\xi_{\rm b}$, to their observed values: $\xi  = 3.43\,{\rm eV}$, $\xi_{\rm b} = 0.54\,{\rm eV}$.  We note that smaller values of $k$ correspond to larger values of $\Lambda$.  (b) The relationship between $\Omega_{\rm m0} \approx 1-\Omega_{\rm \Lambda 0}$ and $-2\Omega_{\rm k0}/\zeta_{\rm b}\Omega_{\rm b0} =  \mathcal{N}(\tau_{0};\Lambda)$ predicted by our model.  When  $\Omega_{\rm m0} \approx 1-\Omega_{\rm \Lambda 0}$, we find that when $\mathcal{N}(\tau_{0};\Lambda)$ is expressed as a function of $\Omega_{\rm \Lambda 0}$, it is almost independent of the observation time determined by $\tau_{0}$. For fixed $\Omega_{\rm b0}/\Omega_{\rm m0}$ and $\zeta_{\rm b}$, $\Omega_{\rm k0}$ corresponds to a specific value of $\Omega_{\rm \Lambda 0}$ independently of $\tau_{0}$.  As in (a), we have taken $\xi  = 3.43\,{\rm eV}$, $\xi_{\rm b} = 0.54\,{\rm eV}$.}
\label{figNK}
\end{figure*}

The quantities $%
k=-\Omega _{k0}H_{0}^{2}a_{0}^{2}$ and $\zeta _{\mathrm{b}}\Omega _{\mathrm{%
baryon0}}a_{0}^{3}H_{0}^{2}$ are fixed by the initial conditions and so this
equation determines $\Lambda $. With all other quantities fixed, Eq. (\ref%
{eq:const}) gives $k=k_{0}(\Lambda )$ where the form of $k_{0}(\Lambda )$
follows from Eq. (\ref{eq:const}). We can invert this to give $\Lambda
=\Lambda _{0}(k)$.   In FIG. \ref{figNK}a and Table \ref{table1} we show the value of $k$ required for different values of $\Lambda$ for an observation time: $t =  t_{\rm U} \approx 13.77\,{\rm Gyrs}$. In both the table and the figure, $k$ is given in units of $a_{\star}^2 H_{\star}^2$ where $1/(a_{\star}H_{\star})$ is a fixed comoving length scale that is equal to $1/(a_{0}H_{0})$ when $\Lambda = \Lambda_{\rm obs}$.  We see that large values of $\Lambda$ require smaller values of $k$.

We find that when $\Omega_{\rm m0} \approx 1-\Omega_{\rm \Lambda 0}$, $\mathcal{N}(\tau_{0};\Lambda) \approx \bar{\mathcal{N}}(\Omega_{\rm \Lambda 0})$ for any $\tau_{\rm 0}$ i.e. $\mathcal{N}$ is determined entirely by $\Omega_{\rm \Lambda 0}$.  Thus, by Eq. (\ref{eq:const}), $\bar{\mathcal{N}}(\Omega_{\rm \Lambda 0}) \approx -2\Omega_{\rm k0}/\zeta_{\rm b}\Omega_{\rm b0}$, and so, given that the ratio of baryons to dark matter is fixed for all $\Lambda$, $\Omega_{\rm b0} \propto (1-\Omega_{\rm \Lambda 0})$, and each value of $\Omega_{\rm k0}$ corresponds to a specific value of $\Omega_{\rm \Lambda 0}$ independently of $\tau_{0}$.   We illustrate this in FIG. \ref{figNK}b, where we plot $-2\Omega_{\rm k0}/\zeta_{\rm b}\Omega_{\rm b0}$ against $\Omega_{\rm m0} \approx 1-\Omega_{\rm \Lambda 0}$.    We note that $O(1)$ values of $\Omega_{\rm m0}$ correspond to $O(1)$ values of $-2\Omega_{\rm k0}/\zeta_{\rm b}\Omega_{\rm b0}$.

\subsubsection{A Prediction for the Spatial Curvature}

In principle, $\Lambda $, $\Omega _{\mathrm{b0}}$, and $\Omega _{k0}$ are
quantities astronomers can measure accurately. We can therefore test the
validity of our model by checking that the consistency equation, Eq. %
(\ref{eq:const}), is indeed consistent with the observational limits on $\Lambda $%
, $\Omega _{\mathrm{b0}}$ and $\Omega _{k0}$. We note that $\mathcal{N}>0$
and so our model requires that $k/\zeta_{\rm b} >0$ or equivalently $\Omega _{k0}/\zeta_{\rm b}<0$.  Our estimate of $\zeta_{\rm b}$ from the chiral bag model of baryons in QCD gives $\zeta_{\rm b} > 0$.
Current observations only bound the value of $\Omega _{k0}$ and those bounds
are consistent with $\Omega _{k0}=0$. The values of $\Lambda $ and $\Omega _{%
\mathrm{b0}}$ are relatively well established. 

The most recent 1$\sigma $ limit on $\Omega _{k0}$ from WMAP 7 combined with
BAO and $H_{0}$ data (and $\Lambda $CDM prior) is \cite{Komatsu:2010fb}: 
\begin{eqnarray}
&\Omega _{k0}=-0.0023_{-0.0056}^{+0.0054}.
\end{eqnarray}

When our model is applied to our universe with $\Omega _{\Lambda 0}=\Lambda
/3H_{0}^{2}=0.73$, $\Omega _{\mathrm{b0}}=0.0423$, as observed at a present
time when the CMB temperature is $2.725\,\mathrm{k}$, Eq. (\ref{eq:const})
predicts the value of $\Omega _{k0}$ to be: 
\begin{eqnarray}
&\Omega _{k0}=-0.0056 \left(\frac{\zeta _{\mathrm{b}}}{1/2}\right)&,
\end{eqnarray}%
which is consistent with the observational limit at 1$\sigma $ for $\zeta _{%
\mathrm{b}}\in (0,0.7]$ and within the $95\%$ confidence limit for all $%
\zeta _{\mathrm{b}}\in (0,1]$. For the estimated value of $\zeta _{\mathrm{b}%
}=1/2$, our model predicts $\Omega _{k0}=-0.0056$. The combination of data
from the Planck CMB survey with current and future measurements of the
Baryon Acoustic Oscillations (BAO) should be able to confirm or refute this
detailed prediction. Therefore, in contrast to other proposals for solving
the CC problems, our model makes a testable prediction and is
falsifiable in the near future.

\subsection{What is a Natural Value of $\Lambda$?}

\label{sec:cosmology:natural} We have seen that, at a fixed time, our model
predicts the value of $\Lambda $ in terms the spatial curvature $k$. In
inflationary models, the magnitude of spatial curvature $k$ inside the past
light cone is determined by the duration of the inflationary period in the
earlier universe, specified by the number of e-folds $N$. In most
inflationary scenario one imagines that there are many different inflating
regions, or \textquotedblleft bubble universes\textquotedblright . In each
bubble the initial conditions for the scalar field will differ. The number
of e-folds of inflation experienced by a bubble universe depends on these
initial conditions in a model-dependent fashion. The value of $N$ (and hence 
$k)$ will therefore be different in each bubble universe. The curvature
parameter $k$ is therefore an environmentally sensitive parameter: it
depends on the part of universe we observe, and will not be the same
everywhere. In our model, when all other quantities are fixed, $\Lambda $ is
given implicitly as a function or $k$ by Eq. (\ref{eq:const}). Hence, the
value of $\Lambda $ that one observes at a given time is also an
environmentally determined parameter. If we existed in a different bubble
universe with a different value of $k$, we would observe a different value
of $\Lambda $. In order for our model to be said to solve the CC problems,
the value of $\Lambda $ that we do observe must be shown to be in some sense
'natural'. This means that, once selection effects such as the requirement
that $\Lambda $ and $k$ are not so large as to prevent the formation of
non-linear structure in the universe have been taken into account, the
observed value of $\Lambda $ should, ideally, be typical amongst all
possible bubble universes. One could then conclude that the observable
universe is no more fine tuned that it must be given that we are here to
observe it.

\begin{table}[tbp]
\caption{$\Lambda $ in the history seen by an observer at time $t_{\mathrm{U}%
}\approx 13.77\,\mathrm{Gyrs}$ for different values of the spatial curvature
parameter, $k$. Here, $\protect\zeta _{\mathrm{b}}$ is a QCD constant
related to baryon structure, and we expect $\protect\zeta _{\mathrm{b}%
}\approx 1/2$; $\Lambda _{\mathrm{obs}}$ is the particular value of $\Lambda 
$ we observe, and $a_{\star }H_{\star }$ is the value of $a(\protect\tau )H(%
\protect\tau )$ today in our visible universe. We have taken the matter
energy per photon to be $\protect\xi =3.43\,\mathrm{eV}$, and the baryon
energy per photon is $\protect\xi _{\mathrm{b}}=0.54\,\mathrm{eV}$. }
\begin{center}
\begin{tabular}{|c|c||c|c|}
\hline
$\frac{k}{2\zeta_{\mathrm{b}}a_{\star}^2H_{\star}^2}$ & $\frac{\Lambda}{%
\Lambda_{\mathrm{obs}}}$ & $\frac{k}{2\zeta_{\mathrm{b}}a_{\star}^2H_{%
\star}^2}$ & $\frac{\Lambda}{\Lambda_{\mathrm{obs}}}$ \\ \hline
0.0060 & 0.00 & 0.0049 & 2.9 \\ \hline
0.0059 & 0.25 & 0.0042 & 5.5 \\ \hline
0.0057 & 0.78 & 0.0034 & 10.6 \\ \hline
0.0056 & 1.0 & 0.0018 & 35.7 \\ \hline
0.0053 & 1.6 & 0.00084 & 200 \\ \hline
\end{tabular}%
\end{center}
\label{table1}
\end{table}

Observers like ourselves require the universe to be old enough for a
sufficient number of collapsed structures such as galaxies to have formed,
and then for heavy elements to have been formed by stars. If $k$ or $\Lambda $
are too large then either the universe will recollapse before these
conditions have been achieved or the growth of structure will have been so
suppressed that even as $t\rightarrow \infty $, galaxies never form \cite%
{btip}.

\subsubsection{The naturalness of $\Lambda $ in string landscape models}

Before addressing the naturalness of the observed value of $\Lambda $ in our
model, we consider the extent to which the string landscape model solves the
CC problems. The string landscape solution to the cosmological constant
problems is totally reliant on anthropic selection effects to determine the
value of $\Lambda $. In that scenario, it is assumed that there are many
different possible vacua, each with a different value of vacuum energy, or
equivalently of $\Lambda $. The probability of a vacuum having a CC in the
interval $[\Lambda ,\Lambda +\,\mathrm{d}\Lambda ]$ is $f_{\mathrm{prior}%
}(\Lambda )\,\mathrm{d}\Lambda $, where $f_{\mathrm{prior}}(\Lambda )$ is
the prior probability distribution of $\Lambda $ and has not been directly
determined by theory. Anthropic selection effects provide the probability, $%
f_{\mathrm{selec}}(\Lambda )$, of being able to observe a universe with a
given value of $\Lambda $. By Bayes' theorem, the unnormalized probability
distribution function of observing a vacuum state with a CC in the interval $%
[\Lambda ,\Lambda +\,\mathrm{d}\Lambda ]$ is: 
\begin{equation*}
f_{\Lambda }(\Lambda )\,\mathrm{d}\Lambda =f_{\mathrm{selec}}(\Lambda )f_{%
\mathrm{prior}}(\Lambda )\,\mathrm{d}\Lambda .
\end{equation*}%
The form of $f_{\mathrm{selec}}(\Lambda )$ can be estimated by taking the
number of galaxies (collapsed structures with a given mass) as a proxy for
the number of observers, see for instance Ref. \cite{Tegmark:2005dy} for
such a calculation. However, without knowing the form of $f_{\mathrm{prior}%
}(\Lambda ),$ it is not possible to say whether or not the observed value of 
$\Lambda $ is natural. Some authors argue that a uniform prior is the most
reasonable for small values of $\Lambda $. If this is the case then, as
shown in Ref. \cite{Tegmark:2005dy}, when all other parameters are
fixed, the observed value of $\Lambda $ is not atypical, although the most
probable values are still an order of magnitude or two larger. Specifically,
with a uniform prior, and $f_{\mathrm{selec}}$ from Ref. \cite%
{Tegmark:2005dy}, one finds
\begin{eqnarray}
2.84 &<&\frac{\Lambda }{\Lambda _{\mathrm{obs}}}<44.63,\qquad \text{(68\%
Confidence)}, \\
0.40 &<&\frac{\Lambda }{\Lambda _{\mathrm{obs}}}<123.68,\qquad \text{(95\%
Confidence)},  \notag
\end{eqnarray}%
where $\Lambda _{\mathrm{obs}}$ is the particular value of $\Lambda $ that
we observe. Hence, with a uniform prior, this value is outside the 68\%
confidence limit by about a factor of 2.8 but inside the 95\% confidence
interval. The observed value of $\Lambda $ is therefore small but not
atypically small here.

A uniform prior is not, however, the only reasonable choice one could make
for $f_{\mathrm{prior}}(\Lambda )$. A log-prior, $f_{\mathrm{prior}}\,%
\mathrm{d}\Lambda \propto \,\mathrm{d}\ln \Lambda ,$ or an exponential,  $f_{%
\mathrm{prior}}\propto \exp (3\pi /G\Lambda ),$ have also be supported by
theoretical arguments and in both cases the most probable values of $\Lambda 
$ would be many orders of magnitude smaller than the observed value. In the
string landscape and other multiverse models, the natural value of $\Lambda $
is crucially dependent on the choice of prior, and until the prior can be
calculated from first principles using the theory it is not clear whether
this model provides an natural explanation for the observed value of $%
\Lambda $.

\subsubsection{The naturalness of $\Lambda $ in our model}

\begin{figure*}[tbh]
\includegraphics[width=8cm]{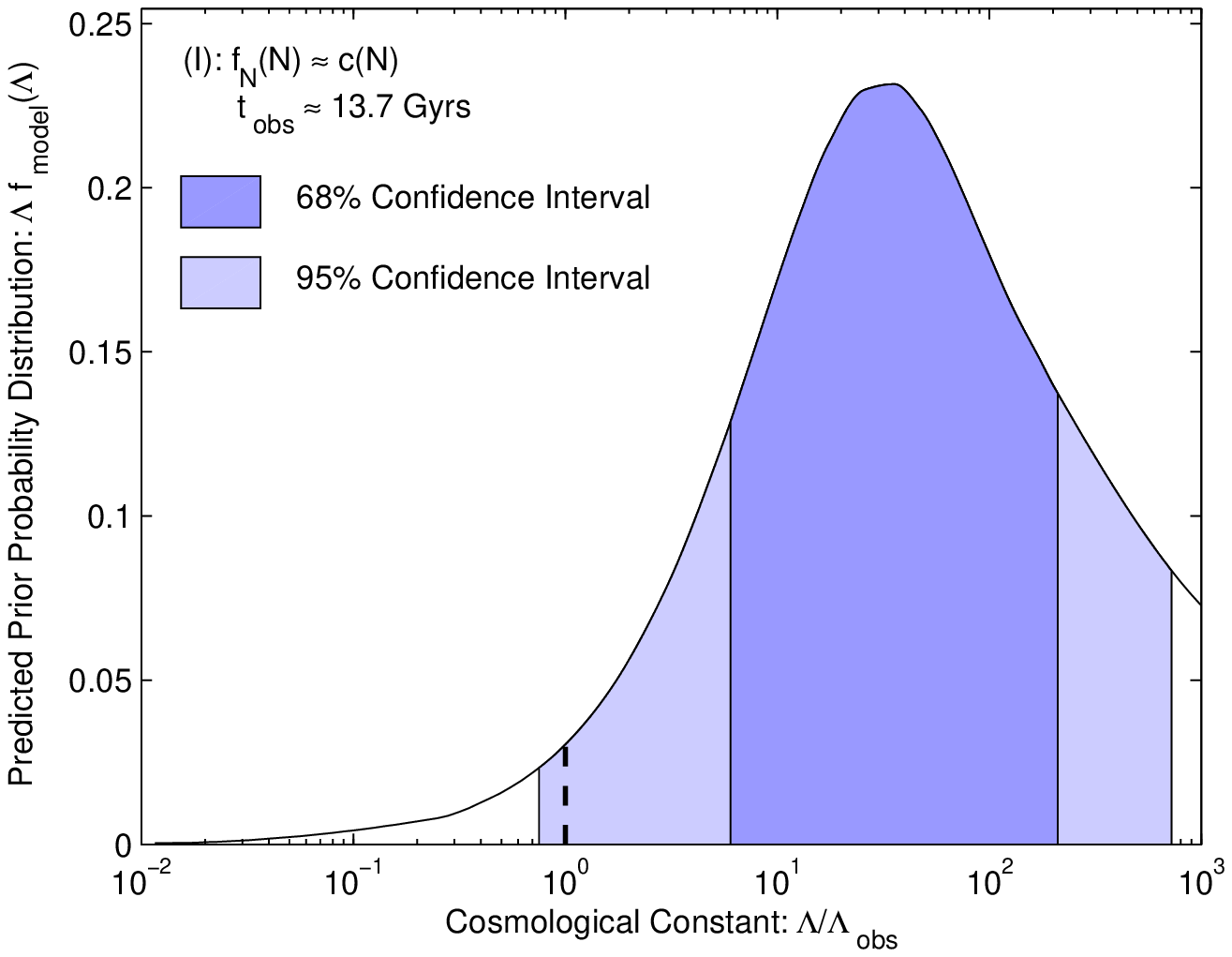}
\includegraphics[width=8cm]{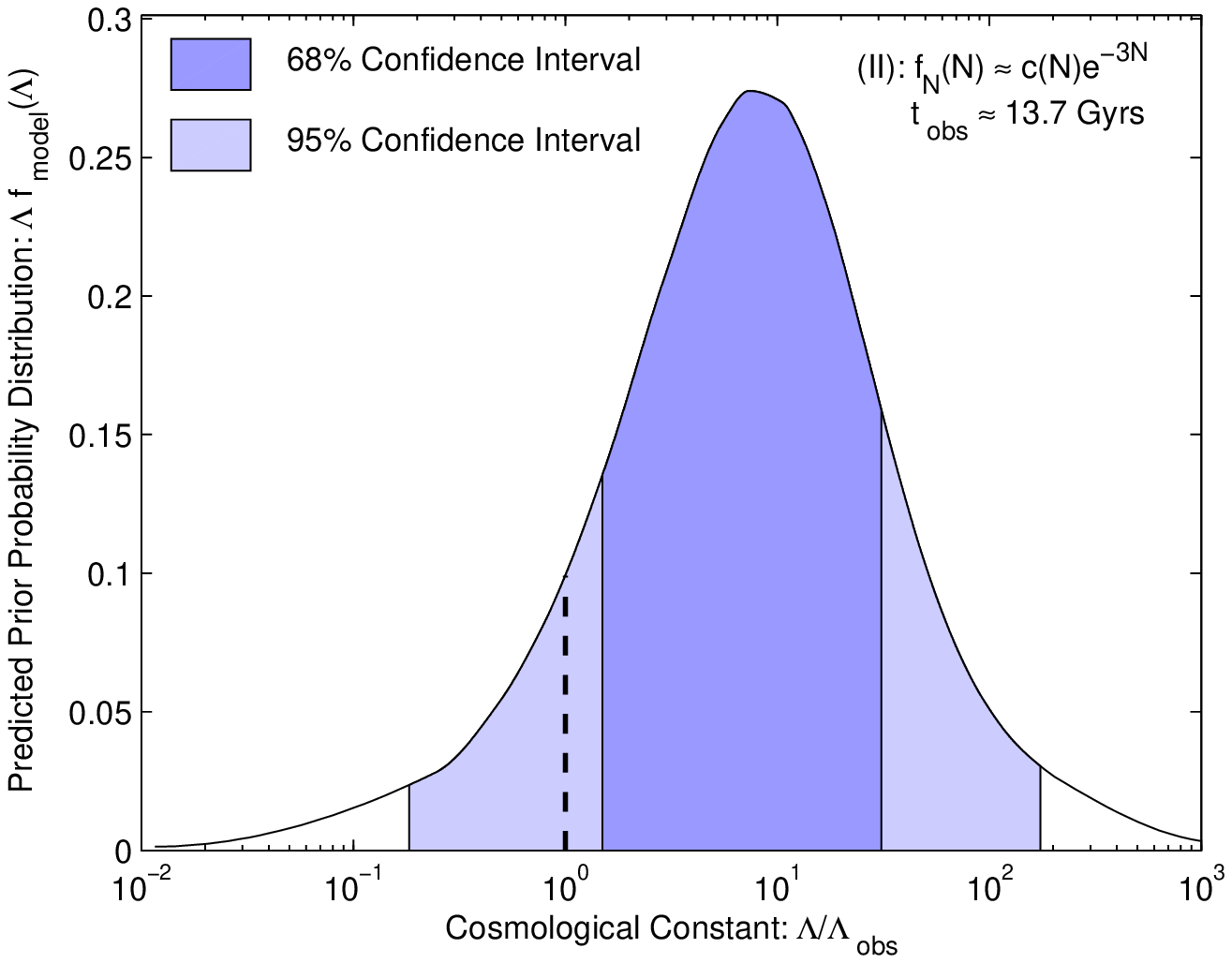}
\caption{Our model's prediction for the (prior) probability distribution, $%
f_{\mathrm{model}}(\Lambda)$, for the value of cosmological constant $\Lambda$
that one would measure at an observation time $t=t_{\mathrm{U}} = 13.77\,%
\mathrm{Gyrs}$. In our theory, the observed value of $\Lambda $ is given as a function of
the spatial curvature, $k$, inside the observer's past light cone. The
curvature parameter $k$ depends on $N$, the number of e-folds of inflation.
Hence, ultimately, the probability of astronomers observing a given value of 
$\Lambda $ depends on the prior probability of living in a bubble universe
where the universe underwent $N$ e-folds of inflation. The probability
distribution function of $N$ is $f_{\mathrm{N}}(N)$. Given an $f_{\mathrm{N}%
}(N),$ our model completely determines the prior probability distribution function, $f_{\rm model}(\Lambda)$.  We have plotted $f_{\rm model}(\Lambda )$ for two different choices of $f_{\mathrm{N}}(N)$. Case I is where $f_{\mathrm{N}}(N) = c(N)$ where $c(N) \approx {\rm const}$ for changes, $\Delta N$, in $N$ over less than $\approx 2.5\%$ (e.g. a power-law $N^{-p}$ for $\vert p \vert < 10$). Case II is where $f_{\mathrm{N}} = c(N)e^{-3N}$ where again $c(N)\approx {\rm const}$ for $\Delta N/N \lesssim 2.5\%$. This latter choice is the one calculated in Ref.  \protect\cite{Gibbons:2006pa} for slow-roll single-field inflation. In our model this prediction for the prior probability is independent of the fundamental prior weighting of different values of $\Lambda$ in the partition function. Our $f_{\mathrm{model}}(\Lambda)$
does not include any observer-dependent selection effects, apart from the
requirement that a classical solution exists.  The observed value of $\Lambda$
is shown by a dotted black line, and the whole shaded region is the
symmetric 95\% confidence interval. The darker shaded area is the symmetric
68\% confidence interval. In Case I we have calculated the confidence intervals by sharply cutting off $f_{\rm model}$ for $\Lambda > 10^{3}\Lambda_{\rm obs}$, since such large values of the CC are incompatible with the existence of galaxies. For $\Lambda < 10^{3}\Lambda_{\rm obs}$ we have \emph{not} included any weighting by probability of living in a universe with a given $\Lambda$.     For these natural choices of $f_{\mathrm{N}}(N)$,
the observed value of $\Lambda$ is within the 95\% confidence interval in both cases, and well within it for $f_{\rm N}=c(N)e^{-3N}$.  In both cases, $\Lambda = \Lambda_{\rm obs}$ is not atypical whatever precise form of the observer conditioned selection effect parametrized by $f_{\rm selec}(\Lambda)$.  When selection effects are included the probability of larger values $\Lambda \gtrsim 100 \Lambda_{\mathrm{obs}}$ is further
suppressed, and the observed value moves within the 1-$\protect\sigma$
confidence interval for Case II, and just outside this interval in Case I. This is shown in FIG \protect\ref{figII} below. We have taken the matter energy per photon to be $\protect\xi = 3.43\,\mathrm{eV}$,
and the baryon energy per photon is $\protect\xi_{\mathrm{b}} = 0.54\,\mathrm{eV}$, as is observed. }
\label{figI}
\end{figure*} 
In our model, we shall see below that, just as in landscape and other
multiverse models, anthropic selection still plays a role in limiting the
maximum allowed values of $\Lambda $. The equivalent prior on $\Lambda $ in
our model is the undetermined measure $\mu \lbrack \Lambda ]$ in the
partition function. Unlike for landscape or multiverse models though, this
unknown prior on $\Lambda $ plays no role. This is because our model
requires $\Lambda =\Lambda _{0}(k)$ and $\Lambda _{0}(k)$ is a function of $k
$ that is given by our model. Thus, whatever the prior on $\Lambda $, the
normalized posterior probability distribution of $\Lambda $ given $k$ is a
delta-function,
\begin{equation*}
f_{\Lambda |k}(\Lambda ;k)=\delta (\Lambda -\Lambda _{0}(k)),
\end{equation*}%
where $\Lambda _{0}(k)$ also depends on the size of the observer's past
light cone, $\mathcal{M}$, and hence observation time. Given that one lives
in a bubble universe with a certain value of $k,$ and observes it at a given
time, in our model there is only one value that $\Lambda $ can take. The
probability of measuring a $\Lambda $ in a given range is then given
entirely by the probability of measuring $k$ in a corresponding range. It is
independent of the measure, or prior, $\mu \lbrack \Lambda ]$.

Now the curvature parameter $k$ is related to the number of e-folds $N$,
since $k=k(N)=\bar{k}e^{2(\bar{N}-N)}$ for some fixed $N_{0}$ and $\bar{k}$.
As above, $1/(a_{\star} H_{\star })$ is a comoving length scale equal to $1/a_0H_0$ in our particular universe. In the expression for $k(N)$, $\bar{N}$ is the
number of e-folds required to bring about the bound $|\Omega _{\mathrm{k}}|<|%
\bar{k}|/H_{\star }^{2}a_{\star }^{2}$ today. We are free to take $|\bar{%
k}|=10^{-2}a_{\star }^{2}H_{\star }^{2}$. Depending on the efficiency of
reheating after the end of inflation, we have $\bar{N}\gtrsim 50-62$ in
realistic inflation models.

The probability distribution for $k$ in different
bubble universes, $f_{k}(k)\,\mathrm{d}k$, is therefore given by  $f_{N}(N)\,\mathrm{d}N$, the
probability distribution of the number of e-folds, $N$. Specifically, we have
\begin{equation*}
f_{k}(k(N))=f_{N}(N)\left\vert \frac{\,\mathrm{d}N}{\,\mathrm{d}k}%
\right\vert =\frac{f_{N}(N(k))}{2k}.
\end{equation*}

\begin{figure*}[tbh]
\includegraphics[width=8cm]{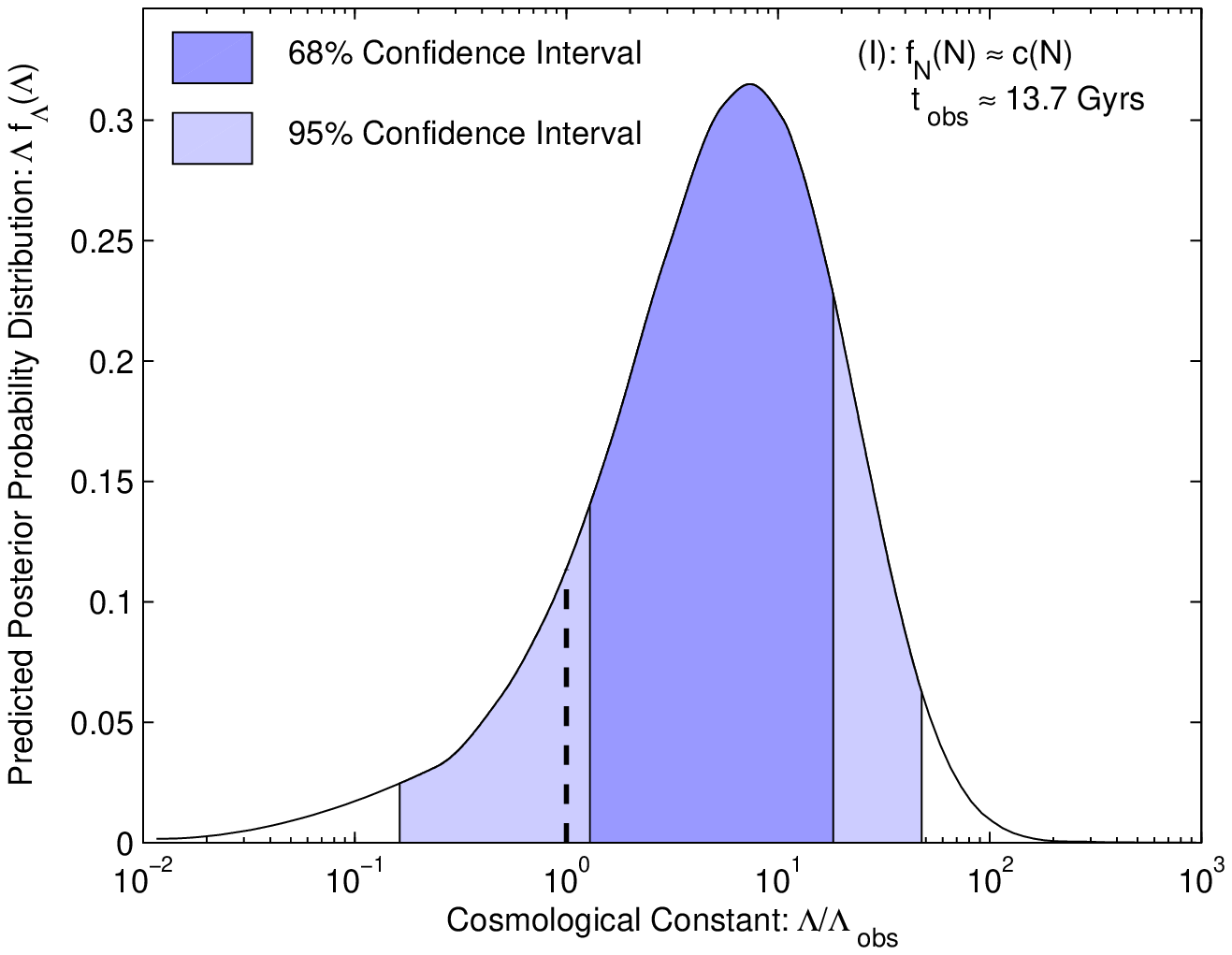} \includegraphics[width=8cm]{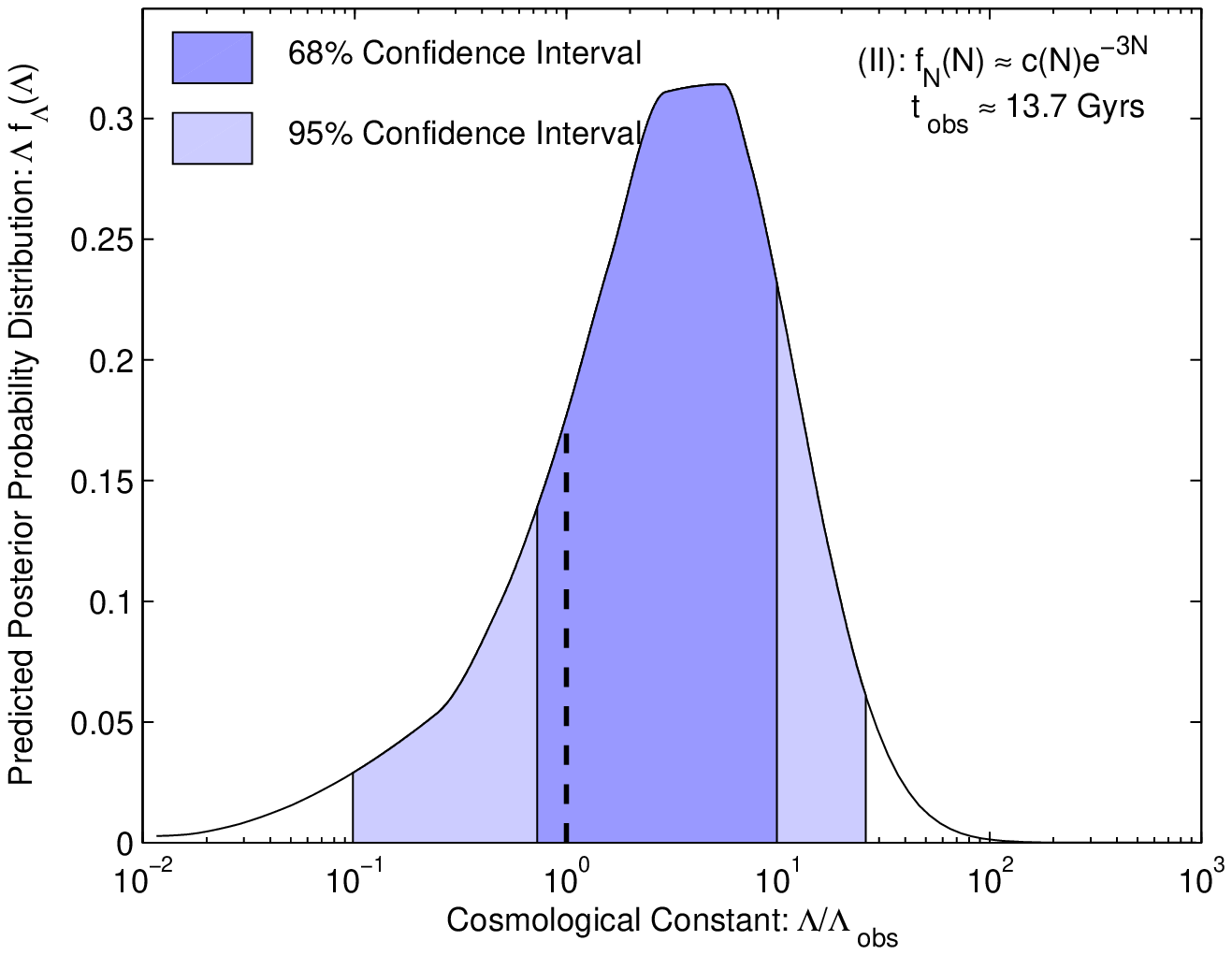}
\caption{The predicted posterior probability distribution, $f_{\Lambda }(\Lambda ) = f_{\rm selec}(\Lambda)f_{\rm model}(\Lambda),$ for the value of cosmological $\Lambda $ that one would measure at an
observation time $t=t_{\mathrm{U}}=13.77\,\mathrm{Gyrs}$. Here, $\Lambda _{%
\mathrm{obs}}$ is the particular value of $\Lambda $ that we have observed.
In our theory, the observed value of $\Lambda $ is given as a function of
the spatial curvature, $k$, inside the observer's past light cone. The
curvature parameter $k$ depends on $N$, the number of e-folds of inflation.
Hence, ultimately, the probability of astronomers observing a given value of 
$\Lambda $ depends on the prior probability of living in a bubble universe
where the universe underwent $N$ e-folds of inflation. The probability
distribution function of $N$ is $f_{\mathrm{N}}(N)$. Given an $f_{\mathrm{N}%
}(N),$ our model completely determines the probability distribution function, $f_{\rm model}(\Lambda)$,
of $\Lambda $ prior to the inclusion of selection effects. In the above
plots we have included the limits on $\Lambda $ due to observational
selection effects using the prescription for $f_{\rm selec}$ given by Tegmark \emph{et al.} in 
\protect\cite{Tegmark:2005dy} which uses the number of galaxies as a proxy for the number of observers. When these are included, the full posterior
probability of living in a bubble universe where one observes a given value
of $\Lambda $ is $f_{\Lambda }(\Lambda )$. Additionally, we find that the
inclusion of selection effects makes $f_{\Lambda }(\Lambda )$ only
relatively weakly dependent on the form of $f_{\mathrm{N}}(N)$ because for
allowed values of $\Lambda $, the required $k$ (and hence $N$) vary only
over a small range. We have plotted $f_{\Lambda }(\Lambda )$ for two
different choices of $f_{\mathrm{N}}(N)$. Case I is where $f_{\mathrm{N}}(N) = c(N)$ where $c(N) \approx {\rm const}$ for changes, $\Delta N$, in $N$ over less than $\approx 2.5\%$ (e.g. a power-law $N^{-p}$ for $\vert p \vert < 10$). Case II is where $f_{\mathrm{N}} = c(N)e^{-3N}$ where again $c(N)\approx {\rm const}$ for $\Delta N/N \lesssim 2.5\%$. This latter choice is the one calculated in Ref.  \protect\cite{Gibbons:2006pa} for slow-roll single-field inflation. In both
cases we see that the observed value of $\Lambda $ (dotted black line) is
well inside the 95\% confidence interval (the shaded areas), and near the
boundary of the 68\% confidence interval (the more darkly shaded area). In
Case I, $\Lambda =\Lambda _{\mathrm{obs}}$ is just outside this interval,
whereas in Case II it is just inside it. In both cases, it is clear that the
observed value of $\Lambda $ is typical and can be explained without the
need for fine tuning. We have taken the matter energy per photon to be $%
\protect\xi =3.43\,\mathrm{eV}$, and the baryon energy per photon is $%
\protect\xi _{\mathrm{b}}=0.54\,\mathrm{eV}$ for all values of $\Lambda$.}
\label{figII}
\end{figure*}

The calculation of $f_{N}(N)\,\mathrm{d}N$, that is of the probability that
the number of e-folds lies in the region $[N,N+\,\mathrm{d}N]$ is the
measure problem for inflation, and has been the subject of a considerable
amount of work and debate as to which is the correct measure. Recently,
Gibbons and Turok \cite{Gibbons:2006pa} used the natural canonical measure
on the space of all classical universes, provided by the Hamiltonian of
general relativity \cite{Gibbons:1986xk}, to show that in single-field,
slow-roll inflation, $f_{N}(N)=c(N)e^{-3N}$, where $c(N)$ is model dependent
but generally changes much more slowly with $N$ than $\exp (-3N)$. The $\exp
(-3N)$ suppression indicates that the region of phase space that results in
a long-lived period of slow-roll inflation is very small (however, the
result is to some extent just an artefact of having no high-energy cut-off
where the theory inevitably breaks down). To find this $f_{\mathrm{N}}$,
Gibbons and Turok (GT) have to regularize the canonical measure by imposing
a cut-off on the curvature and the end of inflation or equivalently on the
scale factor at the end of inflation i.e. $a<a_{\mathrm{max}}$. The
regularized measure is then $\propto a_{\mathrm{max}}^{3}$. If one takes $a_{%
\mathrm{max}}$ to be independent of $N$, then it drops out of the normalized
probability distribution for $N$, and $f_{\mathrm{N}}(N)\propto e^{-3N}$. An
alternative procedure would be to place the cut-off on the curvature at the
beginning of inflation or, equivalently, on the scale-factor there. In this
case, at the end of inflation, $a<a_{\mathrm{max}}\propto e^{N}$. We would
then have $f_{\mathrm{N}}\rightarrow e^{3N}f_{\mathrm{N}}(N),$ and so $f_{%
\mathrm{N}}=c(N)$. This procedure factors in a weighting by volume. Another
example where $f_{\mathrm{N}}=c(N)$ and $c(N)$ varys much more slowly with $%
N$ than $\exp (-3N)$ was estimated by Freivogel, Kleban, Rodriguez Martinez,
and Susskind (FKRMS), and then extended by De Simone and Salem in the
context of eternal inflationary models on the string landscape \cite%
{Freivogel:2005vv,DeSimone:2009dq}. They found $f_{\mathrm{N}}\propto N^{-4}$%
.


Our model unambiguously predicts the prior probability of living in a
universe with effective CC in the interval $[\Lambda ,\Lambda +\,\mathrm{d}%
\Lambda ]$. We define this to be $f_{\mathrm{model}}(\Lambda )\,\mathrm{d}%
\Lambda $, and it is given (up to a calculable normalization factor) by
\begin{eqnarray}
f_{\mathrm{model}}(\Lambda ) &=&\int \,\mathrm{d}k\,f_{\Lambda |k}(\Lambda
;k)f_{k}(k), \\
&=&\frac{f_{\mathrm{N}}(N(k_{0}(\Lambda )))}{2}\left\vert \frac{\,\mathrm{d}%
\ln k_{0}(\Lambda )}{\,\mathrm{d}\Lambda }\right\vert ,  \notag
\end{eqnarray}%
where $k_{0}(\Lambda )$ follows from Eq. (\ref{eq:const}). Note that $%
k_{0}(\Lambda )$ also depends on the observation time, and the matter/baryon
energy per photon. Depending on the form of $f_{\mathrm{N}}(N)$, the
prediction for $f_{\mathrm{model}}(\Lambda )$ may indicate that the observed
value of $\Lambda $ is natural independently of selection effects
conditioned on the existence of observers. We discuss this point below.

The dependence on the precise form of $f_{\mathrm{N}}(N)$ is greatly
weakened when selection effects on $\Lambda $ are included. Then, we have
that, in our model, the (unnormalised) posterior probability of living in a
universe with effective CC in the interval $[\Lambda ,\Lambda +\,\mathrm{d}%
\Lambda ]$ is $f_{\Lambda }(\Lambda )\,\mathrm{d}\Lambda ,$ where 
\begin{equation*}
f_{\Lambda }(\Lambda )=f_{\mathrm{selec}}(\Lambda )f_{\mathrm{model}%
}(\Lambda ).
\end{equation*}%
Roughly, the observer-conditioned selection effects on $\Lambda $ limit its
value to be no more than about $1000$ times that which is observed in our
universe ($\Lambda _{\mathrm{obs}}$). Tegmark \emph{et al.} \cite%
{Tegmark:2005dy} calculated $f_{\mathrm{selec}}(\Lambda )$ by using the
number of galaxies (virialized halos with a mass $\gtrsim 10^{12}\,M_{\odot }
$) as a proxy for the number of observers. We use their form of $f_{\mathrm{%
selec}}$ here when evaluating $f_{\Lambda }(\Lambda )$.

Unlike in the string landscape model, $f_{\Lambda }(\Lambda )$ and $f_{%
\mathrm{model}}(\Lambda )$ have no dependence on the unknown prior weighting
of different values of $\Lambda $. All that is required in order to specify $%
f_{\Lambda }(\Lambda ),$ or $f_{\mathrm{model}}(\Lambda ),$ fully is to
specify the prior probability of the number of e-folds, $f_{\mathrm{N}}(N)$.
At present, much more is known and and is calculable about the form of $f_{%
\mathrm{N}}(N)$ for different inflation models than is known about the
landscape prior on $\Lambda $. Also, we shall see that $f_{\Lambda }(\Lambda
)$ is much less sensitive to the precise form of $f_{\mathrm{N}}(N)$ than
the string landscape model is to $\Lambda $-prior.

In Table \ref{table1} we provide the value of $k$ (in units of $H_{\star
}^{2}a_{\star }^{2}$) required by our model for different values of $\Lambda 
$ at an observational time of $13.77\,\mathrm{Gyrs}$. Larger values of $%
\Lambda $ require a smaller value of $k,$ and hence a larger value of $N$.
Given that $f_{\mathrm{N}}(N)$ is generally estimated to be a decreasing
function of $N,$ this means that the probability of larger values of $%
\Lambda $  will be suppressed relative to smaller values. Anthropic limits
on $\Lambda $ imply that it could not have been more than about 1000 times
larger the value we observe. In this allowed range the required $k$ for a
given $\Lambda $ decreases by less than a factor of $10$. Thus, the required
number of e-folds, $N=N_{0}(\Lambda )=N(k_{0}(\Lambda ),$ changes by less
than $\Delta N\approx \Delta (\ln k)/2\lesssim \ln (10)/2\approx 1.2$. At
the same time, $N_{0}(\Lambda )\gtrsim \bar{N}>50-62$ in realistic models,
and so $\Delta N/N\lesssim 0.025$. So, unless $|\,\mathrm{d}\ln f_{\mathrm{N}%
}(N)/\,\mathrm{d}\ln N|\gtrsim 10$ or so, we have $f_{\mathrm{N}}(N)\approx 
\mathrm{const}$ for anthropically allowed values of $\Lambda $. Such a flat $%
f_{\mathrm{N}}(N)$ emerges if we weight the GT probability distribution for $%
N$ (which is $\propto e^{-3N}$) by the bubble universe 3-volume at any
given time, (which is $\propto e^{3N}$). We then have $f_{\mathrm{N}}=c(N)$
where $c(N)$ is fairly flat (e.g. $c(N)\propto N^{-1/2}$ if the inflationary
potential is $\propto m^{2}\phi ^{2}$ \cite{Gibbons:2006pa}). The FKRMS
estimate of $f_{\mathrm{N}}\propto N^{-4}$ is another example where $f_{%
\mathrm{N}}$ is fairly flat for $\Delta N/N\ll 1$ \cite%
{Freivogel:2005vv,DeSimone:2009dq}. In both these cases $f_{\mathrm{N}%
}\approx \mathrm{const}$ for allowed values of $\Lambda $ and so the precise
form of $f_{\mathrm{N}}(N)$ is unimportant.

The GT measure on inflationary solutions has $f_{\mathrm{N}}\propto e^{-3N}$
and so $|\,\mathrm{d}\ln f_{N}/\,\mathrm{d}\ln N|3N\gg 10$. Thus, if this
measure is correct we should not approximate $f_{\mathrm{N}}$ by a constant.
We therefore consider this and the $f_{\mathrm{N}}\approx \mathrm{const}$
cases separately. 

We note that if $f_{\mathrm{N}}(N)\propto e^{-3N}$, the
suppression of large values of $N,$ and of hence large $\Lambda $ values, is
actually sufficient to place the observed value of $\Lambda $ within the
95\% confidence interval for $\Lambda $ prior to the inclusion of selection
effects. Even if we take $f_{\rm N} = c(N)$, where $c(N)$ is fairly flat, the $\Lambda_{\rm obs}$ is inside 95\% confidence interval of the prior probability distribution function for $\Lambda$, when one imposes a sharp cut-off on $\Lambda > 10^{3}\Lambda_{\rm obs}$. We illustrate this in FIG. \ref{figI} where we have plotted $f_{%
\mathrm{model}}(\Lambda )$ for an observation time of $t_{\mathrm{U}}=13.77\,%
\mathrm{Gyrs}$. The entire lighter shaded region is the symmetric 95\%
confidence interval, and the darker shaded region is the symmetric 68\%
confidence interval. The dotted black line marks the observed value of the
CC, $\Lambda _{\mathrm{obs}}$. We see that even before we have included
selection effects which suppress $\Lambda \gg 100\Lambda _{\mathrm{obs}}$
outcomes, the observed value of $\Lambda $ is not atypical with either general form of $f_{\rm N}(N)$.  For comparison, in the multiverse or landscape model with a uniform prior and a sharp cut-off on $\Lambda >10^{3}\Lambda_{\rm obs}$,  $\Lambda \leq \Lambda_{\rm obs}$ is much less likely and has a probability of only $0.1\%$ prior to the inclusion of selection effects.

Whilst there are anthropic selection effects on $k,$ these are
automatically satisfied when $k$ is small enough for a classical solution to
exist. The existence of a classical solution is therefore by far the
strongest selection effect on $k$. Current observational limits require $%
-0.084<k/a_{\star }^{2}H_{\star }^{2}<0.0133$ at 95\% confidence, where $%
1/a_{\star }H_{\star }$ is the measured value of the comoving Hubble radius, 
$r_{H}=1/aH,$ today. All the values of $k$ in Table \ref{table1} are well
within these limits, and so the existence of a classical solution in our
model is sufficient to explain why we must live in a bubble universe where $k
$ is within the current observational limits, and hence why our observable
universe must have undergone a large number of e-folds of inflation, no
matter how unlikely that is a priori.

We now turn our attention to the posterior probability, $f_{\Lambda
}(\Lambda )\,\mathrm{d}\Lambda ,$ of observing $\Lambda $ in the interval $%
[\Lambda ,\Lambda +\,\mathrm{d}\Lambda ]$ in our model. We consider the
consequences of two general forms of $f_{\mathrm{N}}(N)$: (I) $f_{\mathrm{N}%
}(N)=c(N)$ where $c(N)\approx \mathrm{const}$ in the allowed range (i.e.
less steep than $\sim N^{-10}$) and (II) $f_{\mathrm{N}}(N)\approx
c(N)e^{-3N}$ where again $c(N)\approx \mathrm{const}$ for allowed $\Lambda $
values. Finally, for $f_{\mathrm{selec}}(\Lambda )$ we take the form
calculated by Tegmark \emph{et al.} in Ref. \cite{Tegmark:2005dy}.

In Case I, with $f_{\mathrm{N}}(N)=c(N)\approx \mathrm{const}$, we have 
\begin{equation*}
f_{\Lambda }(\Lambda )\,\mathrm{d}\Lambda \propto f_{\mathrm{selec}}(\Lambda
)\left\vert \frac{\,\mathrm{d}\ln k(\Lambda )}{\,\mathrm{d}\Lambda }\,%
\mathrm{d}\Lambda \right\vert ,
\end{equation*}%
and in case II where $f_{\mathrm{N}}(N)\propto c(N)e^{-3N}$ 
\begin{equation*}
f_{\Lambda }(\Lambda )\,\mathrm{d}\Lambda \propto f_{\mathrm{selec}}(\Lambda
)k^{3/2}(\Lambda )\left\vert \frac{\,\mathrm{d}\ln k(\Lambda )}{\,\mathrm{d}%
\Lambda }\,\mathrm{d}\Lambda \right\vert .
\end{equation*}%
In FIG. \ref{figII}, we plot $\Lambda f_{\Lambda }(\Lambda )$ against $\ln
(\Lambda /\Lambda _{\mathrm{obs}})$ for the two cases given above. We also
show the 68\% and 95\% confidence limits on $\Lambda $ in both cases. In
case I, for $f_{\mathrm{N}}(N)=c(N)\approx \mathrm{const}$, these limits
are: 
\begin{eqnarray}
1.31 &<&\frac{\Lambda }{\Lambda _{\mathrm{obs}}}<18.37,\qquad \text{(68\%
Confidence)}, \\
0.19 &<&\frac{\Lambda }{\Lambda _{\mathrm{obs}}}<48.13,\qquad \text{(95\%
Confidence)}.  \notag
\end{eqnarray}%
In case II, where $f_{\mathrm{N}}(N)=c(N)\exp (-3N)$, ($c(N)\approx \mathrm{%
const}$) we have: 
\begin{eqnarray}
0.76 &<&\frac{\Lambda }{\Lambda _{\mathrm{obs}}}<10.03,\qquad \text{(68\%
Confidence)}, \\
0.11 &<&\frac{\Lambda }{\Lambda _{\mathrm{obs}}}<26.27,\qquad \text{(95\%
Confidence)}.  \notag
\end{eqnarray}%
We note that, with the same selection effects, for both choices of $f_{%
\mathrm{N}}$, our model prefers smaller values of $\Lambda $ than does the
string landscape model with a uniform prior.

In both cases, the observed value of $\Lambda $, $\Lambda _{\mathrm{obs}}$,
is well within the 95\% confidence limit. In case I with a power-law $f_{%
\mathrm{N}}(N)$, $\Lambda =\Lambda _{\mathrm{obs}}$ is just outside the 68\%
confidence limit, whereas with $f_{\mathrm{N}}\propto \exp (-3N)$, it is
just inside this limit. Thus, whichever form $f_{\mathrm{N}}(N)$ takes, the
observed value of $\Lambda $ is typical within our model. Note again that
this conclusion is independent of the precise form of $f_{\mathrm{N}}(N)$,
and totally independent of the prior weighting of different values of $%
\lambda $.

\subsubsection{The Coincidence Problem}

To address the coincidence problem directly we can calculate the probability
that the cosmological timescale $t_{\Lambda }=1/\sqrt{\Lambda }$ introduced
by the CC correlates with the current age of the universe, $t_{\mathrm{U}%
}\approx 13.77\,\mathrm{Gyrs}$. We define $r=|\ln (t_{\mathrm{U}}/t_{\Lambda
})|$, and take, fairly arbitrarily, $r$ to be our measure of the coincidence
in the values of $t_{\Lambda }$ and $t_{\mathrm{U}}$. If $r\ll 1$ there is a
strong coincidence in the two times, whereas if $r\gg 1$ there is not. Using 
$f_{\mathrm{posterior}}(\Lambda )$ as provided by our model, we calculate
probabilities of living in an observable universe where, at a time $t_{%
\mathrm{U}}=13.77\,\mathrm{Gyrs}$, $r<r_{0}$ for different choices of $r_{0}$%
. We find:

\begin{eqnarray}
P(r =|\ln (t_{\mathrm{U}}/t_{\Lambda })|<\tfrac{1}{2}) &=&14\%, \\
P(r<1)&=& 36\%,\qquad \mathrm{(}f_{\mathrm{N}}=c(N)\mathrm{)}, \nonumber \\
P(r <\tfrac{1}{2})&=& 22\%, \\ P(r<1)&=&53\%, \qquad \mathrm{(}f_{\mathrm{N}%
}=c(N)e^{-3N}\mathrm{)}, \nonumber
\end{eqnarray}%
where in both  $c(N)$ cases, $|\,\mathrm{d}\ln c/\,\mathrm{d}\ln \Lambda
|\lesssim 10$. It is clear from these figures that within our model, a
coincidence in the values of $t_{\mathrm{U}}$ and $t_{\Lambda }$ is quite
typical. If we were to do the same calculation for the landscape model with
uniform prior on $\Lambda $, we would find $7.0\%$ and $19.7\%$ respectively
for $P(r<1/2)$ and $P(r<1)$. Thus, even if we have uniform prior on $\Lambda 
$, the probability of $t_{\mathrm{U}}$ and $t_{\Lambda }$ coinciding to
within a given factor is smaller in the landscape model than in our proposal.

An alternative quantitative statement of the coincidence problem is the
probability of observing $O_{0}<\Omega _{\Lambda 0}<1-O_{0}$ for some $O_{0}$%
, e.g. for $O_{0}=0.1$ and $O_{0}=0.05$, we find: 
\begin{eqnarray}
P(\Omega _{\Lambda 0} \in (0.10,0.90)) &=&23\%, \\  P(\Omega _{\Lambda 0} \in
(0.05,0.95)) &=& 31\%,\qquad \mathrm{(}f_{\mathrm{N}}=c(N)\mathrm{)}, \nonumber \\
P(\Omega _{\Lambda 0} \in (0.10,0.90)) &=& 35\%, \\ P(\Omega _{\Lambda 0} \in
(0.05,0.95)) &=& 47\%,\qquad \mathrm{(}f_{\mathrm{N}}=c(N)e^{-3N}\mathrm{)}. \nonumber
\end{eqnarray}%
For comparison, with the same selection effects and a
uniform prior on $\Lambda $, the landscape model gives: 
\begin{eqnarray}
P(0.10<\Omega _{\Lambda 0}<0.90) &=& 11\% \nonumber \\ 
P(0.05<\Omega _{\Lambda 0}<0.95) &=& 16\%. \nonumber
\end{eqnarray}

Again, the observation of a cosmic coincidence in the values of $t_{\Lambda }
$ and $t_{U}$ is not atypical in our model or in the string landscape model
with uniform prior. However, it is significantly more likely in the model we
have proposed, and our model\emph{\ is independent of the choice of prior
for }$\Lambda $.

\section{Concluding Remarks and Possible Questions}

\label{sec:conclusion} The cosmological constant problem and the related
coincidence problem are two of the most important unsolved problems in
cosmology, and are also of importance for high-energy physics and the search
for a complete theory of quantum gravity. So far, cosmologists have only
been able to describe the effects of the cosmological constant by introducing an
arbitrary $\Lambda $ term chosen to have the observed value ($\Lambda _{%
\mathrm{obs}}$), or to model it by a scalar field that evolves so slowly 
that its (dark) energy density  is `almost' a cosmological constant at late times
(as in quintessence models). It is known that the existence of galaxies,
which one may take as a pre-requisite for atom-based observers such as
ourselves, would not be possible if $\Lambda \gtrsim 10^{3}\Lambda _{\mathrm{%
obs}}$. In the context of a multiverse of different universes, each with a
different $\Lambda $, using the anthropic upper limit $\Lambda \lesssim
10^{3}\Lambda _{\mathrm{obs}}$ to explain the observed $\Lambda $ depends
heavily on the prior likelihood of finding different values of $\Lambda $ in
the multiverse. This prior, $f_{\mathrm{prior}}(\Lambda )\,\mathrm{d}\Lambda 
$, is the fraction of all universes with a CC in the region $[\Lambda
,\Lambda +\,\mathrm{d}\Lambda ]$. If, for $\Lambda \lesssim 10^{3}\Lambda _{%
\mathrm{obs}}$, we have $f_{\mathrm{prior}}(\Lambda )\approx \mathrm{const,}$
then the observed value of $\Lambda $ is not atypical in universes
compatible with the anthropic limit. Other plausible forms for the prior
include a uniform prior in log-space, $f_{\mathrm{prior}}\propto 1/\Lambda ,$
or the form $f_{\mathrm{prior}}\propto \exp (3\pi /G\Lambda )$. In either
case, non-zero values of $\Lambda $ would be greatly disfavoured and the
observed value of $\Lambda $ highly unnatural. However, until it is clear
that a uniform prior is (at least approximately) the form of $f_{\mathrm{%
prior}}$ predicted by fundamental theory, the multiverse/anthropic
explanation of $\Lambda $ remains incomplete. Even if it is correct, the
multiverse explanation has not so far made any testable predictions.

We have presented a new proposal for solving the cosmological constant and
coincidence problems. Crucially, in contrast to the multiverse explanation,
our proposal makes a falsifiable prediction. The essence of our new approach
is that the bare cosmological constant $\lambda $ is promoted from a
parameter to a field. The minimisation of the action with respect to $%
\lambda $ then yields an additional field equation, Eq. (\ref{eq:phisimp})
which determines the value of the effective CC, $\Lambda ,$ in the classical
history that dominates the partition (wave) function of the universe, $Z$.
Our proposal is agnostic about the theory of gravity and the number of
space-time dimensions.

In the sense that the cosmological constant is promoted to a field, there is
a superficial similarity between our proposal and quintessence models. In
the latter, the effective cosmological constant depends on a scalar field, $%
\phi (x^{\mu })$, and the variation of the action with respect to $\phi $
gives a local second-order differential equation which determines the
dynamics of $\phi $ up to a specification of two free functions of initial
data. In our proposal, the different values of $\lambda $ are summed over in
the partition function of the universe, and hence $\lambda $ is a field
rather than parameter, but it is not a local scalar field. Whereas a scalar
field, $\phi (x^{\mu }),$ may take a different value at every point in
space-time, $\lambda $ is the same at all points in a given classical history. 
Hence, the additional field equation obtained from the variation of the
action with respect to $\lambda $ is not, as in quintessence theories, a
local second-order differential equation in $\lambda $ (ie $\phi $), but\
is instead an integral equation, Eq. (\ref{eq:phi}), where the domain of
integration is the same as for the action in the partition function, and is
algebraic in $\lambda $. This algebraic property means that there are no
initial or boundary data for $\lambda $ to specify (unlike in the
quintessence case) and our method makes a unique prediction for the value of 
$\lambda $ in terms of the universal configuration of gravitational and
matter fields.

A specific application of our proposal generically results in a testable
prediction. We have taken the action in the wave / partition function, $Z$,
of the universe to be defined on some manifold $\mathcal{M}$. A choice of
definition for $\mathcal{M}$ (e.g. the causal past of the observer) is
required for a specific application of our proposal. Different choices will
result in different predictions for the effective CC, $\Lambda $. With a
given $\mathcal{M}$, the equation for $\lambda $, Eq. (\ref{eq:phisimp}) can
be viewed as a consistency equation which relates the configuration of
metric and matter variables in $\mathcal{M}$ to $\lambda $. Eq. (\ref%
{eq:phisimp}) can be viewed as a consistency condition on the configuration
of the effective CC, $\Lambda ,$ the matter, $\Psi ^{a}$, and metric, $%
g_{\mu \nu }$, fields in $\mathcal{M}$. The consistency condition provided by Eq. (\ref%
{eq:phisimp}) will be violated for the vast majority of potential
configurations $\left\{ g_{\mu \nu },\Psi ^{a},\Lambda \right\} $ (even if
one demands that $g_{\mu \nu }$, $\Psi ^{a}$ obey their respective field
equations). If observations determine a set of $\left\{ g_{\mu \nu },\Psi
^{a},\Lambda \right\} $ for which Eq. (\ref{eq:phisimp}) is violated then
our proposal would be falsified. At the same time, if the observed
configuration is consistent with Eq. (\ref{eq:phi}) to within observational
limits, then our proposal would, for the time being, have passed an important
empirical test and remain a credible solution to the CC problems. In
addition, if one has measured $\Lambda $ but not (or at least not fully),
the $\left\{ g_{\mu \nu },\Psi ^{a}\right\},$ then Eq. (\ref%
{eq:phisimp}) would require that a certain functional of the undetermined $%
\left\{ g_{\mu \nu },\Psi ^{a}\right\} $ vanishes. This would represent
a prediction of our model which could be tested and falsified by subsequent
observations.

In \S \ref{sec:paradigm} we formally described our proposal in its most
general form. In \S \ref{sec:cosmology}, we considered in detail the
specific and simple case where $\mathcal{M}$ is taken to be the causal past
of the observer. With this choice of $\mathcal{M}$, the partition function, $%
Z[\mathcal{M}]$, and $\Lambda $ equation, only depend on those parts of the
universe to which the observer is causally connected. For a given observer,
this choice of $\mathcal{M}$ is well-defined in a coordinate invariant
fashion. Using this choice, we found that Eq. (\ref{eq:phi}) for $\Lambda $
reduces to a simple form which, keeping only the dominant terms, requires a
balance between the spatial curvature and the contribution of baryonic
matter to the matter Lagrangian density, $\mathcal{L}_{\mathrm{matter}}$. We
defined $\mathcal{L}_{\mathrm{matter}}=-\zeta _{\mathrm{b}}\rho _{\mathrm{%
baryon}}$ where $\rho _{\mathrm{baryon}}$ is the density of baryonic matter
and $\zeta _{\mathrm{b}}$ is a constant whose value can in principle be
calculated from QCD. Using an approximate analytical model for baryon
structure (the chiral bag model), we estimated $\zeta _{\mathrm{b}}\approx
1/2$. Given the complexity of modelling baryon structure, we conservatively
estimate that $\zeta _{\mathrm{b}}=1/2$ only to within $\pm 30\%$ or so.

We found that Eq. (\ref{eq:phi}) is consistent with the current observable
limits on $\Lambda $, the spatial curvature, and other observable properties
of our universe. If this application of our theory is correct, then the
spatial curvature of our universe must take a particular value. This value
depends on $\Omega _{\Lambda 0}=\Lambda /3H_{0}^{2}$, and the matter and
baryon energy per photon, $\xi $ and $\xi _{\mathrm{b}}$ respectively as
well as the time at which observation take place which can be parametrized,
for instance by the age of the universe, $t_{\mathrm{U}}$ or the CMB
temperature, $T_{\mathrm{CMB}}$. Taking values of $\Omega _{\Lambda 0}=0.73$%
, $\xi =3.43\,\mathrm{eV}$ and $\xi _{\mathrm{b}}=0.54\,\mathrm{eV}$
(consistent with observations), we found that the observed dimensionless
spatial curvature must be: 
\begin{equation*}
\Omega _{k0}=-0.0056\left(\frac{\zeta _{\mathrm{b}}}{1/2}\right).
\end{equation*}%
For reasonable values of $\zeta _{\mathrm{b}}\sim 1/2$, this is within the
current 68\% confidence limit $\Omega _{k0}=-0.0023_{-0.0056}^{+0.0054}$
from the combination of WMAP7 CMB data, with BAO and $H_{0}$ measurements.
Additionally, the predicted value of $\Omega _{k0}$ should be easily
confirmed or ruled out by future measurements of the CMB, $H_{0}$ and baryon acoustic
oscillations (BAO). For example, a combination of Planck CMB data with the
WFMOS BAO has been estimated to be able to determine $\Omega _{k0}$ to a 1-$%
\sigma $ accuracy of $\ $about $1.76\times 10^{-3}$ \cite{Vardanyan:2009ft}.
With the addition of BAO data from the Square Kilometre Array (SKA) or
something similar, the accuracy could be increased to $5.64\times 10^{-4}$
at 68\% confidence \cite{Vardanyan:2009ft}. This would be more than
sufficient to rule out an $\Omega _{k0}$ at the predicted level whatever the
precise value of $\zeta _{\mathrm{b}}$. This could conclusively test our
model as an explanation of the CC problems in the real universe.

For the time being, our model is consistent with current observations. We
also considered the extent to which the observed value of $\Lambda $ is
typical within our model, and hence whether or not our model can truly be
said to solve the CC problems. We found that in our theory the probability of
living in a region of the universe where one observes a given value of $%
\Lambda $ was independent of the fundamental prior weighting on different
values of $\Lambda $:  instead, it was completely determined by the
probability distribution of the number of e-folds of inflation in the early
universe, $f_{\mathrm{N}}(N)$. Larger values of $\Lambda $ require smaller $k
$ and hence more e-folds of inflation. However, the difference in the value
of $N$ required for $\Lambda =0$ and a CC on the edge of the anthropic upper
limit ($\Lambda \sim 10^{3}\Lambda _{\mathrm{obs}}$) is only $\Delta
N\lesssim 1.2$. At the same time, for a single value of $\Lambda $ to
dominate the partition function, and hence for the universe to behave
classically, one requires a fairly small curvature and hence $N\gtrsim 50-62$
(depending on the efficiency of reheating). Thus, the anthropically allowed
range of values of $\Lambda $ correspond to a range of e-folds with $N\in
\lbrack \bar{N},\bar{N}+\Delta N],$ where $\Delta N/\bar{N}\ll 1$. This
means that the dependence of the posterior probability distribution for $%
\Lambda $ depends only fairly weakly on $f_{\mathrm{N}}(N)$.

Unless there is a strong (exponential) preference for a value of $\bar{N}%
+\Delta N$ over $\bar{N}$ (which would increase the preference for larger $%
\Lambda $) we found that the observed value of $\Lambda $ is indeed typical
within our model. Specifically, the observed value, $\Lambda _{\mathrm{obs}}$%
, lies close (either just inside or just outside depending on $f_{\mathrm{N}}
$) the symmetric 1-$\sigma $ confidence interval for $\Lambda $. We also
found that the probability of us observing the cosmological coincidence
between the value of $t_{\Lambda }$ and the age of the universe is
relatively high ($14-53\%$ depending on how one quantifies what counts as a
coincidence).

Our proposal for solving the CC is similar in certain respects to other
multiverse models such as the string landscape, when $\Lambda $ takes
different values in different vacua/parts of the multiverse. Despite this
similarity, it differs from multiverse / landscape models in three crucial
respects: 

(1) Our model is, unlike multiverse models, independent of the fundamental
prior weighting on different values of $\Lambda $; 

(2) The preference for small $\Lambda $ is our model does not come wholly
from anthropic selection effects as it does in multiverse models. Roughly,
the prior probability for different values of $\Lambda $ is uniform for $%
\Lambda \lesssim O(\mathrm{few})\Lambda _{\mathrm{obs}}$ and approximately
uniform in log-space for larger values. This means that the probability of
observing a value of $\Lambda \sim O(\Lambda _{\mathrm{obs}})$ is typically
higher in our model by a factor of $2-5$; 

(3) Our model makes a testable prediction for $%
\Omega _{k0}$ that can be falsified by upcoming CMB/BAO surveys.

\bigskip \emph{\ }

We now address some possible questions about our proposal:

\begin{enumerate}
\item \emph{\textquotedblleft What is the equation of state of dark energy
in this model?"}: The equation of state is exactly $w=-1$ i.e. a pure
cosmological constant / vacuum energy.  Provided that Eq. (\ref{eq:phi}) has a solution,
observers see a classical history with a single constant value of the
effective CC $\Lambda $ (as determined by Eq. (\ref{eq:phi})). Since the $\Lambda $ is observed to be constant, it has an effective pressure, $P_{\Lambda }$, equal in
magnitude but opposite in sign to its effective energy density, $\rho
_{\Lambda }=\kappa ^{-1}\Lambda $ and hence $w \equiv P_{\Lambda }/\rho _{\Lambda }=-1$.


\item \emph{\textquotedblleft Does the observed effective CC depend the time
of observation?"}: Yes (but see Q. 3). In \S \ref{sec:cosmology}, we the manifold, $\mathcal{M}$, on which the action was defined to be the causal past of the observer, which clearly depends on the time (and position) of the observer. 
The observed $\Lambda$ depends on $\mathcal{M}$ and $\partial \mathcal{M}$ through Eq. (\ref{eq:phi}) and hence also depends on the observation time.  Crucially, however, $\Lambda$ is \emph{not} seen to evolve. The classical history that dominates the wave function, $Z[\mathcal{M}]$, has a single constant value of $\Lambda $ throughout the observer's past, $\mathcal{M}$. Thus all observations are consistent with a single constant $\Lambda $ as given by Eq. (\ref{eq:phi}).  The observation time changes, so does the classical
history that dominants the parition function.  Observers at slightly different times would see slightly different classical histories respectively consistent with slightly different values of $\Lambda$.  
In this way, our proposal is quite unlike the `ever-present lambda' models \cite{ever}
where $\Lambda $ arises as a random space-time fluctuation and is always
inversely proportional to the square root of the space-time 4-volume, so $%
\Lambda \approx Gt^{-2}$ at time $t$. Everpresent-$\Lambda$ models have severe observational
problems \cite{everJDB} and furthermore are only consistent with $\Lambda \sim 1/t_{\rm U}^2$ in $3+1$ dimensions (in D+1 spacetime dimensions $\Lambda \sim t^{-(D+1)/2}_{\rm U}$ in natural units).

\item \emph{\textquotedblleft Can observers at different times establish that they measure different values of $\Lambda$?"}: No (at least not classically). Different values of $\Lambda$ correspond to different classical histories. For an observer at one time to communicate the value of $\Lambda $ that they measure to another observer (or even the same observer) at a later time, and hence reveal that the two values are different, they would
have to find a way of sending information from one classical history to another. At the classical level (at least) this is not possible. Classically, an observer will see only a history consistent with the value
of $\Lambda ,$ equal to $\Lambda _{\star }$ say, for their observation time.
This would include seeing reports of / remembering all previous measurements of $%
\Lambda $ as giving $\Lambda =\Lambda _{\star }$.

\item \emph{\textquotedblleft How does $\Lambda $ change with observation
time for the model presented in \S \ref{sec:cosmology}?"}: It decreases as
the observation time increases. As the observation time, $t_{o}$ increases,
the value of the spatial curvature $k$ that is required (i.e. $k_{0}(\Lambda
,t_{o})$) for a given value of $\Lambda $ decreases i.e. $k_{0}(\Lambda
,t_{1})>k_{0}(\Lambda ,t_{2}>t_{1}),$ so at fixed $\Lambda $, $\left.
\partial k_{0}/\partial t_{o}\right\vert _{\Lambda }<0$. We found that at
fixed $t$, $\left. \partial k_{0}/\partial \Lambda \right\vert _{t}<0$. For
a given observer, $k$ is fixed. Thus, defining the observed CC at $t_{o}$ to
be $\Lambda (t_{o})$, we must have: 
\begin{equation*}
\frac{\,\mathrm{d}k_{0}(\Lambda (t_{o}),t_{o})}{\,\mathrm{d}t_{o}}=0=\frac{\,%
\mathrm{d}\Lambda (t_{o})}{\,\mathrm{d}t_{o}}\frac{\partial k_{0}}{\partial
\Lambda }+\frac{\partial k_{0}}{\partial t_{o}}.
\end{equation*}%
It follows that $\,\mathrm{d}\Lambda (t_{o})/\,\mathrm{d}t_{o}<0$. This also
means that at some point in the future, $k$ will be too large for Eq. (\ref%
{eq:phi}) to admit a classical solution. The universe would then cease to
have a dominant classical history. One could view this as in some sense the
end of the classical universe.

\item \emph{\textquotedblleft Does this proposal require the existence of a
multiverse / landscape?"}: No, not in the sense of an ensemble of
\textquotedblleft parallel\textquotedblright\ universes with different
physical constants. However our model does not exclude it either. Our model
does require that the spatial curvature is different for different causally
disconnected observers, a scenario that is naturally realized in the context
of inflationary theory. Different bubbles of space-time will undergo
different amounts of inflation and hence have a different spatial curvature
at the same time, however these different bubble universes are all part of
the same universe in the sense that a hypothetical tachyonic astronaut could
in principle travel between them.

\item \emph{\textquotedblleft What role does inflation play in this model?"}%
: Inflation in important in this model in two related ways. Firstly, it
makes the spatial curvature a spatially varying quantity which is different
in different super-horizon-sized regions, and dependent on the number of
e-folds of inflation that took place in each causally connected part of the
universe. Provided the whole universe (not just the part we see) is large
enough (or infinite) the values of $k$ required for a classical history to
dominate the wave function in our model will occur at least somewhere. It
seems reasonable to assume that this classicality is a prerequisite for the
existence of observers such as ourselves. Provided this is the case, it is
immediately clear that we could only ever live in those parts of the
universe where the number of e-folds $N,$ and the hence spatial curvature $k,
$ lies in the small range we reviewed where classical solutions exist and
anthropic upper-bounds on $\Lambda $ hold. Secondly, in our model the prior
probability (i.e. prior to the inclusion of the anthropic bounds) of
observing a CC in $[\Lambda ,\Lambda +\,\mathrm{d}\Lambda ]$ is related the
probability of a given spatial curvature and hence to the probability of a
given number of e-folds of inflation.

\item \emph{\textquotedblleft If future observations rule out the predicted
value of $\Omega _{k0}$ does this rule out this proposal?"}: Yes, the
application of our scheme given in \S \ref{sec:cosmology} where we take $%
\mathcal{M}$ to the observer's causal past would be conclusively ruled out.
It may be that one could argue the case for a different choice of $\mathcal{M%
}$, and find another application consistent with observations but this is
not something we have investigated at this time. Certainly the choices we
made for the particular application of our proposal in \S \ref{sec:cosmology}
seem to be the most simple and natural. The only wiggle room is if the QCD
parameter $\zeta _{b}$ is significantly different from 0.5, but the required value would then be determined by observations and could be checked against a detailed QCD calculation of $\zeta_{\rm b}$.

\end{enumerate}

In summary: we have introduced a new approach to solving the cosmological
constant and coincidence problems. The bare CC, $\lambda $, or equivalently
the minimum of the vacuum energy, is allowed to take many possible values in
the wave function, $Z$, of the universe. The value of the effective CC in
the classical history that dominates $Z$ is given by a new integral field
equation, Eq. (\ref{eq:phisimp}). Our scheme is agnostic about the theory
of gravity and the number of space-time dimensions. We have applied it in
its simplest and most natural form to a universe in which gravity is
described by GR. The observed classical history will be completely
consistent with a non-evolving cosmological constant. In an homogeneous and
isotropic model of the universe with realistic matter content we find that
the observed value of the effective CC is typical, as is a coincidence
between $1/\sqrt{\Lambda }$ and the present age of the universe, $t_{U}$.
Unlike explanations of the CC problem that rely only on Bayesian selection
in a multiverse, our model in independent of the unknown prior weighting of
different $\Lambda $ values, and makes a specific numerical prediction for
the observed spatial curvature parameter. Specifically,  we should
observe $\Omega _{k0}=-0.0056(2\zeta _{\mathrm{b}}),$ where the QCD bag
parameter is $\zeta _{b}\simeq 0.5$. This prediction is consistent with
current observations but can be tested by Planck/BAO observations in the
very near future. In conclusion, we have described a new type of solution of the
cosmological constant problems. It is consistent with observation and free
of fine-tunings, requires no new forms of dark energy or modifications to
the low-energy theory of gravity, and is subject to high-precision test by
future observations.

\section*{Acknowledgements}
We would like to thank A-C. Davis, R. Brandenberger, Ph. Brax, G. Efstathiou and R. Tavakol  for helpful discussions and
comments. DJS acknowledges STFC.

\appendix

\section{Connection with Unimodular Gravity}

\label{app:unimodular} Central to the paradigm we have proposed for solving
the CC problems is the promotion of the bare cosmological constant, $%
\lambda $, from a fixed parameter to a field. Hence the wave-function
(partition function) of the universe is a super-position of all possible
values of $\lambda $. This concept is not new, and it has been seen to arise
naturally in the study of unimodular gravity (see Refs. \cite%
{Henneaux:1984ji,Unruh:1988in} and references therein).

\subsection{Unimodular Gravity}

Classically, the field equations of unimodular gravity are equivalent to
those of general relativity but with the cosmological constant undetermined.
Unimodular gravity was first formulated in a non-covariant fashion in terms
of the usual Einstein-Hilbert action for GR (and minimally coupled matter
action) but with the degrees of freedom in the metric, $g_{\mu \nu }$,
restricted by the constraint $\sqrt{-g}=1$. The total action is the
unmodified and, dropping surface terms, for some $\lambda $ is given by: 
\begin{eqnarray}
I_{\mathrm{tot}} &=& \frac{1}{2\kappa }\int_{\mathcal{M}}R(g)\sqrt{-g}\,\mathrm{d%
}^{4}x-\frac{1}{\kappa }\int_{\mathcal{M}}\lambda \sqrt{-g}\,\mathrm{d}%
^{4}x \nonumber \\ &&+\int_{\mathcal{M}}\sqrt{-g}\,\mathrm{d}^{4}x\,\mathcal{L}_{\mathrm{%
matter}}. \nonumber
\end{eqnarray}%
Since $\sqrt{-g}$ the second term on the right-hand side (the bare CC term)
is just a constant and so does not contribute to the field equations found
by requiring $\delta I_{\mathrm{tot}}=0$. Varying this action with respect
to the unit modulus metric, $\tilde{g}_{\mu \nu }=(-g)^{-1/4}g_{\mu \nu }$,
gives the trace-free part of the usual Einstein equation: 
\begin{equation*}
R^{\mu \nu }-\frac{1}{4}Rg^{\mu \nu }=\kappa \left[ T_{\mathrm{matter}}^{\mu
\nu }-\frac{1}{4}g^{\mu \nu }T_{\mathrm{matter}}\right] .
\end{equation*}%
Using $\nabla _{\mu }T_{\mathrm{matter}}^{\mu \nu }=0$ and $\nabla _{\mu
}R^{\mu \nu }=\nabla ^{\nu }R/2$ we have
\begin{equation*}
\nabla ^{\mu }(R+\kappa T_{\mathrm{matter}})=0\rightarrow R=-\kappa T_{%
\mathrm{matter}}+4\Lambda ,
\end{equation*}%
where $\Lambda =\mathrm{const}$ is a constant of integration. Using the
above equation, we then recover the usual Einstein equation with effective
cosmological constant $\Lambda $: 
\begin{equation*}
R^{\mu \nu }-\frac{1}{2}Rg^{\mu \nu }=\kappa T_{\mathrm{matter}}^{\mu \nu
}-\Lambda g^{\mu \nu }.
\end{equation*}%
In this formulation of unimodular gravity, there is no connection between
the effective CC, $\Lambda $, and either the bare cosmological constant $%
\lambda $ or the vacuum energy from matter $\rho _{\mathrm{vac}}$. Since $%
\sqrt{-g}=1$, it follows that in unimodular gravity the 4-volume, $V_{%
\mathcal{M}}$, of $\mathcal{M}$ is fixed where: 
\begin{equation*}
V_{\mathcal{M}}=\int_{\mathcal{M}}\sqrt{-g}\,\mathrm{d}^{4}x.
\end{equation*}%
Fixed $V_{\mathcal{M}}$ is usually taken to be the defining feature of
unimodular gravity.

Since $\Lambda $ can take any possible value, the partition/wave function of
the universe in unimodular gravity includes a sum over all values of $%
\Lambda $ with some unspecified weighting $\mu \lbrack \Lambda ]$: 
\begin{equation*}
Z_{\mathrm{uni}}=\int \mu \lbrack \Lambda ]\,\mathrm{d}\Lambda \mathcal{D}%
g_{\mu \nu }\mathcal{D}\Psi ^{a}e^{iI_{\mathrm{tot}}}.
\end{equation*}

Other than the unspecified weight function, $\mu \lbrack \Lambda ]$, another
issue with the original formulation of unimodular gravity is that the
constraint $\sqrt{-g}=1$ is not diffeomorphism invariant. Henneaux and
Teitelboim \cite{Henneaux:1984ji} found an action for unimodular gravity
that is both diffeomorphism invariant and has the shift symmetry under $%
\Lambda \rightarrow \Lambda +\mathrm{const}$ which fixes $\mu \lbrack
\Lambda ]=\mathrm{const}$.

The Henneaux-Teitelboim action is
\begin{equation*}
I_{\mathrm{HT}}\equiv I_{\mathrm{grav}}+I_{\mathrm{m}}-\frac{1}{\kappa }%
\int_{\mathcal{M}}\lambda \left( \sqrt{-g}-\partial _{\mu }\tilde{v}^{\mu
}\right) \,\mathrm{d}^{4}x,
\end{equation*}%
where $I_{\mathrm{grav}}$ is the Einstein-Hilbert action for gravity plus
surface terms, $I_{\mathrm{m}}$ is the matter action including the
contribution from the vacuum energy; $\tilde{v}^{\mu }$ is a vector-density
field and $\lambda =\lambda (x^{\mu })$ is a scalar field. In this
formulation $\sqrt{-g}$ is not fixed a priori. However, varying the action
with respect to the scalar field $\lambda $ gives $\sqrt{-g}\equiv \partial
_{\mu }\tilde{v}^{\mu }$. It follows that there is a shift symmetry under $%
\lambda \rightarrow \lambda +\delta \lambda $ for constant $\delta \lambda $%
\begin{equation*}
I_{\mathrm{HT}}\rightarrow I_{\mathrm{HT}}-\frac{\delta \lambda }{\kappa }%
\int_{\mathcal{M}}\left( \sqrt{-g}-\partial _{\mu }\tilde{v}^{\mu }\right)
=I_{\mathrm{HT}},
\end{equation*}%
where the last equality comes from $\sqrt{-g}=\nabla _{\mu }\tilde{v}^{\mu }$%
. Varying the action with respect to $\tilde{v}^{\mu }$ gives: 
\begin{equation*}
\delta I_{\mathrm{HT}}=\frac{1}{\kappa }\int_{\mathcal{M}}\left[ \tilde{v}%
^{\mu }\partial _{\mu }\lambda +\partial _{\mu }\left( \lambda \delta \tilde{%
v}^{\mu }\right) \right] .
\end{equation*}%
For the $\partial _{\mu }(\lambda \delta \tilde{v}^{\mu })$ term to vanish,
we must have $\delta \tilde{v}^{\mu }n_{\mu }=0$ on $\partial \mathcal{M}$,
where $\partial \mathcal{M}$ corresponds to $f(x^{\mu })=0$ and $n_{\mu
}\propto \nabla _{\mu }f$. With $\delta \tilde{v}^{\mu }$ so fixed on $%
\partial \mathcal{M}$, $\delta I_{\mathrm{HT}}/\delta \tilde{v}^{\mu }=0$
gives 
\begin{equation*}
\partial _{\mu }\lambda =0\Rightarrow \lambda =\mathrm{const},
\end{equation*}%
and so $\lambda $, which represents the bare cosmological constant, is an
arbitrary space-time constant. Varying the action with respect to $g_{\mu
\nu }$ and requiring that any surface integrals vanish on $\partial \mathcal{%
M}$, gives 
\begin{equation*}
R^{\mu \nu }-\frac{1}{2}Rg^{\mu \nu }=\kappa T_{\mathrm{m}}^{\mu \nu
}-\lambda g^{\mu \nu }=\kappa T_{\mathrm{matter}}^{\mu \nu }-\Lambda g^{\mu
\nu },
\end{equation*}%
where $T_{\mathrm{m}}^{\mu \nu }=T_{\mathrm{matter}}^{\mu \nu }-\rho _{%
\mathrm{vac}}g^{\mu \nu }$ and $\Lambda =\lambda +\kappa \rho _{\mathrm{vac}}
$ is the effective CC. Since $\sqrt{-g}=\partial _{\mu }\tilde{v}^{\mu }$,
requiring $n_{\mu }\delta \tilde{v}^{\mu }=0$ fixes the four-volume, $V_{%
\mathcal{M}}$, and ensures that the HT action really does describe a
unimodular theory of gravity: 
\begin{eqnarray}
V_{\mathcal{M}} &=&\int_{\mathcal{M}}\,\mathrm{d}^{4}x\sqrt{-g}=\int_{%
\mathcal{M}}\,\mathrm{d}^{4}x\,\partial _{\mu }\tilde{v}^{\mu },  \notag \\
\delta V_{\mathcal{M}} &=&\int_{\mathcal{M}}\,\mathrm{d}^{4}x\,\partial
_{\mu }\delta \tilde{v}^{\mu }=0.  \notag
\end{eqnarray}%
The partition function for the HT unimodular action is: 
\begin{equation*}
Z_{\mathrm{HT}}=\int \mathcal{D}\lambda \mathcal{D}\tilde{v}^{\mu }\mathcal{D%
}g_{\mu \nu }\mathcal{D}\Psi ^{a}\mu \lbrack \Lambda ]e^{iI_{\mathrm{HT}}},
\end{equation*}%
where the sum over configurations is for $\tilde{v}^{\mu }$ normal to $%
\partial \mathcal{M}$ and the matter and metric variables are fixed on the
boundary.

\subsection{An Alternative Formulation of Our Model}

Now, with $I_{\mathrm{tot}}=I_{\mathrm{grav}}+I_{\mathrm{m}}+I_{\mathrm{CC}%
}[\lambda ,g_{\mu \nu };\mathcal{M}],$ we define we have $I_{\mathrm{HT}}=I_{%
\mathrm{tot}}+I_{v}[\lambda ,\tilde{v}^{\mu };\mathcal{M}]$ where 
\begin{eqnarray}
I_{v} &=&\frac{1}{\kappa }\int_{\mathcal{M}}\lambda \partial _{\mu }\tilde{v}%
^{\mu }=-\int_{\mathcal{M}}\,\mathrm{d}^{4}x\,\tilde{v}^{\mu }\partial _{\mu
}\lambda \\ &&+\frac{1}{\kappa }\int_{\mathcal{M}}\partial _{\mu }\left( \lambda 
\tilde{v}^{\mu }\right) , \nonumber \\
&=&I_{\delta }+I_{\mathrm{v-surf}},  \notag \\
I_{\delta} &=&-\int_{\mathcal{M}}\,\mathrm{d}^{4}x\,\tilde{v}^{\mu
}\partial _{\mu }\lambda ,\qquad I_{\mathrm{v-surf}}=\frac{1}{\kappa }\int_{%
\mathcal{M}}\partial _{\mu }\left( \lambda \tilde{v}^{\mu }\right) .  \notag
\end{eqnarray}%
Here, $I_{\mathrm{v-surf}}$ is a total derivative and so represents a
surface term in the action and $I_{\delta }$ has the property that 
\begin{equation*}
\int \mathcal{D}\tilde{v}^{\mu }e^{iI_{\delta }}\propto \delta \lbrack
\partial _{\mu }\lambda ],
\end{equation*}%
where $\delta \lbrack \partial _{\mu }\lambda ]$ is a functional $\delta $%
-function peaked about space-time constant configurations of $\lambda $, and
so acts as 
\begin{equation*}
\int \mathcal{D}\lambda \,\delta \lbrack \partial _{\mu }\lambda ]A[\lambda
,\dots ]=\int_{-\infty }^{\infty }\,\mathrm{d}\lambda \,A[\lambda ,\dots ].
\end{equation*}%
It follows that the partition function in our proposal can be rewritten (up
to an arbitrary and irrelevant overall constant) as 
\begin{eqnarray}
Z[\mathcal{M}] &=&\int \,\mathrm{d}\lambda \mathcal{D}g_{\mu \nu }\mathcal{D}%
\Psi ^{a}\mu \lbrack \Lambda ]e^{iI_{\mathrm{tot}}} \label{eq:HTlink} \\ &=&\int \mathcal{D}\lambda
\,\delta \lbrack \partial _{\mu }\lambda ]\int \mathcal{D}g_{\mu \nu }%
\mathcal{D}\Psi ^{a}\mu \lbrack \Lambda ]e^{iI_{\mathrm{tot}}}, \nonumber
 \\
&=&\int \mathcal{D}\lambda \mathcal{D}\tilde{v}^{\mu }\mathcal{D}g_{\mu \nu }%
\mathcal{D}\Psi ^{a}\mu \lbrack \Lambda ]e^{i(I_{\mathrm{tot}}+I_{\delta
}=I_{\mathrm{HT}}-I_{\mathrm{v-surf}})},  \notag
\end{eqnarray}%
where in our proposal the sum over configurations is for some fixed set $%
\left\{ Q^{A}\right\} $ of the metric and matter variables fixed on the
boundary; $\tilde{v}^{\mu }$ is just a Lagrange multiplier field here and so
is not assumed to be fixed anywhere.

It is clear from the second line of Ref. (\ref{eq:HTlink}) that there is an
aesthetic similarity between the partition function in our proposal and that
in the HT formulation of unimodular gravity. Both theories can be formulated
in terms of a scalar field $\lambda (x^{\mu })$ and vector-density $\tilde{v}%
^{\mu }$ in addition to the usual metric and matter fields. When written in
this way, the action in our proposal is $I_{\mathrm{tot}}+I_{\delta }$ with
differs from the action in the HT proposal by a surface term $I_{\mathrm{%
v-surf}}$. The two formulations also differ in terms of what is taken to be
fixed on the boundary, with the main difference being that in HT unimodular
gravity, $\tilde{v}^{\mu }$ is fixed normal to $\partial \mathcal{M}$, which
in turn fixes the 4-volume, $V_{\mathcal{M}}$, of $\mathcal{M}$. In our
formulation the addition of the subtraction of the surface term $I_{\mathrm{%
v-surf}}$ relative to $I_{\mathrm{HT}}$, means that one no longer needs to
require $\delta \tilde{v}^{\mu }n_{\mu }=0$ on $\partial \mathcal{M,}$ and
hence the $V_{\mathcal{M}}$ is not fixed in our proposal and it is not a
unimodular gravity theory, despite its similarities to the HT theory.

Varying the action in our model, produces terms proportional to $\delta
\lambda $. Defining $I_{\mathrm{full}}=I_{\mathrm{tot}}+I_{\delta }$, and
assuming a GHY surface term for gravity for illustrative purposes 
\begin{eqnarray}
\delta I_{\mathrm{full}} &=&\int_{\mathcal{M}}\,\mathrm{d}^{4}x\left[ \frac{1%
}{2\kappa }\tilde{E}^{\mu \nu }\delta g_{\mu \nu }+\tilde{\Phi}_{a}\delta
\Psi ^{a}-\kappa ^{-1}\delta \tilde{v}^{\mu }\partial _{\mu }\lambda \right. \nonumber \\ && \left.-\kappa
^{-1}\delta \lambda \left( \sqrt{-g}-\partial _{\mu }\tilde{v}^{\mu }\right) %
\right] \nonumber  \\
&+&\int_{\partial \mathcal{M}}\sqrt{|\gamma |}\,\mathrm{d}^{3}x\left[ \kappa
^{-1}N^{\mu \nu }\delta \gamma _{\mu \nu }+\Sigma _{a}\delta \Psi
^{a} \right. \nonumber \\ && \left. -\kappa ^{-1}\delta \lambda f_{\mu }\tilde{v}^{\mu }\right] ,  \notag
\end{eqnarray}%
where $\partial \mathcal{M}$ corresponds to $f(x^{\mu })=0$, $f<0$ in $%
\mathcal{M}$, and $f_{\mu }=\nabla _{\mu }f$. Here, $\gamma _{\mu \nu }$ is
the induced metric on $\partial \mathcal{M}$, and $\tilde{E}^{\mu \nu
}=E^{\mu \nu }=\kappa T_{\mathrm{matter}}^{\mu \nu }-G^{\mu \nu }-\Lambda
g^{\mu \nu }$; $\tilde{\Phi}_{a}=\sqrt{-g}\Phi _{a}$ and $\Lambda =\lambda
+\rho _{\mathrm{vac}}$ is the effective CC. The classical field equations
for $g_{\mu \nu }$ and $\Psi ^{a}$ are then $E^{\mu \nu }=\Phi _{a}=0$ and
these cause the variation of $I_{\mathrm{full}}$ with respect to $g_{\mu \nu
}$ and $\Psi ^{a}$ to vanish in the bulk (i.e. in $\mathcal{M}$). Similarly,
requiring $\delta I_{\mathrm{full}}=0$ with respect to variations of $%
\lambda $ and $\tilde{v}^{\mu }$ in the bulk gives: 
\begin{equation*}
\sqrt{-g}=\partial _{\mu }\tilde{v}^{\mu },\qquad \partial _{\mu }\lambda =0.
\end{equation*}%
When these field equations hold in the bulk, $\delta I_{\mathrm{full}}$
reduces to surface integrals over $\partial \mathcal{M}$. Since $\partial
_{\mu }\lambda =0$, the allowed variations of $\lambda $ are those for which 
$\delta \lambda $ is a space-time constant. Hence, 
\begin{eqnarray}
\delta I_{\mathrm{full}} &=&\int_{\partial \mathcal{M}}\,\mathrm{d}^{3}x%
\sqrt{\gamma }\left[ \kappa ^{-1}\tilde{N}^{\mu \nu }\delta \gamma _{\mu \nu
}+\tilde{\Sigma}_{a}\delta \Psi ^{a} \right.  \nonumber \\ && \left.-\delta \lambda \int_{\partial \mathcal{M%
}}\,\mathrm{d}^{3}x\,\kappa ^{-1}f_{\mu }\tilde{v}^{\mu }\right] ,  \notag \\
&=&\int_{\partial \mathcal{M}}\,\mathrm{d}^{3}x\sqrt{\gamma }\left[ \kappa
^{-1}\tilde{N}^{\mu \nu }\delta \gamma _{\mu \nu }+\tilde{\Sigma}_{a}\delta
\Psi ^{a}\right] \nonumber \\ && -\delta \lambda \frac{1}{\kappa }\int_{\mathcal{M}}\sqrt{-g}%
\,\mathrm{d}^{4}x,  \notag
\end{eqnarray}%
where in the second line we have used $\sqrt{-g}=\partial _{\mu }\tilde{v}%
^{\mu }$ to eliminate all appearances of the Lagrange multiplier field $%
\tilde{v}^{\mu }$.

We wish to have $\delta I_{\mathrm{full}}=0$ for the classical solution.
This could be achieved by taking $\delta \gamma _{\mu \nu }=\delta \Psi
^{a}=0$ (for all $\lambda $) on $\partial \mathcal{M}$ and $\delta \lambda =0
$. Indeed fixing $\gamma _{\mu \nu }$ and $\Psi ^{a}$ on $\partial \mathcal{M%
}$ would, modulo the field equations, generally fix $\lambda $ and set $%
\delta \lambda =0$. However, fixing $\lambda $ in $\mathcal{M}$ returns us
to the usual action of general relativity where the bare CC is some fixed
external parameter. Thus, to preserve the nature of $\lambda $ as a
configuration variable that is integrated over the partition function, we
cannot take either $\delta \lambda =0$ or $\delta \gamma _{\mu \nu }=\delta
\Psi ^{a}=0$ (for all $\lambda $).

In our scheme for solving the CC problems, we propose making a different
ansatz: $\delta \gamma _{\mu \nu }=\mathcal{H}_{\mu \nu }\delta \lambda $
and $\delta \Psi ^{a}=\mathcal{P}^{a}\delta \lambda $ where the form of $%
\mathcal{H}_{\mu \nu }$ and $\mathcal{P}^{a}$ must be consistent with the
classical field equations. This is equivalent to fixing $\gamma _{\mu \nu }$
and $\mathcal{P}^{a}$ only for each value of $\lambda $ rather than for all $%
\lambda $. Then, we have 
\begin{eqnarray}
\delta I_{\mathrm{full}} &=& \delta \lambda \left( \int_{\partial \mathcal{M}}%
\sqrt{\gamma }\,\mathrm{d}^{3}x\left[ \kappa ^{-1}N^{\mu \nu }\mathcal{H}%
_{\mu \nu }+\Sigma _{a}\mathcal{P}^{a}\right] \right. \nonumber \\ &&\left.-\kappa ^{-1}\int_{\partial 
\mathcal{M}}\,\mathrm{d}^{3}x\,f_{\mu }\tilde{v}^{\mu }\right) , \nonumber
\end{eqnarray}%
and so we can have classical solutions where $\delta I_{\mathrm{full}}=0$
without having to externally fix $\lambda $ (i.e. set $\delta \lambda =0$).
Quantum mechanically, $\lambda $ can take all possible values and the
partition/wave function is a super-position over histories with all possible
values of $\lambda $. Classically, the dominant history is the one where the
value of $\lambda $ is such that $\delta I_{\mathrm{full}}=0$ i.e.: 
\begin{eqnarray}
\int_{\partial \mathcal{M}}\sqrt{\gamma }\,\mathrm{d}^{3}x\left[ \kappa
^{-1}N^{\mu \nu }\mathcal{H}_{\mu \nu }+\Sigma _{a}\mathcal{P}^{a}\right] &=&%
\frac{1}{\kappa }\int_{\partial \mathcal{M}}\,\mathrm{d}^{3}x\,f_{\mu }%
\tilde{v}^{\mu } \nonumber \\ &=&\frac{1}{\kappa }\int_{\mathcal{M}}\sqrt{-g}\,\mathrm{d}%
^{4}x.\nonumber
\end{eqnarray}%
In \S \ref{sec:prop:Leqn}, we showed that equation of $\Lambda$ in our theory is entirely equivalent to 
\begin{equation*}
\frac{\,\mathrm{d}I_{\mathrm{class}}}{\,\mathrm{d}\lambda }=0,
\end{equation*}%
where $I_{\mathrm{class}}$ is the value of $I_{\mathrm{tot}}$ when the
classical field equations hold for the matter and the metric.

Since $\lambda $ only ever appears in the other field equations in the
combination $\lambda +\kappa \rho _{\mathrm{vac}}=\Lambda $, this is a field
equation for the effective CC, $\Lambda $. We note that by introducing the
Lagrange multiplier field $\tilde{v}^{\mu }$, this field equation is
equivalent to the vanishing of a surface-integral over the boundary $%
\partial \mathcal{M}$. In this sense, it can be said to be holographic.

\subsection{Summary}

In this appendix, we have seen that our proposal can alternatively be
formulated in terms of the action $I_{\mathrm{full}} =I_{\mathrm{tot}%
}+I_{\delta}$ and an ansatz about how the boundary metric and matter fields
depend on $\Lambda$ which is required to preserve the freedom to vary the
bare cosmological constant $\lambda$.

We noted that $I_{\mathrm{full}}$ is almost equivalent to the
Henneaux-Teitelboim action, $I_{\mathrm{HT}}$ for unimodular gravity with
the only difference between a surface term, $I_{\mathrm{v-surf}}$. Despite
this similarity, the subtraction of $I_{\mathrm{v-surf}}$ from $I_{\mathrm{HT%
}}$ to $I_{\mathrm{full}}$ greatly alters the properties of the theory as in
our model one does not need to fix the four-volume, $V_{\mathcal{M}}$,
whereas in the HT model, as a unimodular gravity theory, $V_{\mathcal{M}}$
must be held fixed to return the usual classical field equations. Both
unimodular gravity and our proposal feature a sum over all possible values
of the bare CC, $\lambda $, in the partition function. This sum includes an
unspecified weight, or prior, on $\Lambda $: $\mu \lbrack \Lambda ]$. In
unimodular gravity there is no accompanying classical field equation for $%
\lambda $ and so it remains completely unspecified. The weighting $\mu
\lbrack \Lambda ]$ then plays an important role in determining the relative
contributions of different values of $\lambda $ to the partition function.
In our model, the subtraction of a surface term from the unimodular action,
combined with the ansatz about the dependence of boundary quantities on $%
\lambda $, provides a field equation which determines the classical value of
the effective CC. The partition function is strongly peaked about the value
of $\lambda $ for which this field equation holds. In this classical limit,
only this value of $\lambda $ contributes to the partition function and $\mu
\lbrack \Lambda ]$ simply becomes an irrelevant overall constant multiplying
the partition function. Its form is no longer important. Physics in our
model is independent of the prior weighting function $\mu \lbrack \Lambda ]$
to an excellent approximation whereas in unimodular gravity it is not. We
also saw that the $\Lambda $ equation in our model can be written in a
holographic fashion, as the vanishing of an integral over the boundary $%
\partial \mathcal{M}$.

\section{Surfaces Terms in General Relativity}

\label{app:surface} In this appendix we rederive, for completeness, the form of the surface terms which must be added to the usual
Einstein-Hilbert action, $I_{\mathrm{EH}}$, to make it first order in
derivatives of the metric. The need for these boundary terms was first
realized by York \cite{York:1972sj}, and then rediscovered and refined by
Gibbons and Hawking \cite{Gibbons:1976ue}. York, and then Gibbons and
Hawking, explicitly derived the form of the required surface term for a
non-null boundary. The equivalent surface terms for null boundaries follow
from a double null decomposition of the Einstein field equations, see Refs. 
\cite{Brady:1995na,Geroch:1973am}, although this has rarely been explicitly
stated. We detail the derivation of the Gibbons-Hawking-York (GHY) surface
terms, $I_{\mathrm{GHY}}$, for a `cosmological' boundary defined to be the
union of the surface of past-light cone of a given observer, $\partial 
\mathcal{M}_{u}$ boundary and some initial hypersurface $\partial \mathcal{M}%
_{I}$ with timelike normal. For this setting, we explicitly state how the
variation of $I_{\mathrm{grav}}=I_{\mathrm{EH}}+I_{\mathrm{GHY}}$ depends on
the metric on the boundary. We also restate the definition of York's
cosmological surface term, $I_{\mathrm{YC}}$, defined in Ref. \cite%
{York:1972sj}, since this is relevant for boundaries such as the initial
singularity.

We take $\mathcal{M}$ to be the manifold where $u(x^{\mu })<0$ and $0<\tau
(x^{\mu })<\tau _{0}$ for some $u(x^{\mu })$ and $\tau (x^{\mu })$. We
define $u(x^{\mu })=\tau_{0}$ on the past-light of an observer (at $\tau =\tau
_{0}$) and $w(x^{\mu })$ to be a null coordinate that lies perpendicular to $%
u$, defined so that $\tau =(u+w)/2$; $\tau $ is a timelike coordinate i.e. $%
\nabla _{\mu }\tau \nabla ^{\mu }\tau <0$.

An integral over $\mathcal{M}$ is equivalent to an integral over the whole
space-time weighted by $H(-u)H(\tau )H(\tau _{0}-\tau )$ where $H(y)$ is the
Heaviside function which is unity for $y>0$ and vanishes for $y<0$; $\,%
\mathrm{d}H(y)/\,\mathrm{d}y=\delta (y)$ where $\delta (y)$ is the Dirac
delta function.

We therefore write: 
\begin{equation*}
\mathcal{M}=\left\{ x^{\mu }:H(\tau_{0}-u)H(\tau )H(\tau _{0}-\tau )>0\right\} ,
\end{equation*}%
and $\partial \mathcal{M}=\partial \mathcal{M}_{u}\cup \partial \mathcal{M}%
_{I}$ where $\partial \mathcal{M}_{u}$ is $\{u=\tau_{0},0<\tau <\tau _{0}\}$
and $\partial \mathcal{M}_{I}$ is $\{u<\tau_{0},\tau =0\}$. We define $u_{\mu
}=\nabla _{\mu }u$ and $w_{\mu }=\nabla _{\mu }w$. Now $u$ and $w$ are null
coordinates so $u^{\mu }u_{\mu }=w_{\mu }w^{\mu }=0$ and we define $u_{\mu
}w^{\mu }=-2e^{-2\sigma }$. We also define $\tau _{\mu }=\nabla _{\mu }\tau $%
. Since $\tau =(u+w)/2$ it follows that $\tau _{\mu }\tau ^{\mu
}=-e^{-2\sigma }$. We define $m_{\mu }=e^{\sigma }\tau _{\mu }$ so that $%
m_{\mu }m^{\mu }=-1$ and then $n_{\mu }=e^{\sigma }u_{\mu }$. We then have $%
m_{\mu }n^{\mu }=1$. Finally, we define $\{\theta ^{i}\}$, $i=1,2$, to be
intrinsic coordinates on the surfaces, $S$, of constant $\tau $ and $u$. We
define $e_{\mu }^{i}=\partial _{\mu }\theta ^{i}$. The metric $g_{\mu \nu }$
can then be decomposed thus: 
\begin{equation}
g_{\mu \nu }=n_{\mu }n_{\nu }+2n_{(\nu }m_{\mu )}+h_{\mu \nu },\label{eq:g:decomp1}
\end{equation}%
where $h_{\mu \nu }n^{\mu }=h_{\mu \nu }m^{\mu }=0$.

The Einstein-Hilbert action for General Relativity is: 
\begin{eqnarray}
I_{\mathrm{EH}} = \frac{1}{2\kappa} \int_{\mathcal{M}} \sqrt{-g}R(g) \, 
\mathrm{d}^4 x,  \notag
\end{eqnarray}
where $R(g)$ is the Ricci scalar curvature of $g_{\mu \nu}$. This is second
order in derivatives of the metric.

\subsection{GHY Surface Term}

On a non-null boundary, the Gibbons-Hawking-York surface term, $I_{\mathrm{%
GHY}}$, is the surface term that must be added $I_{\mathrm{EH}}$ to make $I_{%
\mathrm{EH}}+I_{\mathrm{GHY}}$ first order in derivatives of the metric. It
is natural to extend this definition to null boundaries, so that on a
general boundary $I_{\mathrm{EH}}+I_{\mathrm{GHY}}$ is first order in
derivatives of the metric. We use this definition to find the form of $I_{%
\mathrm{GHY}}$ for our $\partial \mathcal{M}$. This is more simply and
clearly done by writing the metric in terms of a vierbein $E_{\mu }^{I}$, $%
I=1,2,3,4,$ where $g_{\mu \nu }=E_{\mu }^{I}E_{\nu }^{J}\eta _{IJ}$ for some
fixed $\eta _{IJ}$; $\mathrm{det}\eta =-1$. We choose a form for $\eta _{IJ}$
suited for a decomposition of the metric along a null and a time-like
direction: 
\begin{eqnarray}
\eta _{IJ} &=&\ba\mathbb{I}_{2\times 2} &0 \\
0 &\mathbb{N}_{2\times 2}\ea,\qquad \mathbb{N}_{2\times 2}=\ba1&1 \\
1 &0\ea,\nonumber \\ \mathbb{I}_{2\times 2}&=&\mathrm{diag}(1,1).  \notag
\end{eqnarray}%
We use $\eta _{IJ}$ and its inverse $\eta ^{IJ}$ to raise and lower indices $%
I$.

It follows from $g_{\mu \nu} = E_{\mu}^{I}E_{\nu}^{J}\eta_{IJ}$ that
\begin{eqnarray}
g_{\mu \nu} = E_{\mu}^{3}E_{\nu}^{3} + 2E_{(\mu}^{3}E_{\nu)}^4 +
\sum_{i=1}^{2}E_{\mu}^{i}E_{\nu}^{j}.  \label{eq:g:decomp2}
\end{eqnarray}
Comparing Eqs. (\ref{eq:g:decomp1}) and (\ref{eq:g:decomp2}), we see that we
can take $E_{\mu}^{3}=n_{\mu}$, $E_{\mu}^{4} = m_{\mu}$ and then $h_{\mu
\nu} = E_{\mu}^{1}E_{\nu}^{1}+E_{\mu}^{2}E_{\nu}^{2}$. With intrinsic
coordinates $\lbrace \theta^{i}\rbrace$ on $\lbrace u, \tau\rbrace = \mathrm{%
const}$ surfaces, $S$, we have: $E_{\mu}^{i} = a^{i}{}_{j}\left[%
e_{\mu}^{i}+K^{i}u_{\mu}+\beta^{j}\tau_{\mu}\right]$ and let $h_{ij} =
\sum_{l} a^{l}{}_{i} a^{l}{}_{j}$. $h_{ij}$ is then the induced 2-metric on $%
S$ and $K^{i}$ and $\beta^{i}$ are shift 2-vectors. We also define a
`radial' coordinate, $r$, on surfaces of constant $\tau$ by $r = (u-w)/2$.

Thus, we have 
\begin{eqnarray}
g_{\mu \nu }\,\mathrm{d}x^{\mu }\,\mathrm{d}x^{\nu } &=& e^{2\sigma }\,\mathrm{d}%
u^{2}-2e^{\sigma }N\,\mathrm{d}u\,\mathrm{d}\tau +h_{ij} \mathrm{D}\theta^{i}\mathrm{D}\theta^{j},\nonumber \\
\mathrm{D}\theta^{i} &=& \mathrm{d}%
\theta ^{i}+K^{i}\,\mathrm{d}u+\beta ^{i}\,\mathrm{d}\tau. \nonumber
\end{eqnarray}

Now $E^{\mu I}$ is a 4-vector and so 
\begin{eqnarray}
R(g) &=& E^{\nu }{}_{I}\left[ \nabla _{\mu },\nabla _{\nu }\right] E^{\mu I}
=\nabla _{\mu }\left[ E^{\nu }{}_{I}\nabla _{\nu }E^{\mu I}-E^{\mu
}{}_{I}\nabla _{\nu }E^{\nu I}\right]   \notag \\
&&+\nabla _{\mu }E^{\mu }{}_{I}\nabla _{\nu }E^{\nu I}-\nabla _{\mu }E^{\nu
}{}_{I}\nabla _{\nu }E^{\mu I}.  \notag
\end{eqnarray}%
We define $\omega ^{\mu \nu \rho }=-\omega ^{\mu \rho \nu }=E_{I}^{\rho
}\nabla ^{\mu }E^{\nu I}$. In terms of the vierbein, we can rewrite $I_{%
\mathrm{EH}}$ in the following form: 
\begin{eqnarray}
I_{\mathrm{EH}} &=&\frac{1}{2\kappa }\int_{\mathcal{M}}\sqrt{-g}\,\mathrm{d}%
^{4}x\,R(g)=-\frac{1}{\kappa }\int_{\mathcal{M}}\partial _{\rho }\left[ 
\sqrt{-g}\omega _{\mu }{}^{\mu \rho }\right] \notag  \\
&&+\frac{1}{2\kappa }\int_{\mathcal{M}}\sqrt{-g}\,\mathrm{d}^{4}x\,\left[
\omega _{\nu }{}^{\nu \rho }\omega ^{\mu }{}_{\mu \rho }-\omega ^{\mu \nu
\rho }\omega _{\nu \mu \rho }\right] .  \label{eq:IEH:rewrite} 
\end{eqnarray}%
All second derivatives of the vierbein, and hence also of the metric, are
contained in the term 
\begin{equation*}
-\frac{1}{\kappa }\int_{\mathcal{M}}\partial _{\rho }\left[ \sqrt{-g}\omega
_{\mu }{}^{\mu \rho }\right] .
\end{equation*}%
Since this term is a total derivative it is equivalent to a surface integral
over $\partial \mathcal{M}$. $I_{\mathrm{GHY}}$ is the surface term which
must be added to the action to remove second derivatives of the metric, it
is clear from Eq. (\ref{eq:IEH:rewrite}) that: 
\begin{eqnarray}
I_{\mathrm{GHY}} &=&\frac{1}{\kappa }\int_{\mathcal{M}}\partial _{\rho }%
\left[ \sqrt{-g}\omega _{\mu }{}^{\mu \rho }\right] ==I_{\mathrm{GHY}%
}^{(u)}+I_{\mathrm{GHY}}^{(I)}, \\
I_{\mathrm{GHY}}^{(u)} &=&\frac{1}{\kappa }\int_{\partial \mathcal{M}%
_{u}}e^{\sigma }\sqrt{h}\,\mathrm{d}\tau \,\mathrm{d}^{2}\theta \,\left[
-n_{\rho }\omega _{\mu }{}^{\mu \rho }\right] , \\
I_{\mathrm{GHY}}^{(I)} &=&\frac{1}{\kappa }\int_{\partial \mathcal{M}%
_{I}}e^{\sigma }\sqrt{h}\,\mathrm{d}r\,\mathrm{d}^{2}\theta \,\left[
-m_{\rho }\omega _{\mu }{}^{\mu \rho }\right] .
\end{eqnarray}%
Then, with $G^{\mu \nu }=R^{\mu \nu }-\frac{1}{2}Rg^{\mu \nu }$, varying $I_{%
\mathrm{EH}}+I_{\mathrm{GHY}}$ with respect to the vierbein gives: 
\begin{eqnarray}
\delta (I_{\mathrm{EH}}+I_{\mathrm{GHY}}) &=&-\frac{1}{2\kappa }\int_{\mathcal{M%
}}\,\mathrm{d}^{4}x\,\sqrt{-g}G^{\mu \nu }\delta g_{\mu \nu } \nonumber \\ &&-\frac{1}{%
2\kappa }\int_{\mathcal{M}}\,\mathrm{d}^{4}x\,\partial _{\rho }\left[ \sqrt{%
-g}S^{\mu \nu \rho }2E_{\nu I}\delta E_{\mu }^{I}\right] .\nonumber
\end{eqnarray}%
where 
\begin{equation*}
S^{\mu \nu \rho }=\omega ^{\nu \mu \rho }-\omega _{\sigma }{}^{\sigma \rho
}g^{\mu \nu }+\omega _{\sigma }{}^{\sigma \mu }g^{\nu \rho },
\end{equation*}%
We note that $S^{\mu \nu \rho }V_{\mu }V_{\rho }=0$ for any $V_{\mu }$.

We define: 
\begin{eqnarray}
\delta I^{(u)}+\delta I^{(I)} &=&-\frac{1}{2\kappa }\int_{\mathcal{M}}\,%
\mathrm{d}^{4}x\,\partial _{\rho }\left[ \sqrt{-g}S^{\mu \nu \rho }2E_{\nu
I}\delta E_{\mu }^{I}\right] ,  \notag \\
\delta I^{(u)} &=&\frac{1}{2\kappa }\int_{\partial \mathcal{M}_{u}}e^{\sigma
}\sqrt{h}\,\mathrm{d}\tau \,\mathrm{d}^{2}\theta \,\left[ n_{\rho }S^{\mu
\nu \rho }2E_{\nu I}\delta E_{\mu }^{I}\right] ,  \notag \\
\delta I^{(I)} &=&\frac{1}{2\kappa }\int_{\partial \mathcal{M}_{I}}e^{\sigma
}\sqrt{h}\,\mathrm{d}r\,\mathrm{d}^{2}\theta \,\left[ m_{\rho }S^{\mu \nu
\rho }2E_{\nu I}\delta E_{\mu }^{I}\right] .  \notag
\end{eqnarray}%
We now re-express the $I_{\mathrm{GHY}}^{(u)}$, $I_{\mathrm{GHY}}^{(I)}$, $%
\delta I^{(u)}$ and $\delta I^{(I)}$ in a more familiar form in terms of the
geometry of the boundaries $\partial \mathcal{M}_{u}$ and $\partial \mathcal{%
M}_{I}$. We note that $E_{\rho }^{I}\omega ^{\mu \nu \rho }=\nabla ^{\mu
}E^{\nu I}$ and so 
\begin{eqnarray}
-n_{\rho }\omega ^{\mu \nu \rho } &=&-E_{\rho }^{3}\omega ^{\mu \nu \rho
}=-\nabla ^{\mu }n^{\nu },  \notag \\
-m_{\rho }\omega ^{\mu \nu \rho } &=&-E_{\rho }^{4}\omega ^{\mu \nu \rho
}=-\nabla ^{\mu }m^{\nu }.  \notag
\end{eqnarray}

\paragraph{Null Boundary:}

We begin by considering the null boundary, $\partial \mathcal{M}_{u}$, given
by $u=0$. We define $\mathcal{K}^{\mu \nu }$ to be the extrinsic curvature
of $h_{\mu \nu }$ along $n^{\mu }$: 
\begin{equation*}
\mathcal{K}^{\mu \nu }=\mathcal{K}^{\nu \mu }=-\frac{1}{2}h^{\mu \rho
}h^{\nu \sigma }\mathcal{L}_{n}h_{\rho \sigma }=-h^{\mu \rho }h^{\nu \sigma
}\nabla _{\rho }n_{\sigma }.
\end{equation*}%
Thus, with $\mathcal{K}=\mathcal{K}^{\mu \nu }g_{\mu \nu }=\mathcal{K}^{\mu
\nu }h_{\mu \nu }$ and using $n^{\rho }n_{\rho }=0$, 
\begin{equation*}
-n_{\rho }\omega _{\mu }{}^{\mu \rho }=-\nabla _{\rho }n^{\rho }=\mathcal{K}%
-n_{\mu }m_{\nu }\nabla ^{\mu }n^{\nu }=\mathcal{K}+\nu ,
\end{equation*}%
where 
\begin{equation*}
\nu =-n_{\mu }m_{\nu }\nabla ^{\mu }n^{\nu }=-\mathcal{L}_{n}\sigma =-n^{\mu
}\nabla _{\mu }\sigma 
\end{equation*}%
is the in-affinity. Using this, the GHY term on $\partial \mathcal{M}_{u}$
can be written succinctly as: 
\begin{equation*}
I_{\mathrm{GHY}}^{(u)}=\frac{1}{\kappa }\int_{\partial \mathcal{M}%
_{u}}e^{\sigma }\sqrt{h}\,\mathrm{d}\tau \,\mathrm{d}^{2}\theta \,\left[ 
\mathcal{K}+\nu \right] .
\end{equation*}

We now consider $\delta I^{(u)}$ with $u$ and $\tau $ fixed $\delta E_{\mu
}^{3}=\delta n_{\mu }=n_{\mu }\delta \sigma $ and $\delta E_{\mu
}^{4}=\delta m_{\mu }=m_{\mu }\delta \sigma $. Using these equations and $%
S^{\mu \nu \rho }n_{\mu }n_{\rho }=0$ 
\begin{equation*}
n_{\rho }S^{\mu \nu \rho }2E_{\nu I}\delta E_{\mu
}^{I}=\sum_{i=1}^{2}n_{\rho }S^{\mu \nu \rho }2E_{\nu i}\delta E_{\mu
}^{i}+2n_{\rho }n_{\nu }m_{\mu }S^{\mu \nu \rho }\delta \sigma ,
\end{equation*}%
and 
\begin{equation*}
n_{\rho }S^{\mu \nu \rho }=\nabla ^{\nu }n^{\mu }-g^{\mu \nu }\nabla _{\rho
}n^{\rho }+\omega _{\sigma }^{\sigma \mu }n^{\nu },
\end{equation*}%
$h^{\mu \rho }\nabla _{\rho }n^{\nu }$ can be decomposed as 
\begin{eqnarray}
h^{\mu \rho }\nabla _{\rho }n^{\nu } &=&-\mathcal{K}^{\mu \nu }+n^{\nu
}m_{\sigma }h^{\mu \rho }\nabla _{\rho }n^{\sigma } \\ &=&-K^{\mu \nu }+n^{\mu
}\omega ^{\mu }, \nonumber \\
\omega ^{\mu } &=&h^{\mu \rho }m_{\sigma }\nabla _{\rho }n^{\sigma },  \notag
\end{eqnarray}%
and we define $\mathcal{K}^{ij}=\mathcal{K}^{\mu \nu }e_{\mu }^{i}e_{\nu
}^{j}$ and $\omega ^{i}=\omega ^{\mu }e_{\mu }^{i}$; $\omega _{i}=\omega
^{j}h_{ij}$.

Then, we find: 
\begin{eqnarray}
n_{\rho }n_{\nu }m_{\mu }S^{\mu \nu \rho } &=&n_{\nu }m_{\mu }\nabla ^{\nu
}n^{\mu }-\nabla _{\rho }n^{\rho } \nonumber \\ &=&-\nu +(\mathcal{K}+\nu )=\mathcal{K}, 
\notag \\
\sum_{i=1}^{2}n_{\rho }S^{\mu \nu \rho }2E_{\nu i}\delta E_{\mu }^{i} &=&
\left[ (\mathcal{K}+\nu )h^{\mu \nu }-\mathcal{K}^{\mu \nu }\right] \delta
h_{ij}  \nonumber\\ &&+2e^{-\sigma }\omega _{i}\delta \beta ^{i}.  \notag
\end{eqnarray}

Finally, we have: 
\begin{eqnarray}
\delta I^{(u)} &=& \frac{1}{2\kappa} \int_{\partial\mathcal{M}_u} N\sqrt{h}
\, \mathrm{d} \tau \, \mathrm{d}^2 \theta \, \left[ n_{\rho} S^{\mu \nu
\rho} 2E_{\nu I}\delta E_{\mu}^{I}\right] \\
&=& \frac{1}{2\kappa} \int_{\partial\mathcal{M}_u} N\sqrt{h} \, \mathrm{d}
\tau \, \mathrm{d}^2 \theta\, \left[(\mathcal{K}^{ij}-\mathcal{K}
h^{ij})\delta h_{ij} \right. \nonumber \\ && \left.+ 2\mathcal{K}\delta \sigma +
2e^{-\sigma}\omega_{i}\delta \beta^{i}\right].  \notag
\end{eqnarray}

\paragraph{Initial hypersurface:}

On the initial hypersurface $\partial \mathcal{M}_{I}$ is given by $\tau =0$
and hence $m^{\mu }$ is the unit normal to $\partial \mathcal{M}_{I}$. The
induced 3-metric on surfaces of constant $\tau $ is $\gamma _{\mu \nu
}=g_{\mu \nu }+m_{\mu }m_{\nu }$ where 
\begin{equation*}
\gamma _{\mu \nu }\,\mathrm{d}x^{\mu }\,\mathrm{d}x^{\nu }=e^{2\sigma }\,%
\mathrm{d}r^{2}+h_{ij}\left[ \,\mathrm{d}\theta ^{i}+K^{i}\,\mathrm{d}u%
\right] \left[ \,\mathrm{d}\theta ^{j}+K^{j}\,\mathrm{d}u\right] ,
\end{equation*}%
and $r=(u-w)/2$. The extrinsic curvature of $\gamma _{\mu \nu }$ is 
\begin{equation*}
K^{\mu \nu }=-\frac{1}{2}\gamma ^{\mu \rho }\gamma ^{\nu \sigma }\mathcal{L}%
_{m}\gamma _{\rho \sigma }=-\gamma ^{\mu \rho }\nabla _{\rho }m^{\nu },
\end{equation*}%
and $K^{\mu \nu }=K^{\nu \mu }$; $K^{\mu \nu }m_{\mu }=0$. Since $m^{\mu
}=E^{\mu 4}$: 
\begin{equation*}
\omega ^{\nu \mu \rho }m_{\rho }=\nabla ^{\nu }m^{\mu }=-K^{\mu \nu }-m^{\nu
}a^{\mu },
\end{equation*}%
where $a^{\mu }=m^{\nu }\nabla _{\nu }m^{\mu }$ is the acceleration; $a^{\mu
}m_{\mu }=0$. It follows that $\nabla _{\mu }m^{\mu }=-K=-K^{\mu \nu }\gamma
_{\mu \nu }$, and so using $e^{\sigma }\sqrt{h}=\sqrt{\gamma }$ and $\,%
\mathrm{d}^{3}x=\,\mathrm{d}r\,\mathrm{d}^{2}\theta $: 
\begin{eqnarray}
I_{\mathrm{GHY}}^{(I)} &=& \frac{1}{\kappa }\int_{\partial \mathcal{M}%
_{I}}e^{\sigma }\sqrt{h}\,\mathrm{d}r\,\mathrm{d}^{2}\theta \,\left[
-m_{\rho }\omega _{\mu }{}^{\mu \rho }\right] \nonumber \\ &=&\frac{1}{\kappa }%
\int_{\partial \mathcal{M}_{I}}\sqrt{\gamma }\,\mathrm{d}^{3}x\,K. \nonumber
\end{eqnarray}%
We also find that: 
\begin{equation*}
S^{\mu \nu \rho }m_{\rho }=-K^{\mu \nu }+K\gamma ^{\mu \nu }-m^{\nu }A^{\mu
}.
\end{equation*}%
where $A^{\mu }=\left[ a^{\mu }-\gamma ^{\mu }{}_{\rho }\omega _{\sigma
}{}^{\sigma \rho }\right] $ and so $A^{\mu }m_{\mu }=0$.

Thus, we have, 
\begin{eqnarray}
S^{\mu \nu \rho }m_{\rho }2E_{\nu I}\delta E_{\mu }^{I}&=&\left[ K\gamma ^{\mu
\nu }-K^{\mu \nu }\right] 2E_{(\nu I}\delta E_{\mu )}^{I} \nonumber \\ &=&\left[ K\gamma
^{\mu \nu }-K^{\mu \nu }\right] \delta \gamma _{\mu \nu },\nonumber
\end{eqnarray}%
where we have used $\delta m^{\mu }=m^{\mu }\delta \sigma $. The
contribution, $\delta I^{(I)}$, to the variation surface term from $\partial 
\mathcal{M}_{I}$ is therefore: 
\begin{eqnarray}
\delta I^{(I)} &=&\frac{1}{2\kappa }\int_{\partial \mathcal{M}_{I}}e^{\sigma
}\sqrt{h}\,\mathrm{d}u\,\mathrm{d}^{2}\theta \,\left[ m_{\rho }S^{\mu \nu
\rho }2E_{\nu I}\delta E_{\mu }^{I}\right] , \nonumber\\
&=&\frac{1}{2\kappa }\int_{\partial \mathcal{M}_{I}}\sqrt{\gamma }\,\mathrm{d%
}^{3}x\left[ K\gamma ^{\mu \nu }-K^{\mu \nu }\right] \delta \gamma _{\mu \nu
}.  \notag
\end{eqnarray}

\paragraph{GHY Term for the Full Boundary:}

Using the results derived above, the full GHY surface term for $\partial%
\mathcal{M} = \partial\mathcal{M}_{u}\cup \partial\mathcal{M}_{I}$ is: 
\begin{eqnarray}
I_{\mathrm{GHY}} &=& \frac{1}{\kappa}\int_{\partial\mathcal{M}_{u}}N\sqrt{h}\, 
\mathrm{d} \tau \, \mathrm{d}^2 \theta\,\left[\mathcal{K}+\nu\right] \\ &&+ \frac{1%
}{\kappa} \int_{\partial\mathcal{M}_{I}}\sqrt{\gamma}\, \mathrm{d}^3 x\, K, \nonumber
\end{eqnarray}
and 
\begin{eqnarray}
\delta (I_{EH}+I_{\mathrm{GHY}}) &=& -\frac{1}{2\kappa} \int_{\mathcal{M}} 
\sqrt{-g}\, \mathrm{d}^4 x\, G^{\mu\nu}\delta g_{\mu \nu} \\ &+& \frac{1}{2\kappa}
\int_{\partial\mathcal{M}_I} \sqrt{\gamma} \, \mathrm{d}^3 x \left[K
\gamma^{\mu \nu}-K^{\mu \nu}\right] \delta \gamma_{\mu \nu},  \notag \\
&+& \frac{1}{2\kappa} \int_{\partial\mathcal{M}_u} e^{\sigma}\sqrt{h} \, 
\mathrm{d} \tau \, \mathrm{d}^2 \theta\, \left[(\mathcal{K}^{ij}-\mathcal{K}
h^{ij})\delta h_{ij}\right. \nonumber \\ && \left. + 2\mathcal{K}\delta \sigma +
2e^{-\sigma}\omega_{i}\delta \beta^{i}\right].  \notag
\end{eqnarray}

\vspace{1em}
\subsection{York's Cosmological (YC) Surface Term}

In Ref. \cite{York:1972sj}, York also considers a `cosmological' surface
term, $I_{\mathrm{YC}}$, in the action. On $\partial \mathcal{M}_{I}$ this
is: 
\begin{equation*}
I_{\mathrm{YC}}^{(I)}=\frac{1}{3\kappa }\int_{\partial \mathcal{M}_{I}}\sqrt{%
\gamma }\,\mathrm{d}^{3}x\,K=I_{\mathrm{GHY}}^{(I)}-\frac{2}{3\kappa }%
\int_{\partial \mathcal{M}_{I}}\sqrt{\gamma }\,\mathrm{d}^{3}x\,K.
\end{equation*}%
Thus, we have 
\begin{equation*}
I_{\mathrm{YC}}^{(I)}=\delta I_{\mathrm{GHY}}^{(I)}-\frac{1}{2\kappa }%
\int_{\partial \mathcal{M}_{I}}\sqrt{\gamma }\,\mathrm{d}^{3}x\,\left[ \frac{%
2}{3}K\gamma ^{\mu \nu }+\frac{4}{3}\delta K\right] .
\end{equation*}%
Hence, if we define $I_{\mathrm{grav}}=I_{\mathrm{EH}}+I_{\mathrm{GHY}%
}^{(u)}+I_{\mathrm{GHY}}^{(I)}$ we have: 
\begin{eqnarray}
\delta I_{\mathrm{grav}} &=&-\frac{1}{2\kappa }\int_{\mathcal{M}}\sqrt{-g}\,%
\mathrm{d}^{4}x\,G^{\mu \nu }\delta g_{\mu \nu } \\ &&-\frac{1}{2\kappa }%
\int_{\partial \mathcal{M}_{I}}\sqrt{\gamma }\,\mathrm{d}^{3}x\,\left[ 
\tilde{P}^{\mu \nu }\delta \tilde{\gamma}_{\mu \nu }+\frac{4}{3}\delta K%
\right] , \nonumber \\
&&+\frac{1}{2\kappa }\int_{\partial \mathcal{M}_{u}}e^{\sigma }\sqrt{h}\,%
\mathrm{d}\tau \,\mathrm{d}^{2}\theta \,\left[ (\mathcal{K}^{ij}-\mathcal{K}%
h^{ij})\delta h_{ij}\right. \nonumber \\ && \left.+2\mathcal{K}\delta \sigma +2e^{-\sigma }\omega
_{i}\delta \beta ^{i}\right] ,  \notag
\end{eqnarray}%
where 
\begin{eqnarray}
\tilde{P}^{\mu \nu }&=&(\mathrm{det}\,\gamma )^{5/6}\left[ K^{\mu \nu }-\frac{1%
}{3}K\gamma ^{\mu \nu }\right] ,\nonumber \\ \tilde{\gamma}_{\mu \nu }&=&(\mathrm{det%
}\,\gamma )^{-1/3}\gamma _{\mu \nu }. \nonumber
\end{eqnarray}

\end{document}